\documentstyle[preprint,amsfonts,amssymb,aps,12pt]{revtex}
\input{epsf}
\newcommand{\drawxyzaxes}[2]{
\setlength{\unitlength}{0.5mm}
\begin{picture}(0, 0)(-#1, -#2)
\thicklines
\put(2,40){\Large $x$}
\put(0,0){\line(0,1){40}}

\put(-16.66666,14){\Large $y$}
\put(0,0){\line(-3,2){20}}

\put(-40,-10){\Large $z$}
\put(0,0){\line(-5,-2){40}}
\end{picture}
}
\tighten

\begin{document}
\draft
%
%
\title{Recent MD Results on Supercooled Thin Polymer Films}
\author{F. Varnik$^1$%
\footnote{To whom correspondence should be addressed. Email:
{\sf varnik@mail.uni-mainz.de}},
J. Baschnagel$^2$, K. Binder$^1$\\[2mm]}
\address{$^1$Institut f\"ur Physik, Johannes-Gutenberg Universit\"at, D-55099 Mainz, Germany}
\address{$^2$Institut Charles Sadron, 6 rue Boussingault, F-67083, Strasbourg Cedex, France}
\maketitle
%
%
%
%
\newcommand{\mbf}[1]{{\mbox{\boldmath$#1$\unboldmath}}}
\newcommand{\qpar}{{\mbf{q}_{\parallel}}}
\newcommand{\ripar}{{\mbf{r}_{i, \parallel}}}
\newcommand{\riper}{{\mbf{r}_{i, \bot}}}
\newcommand{\rpar}{{\mbf{r}_{\parallel}}}
\newcommand{\rper}{{\mbf{r}_{\bot}}}
\newcommand{\myfrac}[2]{{\frac{ \displaystyle {#1}} {\displaystyle {#2}}}}
\newcommand{\mr}[1]{{\rm #1}}
\newcommand{\mrm}[1]{{\rm #1}}
\newcommand{\myeq}{\!=\!}
\newcommand{\myapprox}{\!\approx\!}
\newcommand{\kB}{k_{\rm B}}
\newcommand{\Tc}{T_{\rm c}}
\newcommand{\Tf}{T_{\rm f}}
\newcommand{\D}{{\rm d}}
\newcommand{\Tnull}{T_{\rm 0}}
\newcommand{\Tcbulk}{T^{\rm bulk}_{\rm c}}
\newcommand{\Tnullbulk}{T^{\rm bulk}_{\rm 0}}
\newcommand{\Tg}{T_{\rm g}}
\newcommand{\Np}{N_{\rm p}}
\newcommand{\Omegas}{\Omega_{\rm s}}
\newcommand{\phis}{\phi^{\rm s}_{q}}
\newcommand{\fqs}{f^{\rm s}_{q}}
\newcommand{\hqs}{h^{\rm s}_{q}}
\newcommand{\betaK}{\beta^{\rm K}}
\newcommand{\gammag}{\gamma_{\rm g}}
\newcommand{\gammaq}{\gamma_{\rm q}}
\newcommand{\RgDebye}{R_{\rm g}^{\rm Debye}}
\newcommand{\RgDebyesquare}{R_{\rm g}^{2~\rm Debye}}
\newcommand{\Rg}{R_{\rm g}}
\newcommand{\Rgbulksquare}{R_{\rm g}^{2~\rm bulk}}
\newcommand{\Rgbulk}{R_{\rm g}^{\rm bulk}}
\newcommand{\Reebulk}{R_{\rm ee}^{\rm bulk}}
\newcommand{\Reebulksquare}{R_{\rm ee}^{2~\rm bulk}}
\newcommand{\lambdapar}{\lambda_{\parallel}}
\newcommand{\lambdaper}{\lambda_{\bot}}
\newcommand{\Rgpar}{R^{2}_{\rm g,\parallel}}
\newcommand{\Rgper}{R^{2}_{\rm g,\bot}}
\newcommand{\Sc}{S^{\rm c}}
\newcommand{\Rgsquare}{R^2_{\rm g}}
\newcommand{\Ree}{R_{\rm ee}}
\newcommand{\Reepar}{R^{2}_{\rm ee,\parallel}}
\newcommand{\Reeper}{R^{2}_{\rm ee,\bot}}
\newcommand{\Reesquare}{R^2_{\rm ee}}
\newcommand{\tauq}{\tau_{\rm q}}
\newcommand{\Fq}{F_{\rm q}}
\newcommand{\PNext}{P_{\rm N,ext}}
\newcommand{\PText}{P_{\rm T,ext}}
\renewcommand{\vec}[1]{\mbox{\boldmath$#1$\unboldmath}}
%
%
\begin{abstract}
The dynamic and static properties of a supercooled (non-entangled) 
polymer melt are investigated via molecular dynamics (MD) simulations. 
The system is confined between two completely smooth and purely 
repulsive walls. The wall-to-wall separation (film thickness),
$D$, is varied from  about 3 to about 14 times the bulk radius of 
gyration. Despite the geometric confinement, the supercooled films
exhibit many qualitative features which were also observed in the bulk 
and could be analyzed in terms of mode-coupling theory (MCT). Examples are
the two-step relaxation of the incoherent intermediate scattering 
function, the time-temperature superposition property of the late time
$\alpha$-process and the space-time factorization of the scattering
function on the intermediate time scale of the MCT $\beta$-process. An
analysis of the temperature dependence of the $\alpha$-relaxation time 
suggests that the critical temperature, $\Tc$, of MCT decreases with $D$.
If the confinement is not too strong ($D \ge 10 \,\mbox{monomer diameter}$)
the static structure factor of the film coincides with that of the bulk
when compared for the same distance, $T-\Tc(D)$, to the critical temperature.
This suggests that $T-\Tc(D)$ is an important temperature scale of our
model both in the bulk and in the films.
\end{abstract}
%
%
\pacs{{\sf PACS}: 61.20.Ja,61.25.Hq,64.70.Pf\\[1mm]%
contribution to {\em EPJE Special Issue on Properties of Thin Polymer Film}\\%
(guest editor: James Forrest)
}
\section{Introduction}
\label{section::introduction}
Glass forming materials have been used by mankind since the very 
early days of civilization. An example are early ceramics from the neolithic
period dating back to $5000$ B.C. In addition to conventional glasses 
like those used for windows, bottles, etc., polymers represent a new class 
of glassy systems with a large variety of thermal and elastic properties.
Due to their (in general) low thermal conductivity, polymers are utilized as
protective coatings in (micro-) electronic devices, optical fibres and 
other thermally fragile materials~\cite{Zhang::PolEngSci::1999,%
Rayss::JApplPolSci::1993,%
Armstrong::ElectrochemicaActa::1993}. Furthermore, the
high dielectric constant of some polymeric compounds suggests
their application as porter of electric circuits, a new field under 
evolution~\cite{Garbassi-Morra-Occhiello::POLYMER::SURFACES,%
Laetsch-Hiraoka-Baron::MRS::385::1995}. In all these applications, 
the polymer is in contact with another material. This gives rise
to an interface. It is therefore important to know whether and how
the interface alters the properties of the system. In particular, one would 
like to understand to what extent the glass transition temperature is 
influenced by the polymer-substrate interactions.
Hence, the study of the glass transition in thin polymer films 
finds a strong motivation from the technological side.

In addition to the technological importance, the study of the
glass transition in thin polymer films is also of fundamental interest.
It could help to elucidate the nature of this phenomenon. According to 
Adam and Gibbs~\cite{Adam-Gibbs::JCP::43}, structural relaxation near the glass 
transition can take place only if many particles move in a 
correlated way to allow a collective rearrangement.
The average size of such a ``cooperatively rearranging region''
defines a dynamic correlation length $\xi$. The larger $\xi$, the smaller the 
probability of a cooperative motion and thus the longer the structural relaxation time
are supposed to be. If one assumes that $\xi$ diverges at some temperature (Vogel-Fulcher-Tammann
temperature) below $\Tg$, the glass transition can be considered as a thermodynamic second 
order phase transition.

However, the cooperatively rearranging regions are not directly correlated with static density fluctuations.  The structure of a glass former changes only slightly upon cooling contrary to the hypothesized strong increase of $\xi$.  As a consequence, scattering experiments cannot be used to determine $\xi(T)$~\cite{Forrest::2000}.  One has to resort to indirect investigations.  A possibility consists in studying confined systems.  If the system size, $L$, is finite, one could expect that the cooperatively rearranging region cannot grow beyond any bound, its largest extension being $\xi\myeq L< \infty$.  Therefore, the glass transition temperature should depend on $L$ and decrease with system size. 

In thin polymer films it is the film thickness, $D$, which is finite
and thus one expects $\Tg\myeq \Tg(D)$. Indeed, one finds a dependence 
of $\Tg$ on $D$ in both experiments~\cite{Forrest::PRL77::page2002,%
Keddie::Jones::Cory::EuroPhysLett28,%
Keddie::FaradayDiscuss98,%
Forrest::PRL77::page4108,%
ForrestJones2000,%
Forrest::PRE56} and computer 
simulations~\cite{Torres-Nealey-dePablo::PRL85}.
However, the observed change of $\Tg$ with $D$ 
strongly depends on the system under consideration.
If the interaction between the polymers and the substrate is attractive, 
the glass transition temperature of the films becomes 
larger than the bulk value for small film
thicknesses~\cite{Keddie::FaradayDiscuss98}.
Intuitively, this effect can be attributed to chains which are close 
enough to the substrate to `feel' the attractive interaction.
The motion of these chains should be slowed down with respect 
to the bulk. In a thin film almost all chains touch the attractive 
substrate. So, $T_\mr{g}$ should increase.

On the other hand, measurements (by ellipsometry) of polystyrene (PS) films (of rather large 
molecular weights) on a silicon substrate showed a significant decrease of $\Tg$ 
from $375\,{\rm K}$ down to $345\,{\rm K}$ for the smallest film 
thickness of $10$ nm, i.e., a relative change of $10\%$ in $\Tg$ was 
observed~\cite{Keddie::Jones::Cory::EuroPhysLett28}.  There have also been 
many experiments in recent years on freely standing 
polystyrene films (i.e., no solid substrate, but two polymer-air 
interfaces)~\cite{%
Forrest::PRL77::page2002,%
Forrest::PRL77::page4108,%
ForrestJones2000,%
Forrest::PRE56} 
exhibiting a dramatic decrease of $T_\mr{g}$ by up to $20\%$ if the film thickness 
becomes much smaller than the chain size. 
An interesting explanation of this observation in terms of an interplay 
between polymer-specific properties and free-volume concepts has been 
proposed~\cite{deGennses::EurPhysJE2000}.

This sensitive dependence of $\Tg$ on the polymer-substrate interaction was also obtained from
computer simulations~\cite{Torres-Nealey-dePablo::PRL85}.  One finds that a strongly attractive 
wall leads to an increase of $\Tg$, whereas a weaker attraction has the opposite
effect. Therefore, for repulsive walls, a decrease of $\Tg$ can be expected, provided that 
other parameters of the system, in particular the average density, are 
unaffected by the confinement.  As the simulations of our system in the bulk were done at
constant external pressure~\cite{%
Bennemann-Paul-Binder-Duenweg::PRE57,%
Bennemann-Baschnagel-Paul::EPJB10}, we also carried out the simulations of the films at constant 
normal pressure in order to obtain an average density that lies as close as possible to the bulk 
density at the same temperature. This allows to separate the effect of confinement from that of 
the density.

Note however that the glass transition temperature is an empirical quantity.  It is usually defined 
as the temperature at which the viscosity reaches a value of $10^{13}$ poise.
So, the choice of another number would give a different value of $\Tg$. Furthermore, 
$\Tg$ depends on the cooling rate so that it is not a temperature in a strict thermodynamic sense. 
Therefore, other temperatures have been introduced to characterize the glass transition.
One example is the so-called Vogel-Fulcher-Tammann 
(VFT)-temperature, $T_0$, at which the system viscosity seems to
diverge. This quantity is obtained by fitting the 
viscosity, $\eta(T)$, to the empirical function, 
$\eta(T) \myeq \eta_0\exp[E/(T-T_0)]$.  An attempt to rationalize the VFT-law
is the free volume theory~\cite{Cohen-Turnbull::JCP::31,%
Cohen-Turnbull::JCP::34,%
Cohen-Turnbull::JCP::52,%
Cohen-Grest::PhysRevE1979}.
The main idea of this approach is that a tagged particle can leave its 
initial position only if it finds a ``free volume'' of size 
$v_{\mrm{f}} \! \ge \! v_{\mrm{c}}$ in its neighborhood
($v_{\mrm{c}}$ being some critical volume of the order 
of the size of a molecule).
It is further supposed that the average free volume
vanishes at $T_0$. Assuming statistical independence of the free volumes
and using a Taylor expansion of the average free volume
around $T_0$, one obtains the VFT-law for transport coefficients
like the viscosity or the (inverse) diffusion coefficient.
This brief description illustrates that the free volume theory has a  
phenomenological character. The precise meaning of the free volume
is unclear and the existence of $T_0$, where $v_{\rm f}$ is supposed to vanish,
is not derived, but postulated.

Contrary to the VFT-temperature, the critical temperature of 
the so-called mode-coupling theory (MCT) results from a microscopic approach
to the dynamics of supercooled (simple) liquids~\cite{Goetze::LesHouches::1989,%
Goetze-Sjoegren::TransportTheoryStatPhys,%
Goetze-Sjoegren::RepProgPhys55}.
Within the idealized version of the theory, the structural relaxation time diverges at
a critical temperature $\Tc$, while the static properties of the system remain liquid-like. 
This implies that the system vitrifies at $\Tc$. Thus, from the point of view of MCT, the glass 
transition is a purely dynamic phenomenon.  However, comparisons between the 
theory and experiments \cite{Goetze-Sjoegren::RepProgPhys55,Goetze::JPCM::1999} reveal that
$\Tc$ lies in the region of the supercooled liquid, where the glass former is only moderately
viscous and no absolute freezing occurs.  In reality, there are additional relaxation 
mechanisms which are not incorporated in the idealized MCT and begin to dominate close to and
especially below $\Tc$.  An attempt was made to approximately include these relaxation processes 
in an extended version of MCT \cite{Goetze-Sjoegren::TransportTheoryStatPhys}, but the validity
of this extension is currently unclear \cite{GoetzeVoigtmann::PRE2000}.  Nevertheless, the 
idealized MCT derives several empirically known phenomena, such as the stretching of the 
$\alpha$-relaxation, which is often described by the Kohlrausch-Williams-Watts (KWW)-law 
($\propto \exp[-(t/\tau)^{\beta}]$, where $\beta\!<\!1$), or the time-temperature superposition 
principle. Furthermore, it also makes new predictions, such as the space-time factorization 
property, which have been tested both in experiments 
\cite{Goetze-Sjoegren::RepProgPhys55,Goetze::JPCM::1999} and computer simulations 
\cite{Kob::JPhysCondensedMatt::1999}.

An application of MCT to our polymer model in the bulk \cite{Bennemann-Baschnagel-Paul::EPJB10,%
Bennemann-Paul-Baschnagel-Binder::JPCM11,%
Bennemann-Baschnagel-Paul-Binder,%
Aichele-Baschnagel::EurPhysJE::I::D,%
Aichele-Baschnagel::EurPhysJE::II::D}
revealed that the theory represents a suitable framework to analyze the dynamics of the supercooled melt.  Therefore, we attempt to test whether MCT can also be applied to the supercooled polymer films.  We find that even for an extremely thin film of three monomer layers, some features of the dynamics at low temperatures can be successfully described by the MCT.  This analysis yields $\Tc$ as a function of film thickness and shows that $\Tc$ decreases with $D$.  Within the error bars the Vogel-Fulcher-Tammann temperature exhibits the same $D$-dependence so that we also expect the glass transition temperature of our model to behave analogously due to $T_0(D) \le \Tg(D) \le \Tc(D)$.

The paper is organized as follows: After a presentation of the model in section~\ref{section::model}, 
section~\ref{section:statics} focuses on the influence of the confinement on static properties of the system.  A discussion of the dynamics is the subject of section~\ref{section:Dynamics}. In section~\ref{section:Determination::of::Tc} we determine the dependence of $\Tc$ and $T_0$ on film thickness and the last section summarizes our conclusions.
\section{Model}
\label{section::model}
We study a Lennard-Jones (LJ) model for a dense polymer
melt~\cite{Bennemann-Paul-Binder-Duenweg::PRE57,Kremer-Grest::JCP92}
of short chains (each consisting of 10 monomers) embedded between two completely
smooth, impenetrable 
walls~\cite{Varnik-Baschnagel-Binder::JPhysIV10::ConfProc,%
Varnik-Baschnagel-Binder::JCP113::2000}.
Two potentials are used for the interaction between particles. The first one is a 
truncated and shifted LJ-potential which acts between all pairs of particles
regardless of whether they are connected or not,
$$
U_{\mbox{\scriptsize LJ-ts}}(r)=\left\{ 
\begin{array}{ll}
U_{\mr{LJ}}(r)-U_{\mr{LJ}}(r_{\mr{c}}) & \mbox{if $r<r_{\mathrm c}$}\;, \\
0 & \mbox{otherwise}\;,
\end{array}
\right.
$$
where 
$U_{\mr{LJ}}(r)\myeq  4 \epsilon \Big[ (\sigma/r)^{12} -(\sigma/r)^{6}\Big]$
and $r_{\mr{c}}\myeq 2\times 2^{1/6}\sigma$.
The connectivity between adjacent monomers of a chain is ensured by a
FENE-potential~\cite{Kremer-Grest::JCP92},
\begin{equation}
U_{\mathrm{FENE}}(r)=-\frac{k}{2} R^2_0 \ln \bigg[1-\Big 
(\frac{r}{R_0}\Big)^2\bigg]
\;,
\label{eq:FENE::potential::def}
\end{equation}
where $k\myeq 30 \epsilon/\sigma^2$ is the strength factor 
and  $R_0\myeq 1.5\sigma$ the maximum 
allowed length of a bond. The wall potential was chosen as
\begin{equation}
U_{\mathrm W}(z)=\epsilon \bigg( \frac{\sigma}{z}\bigg )^9 \; ,
\label{eq:def:uw}
\end{equation}
where $z\myeq |z_{\mbox{\scriptsize particle}}-z_{\mbox{\scriptsize wall}}|$
($z_{\mbox{\scriptsize wall}}\myeq  \pm D/2$). This corresponds to 
an infinitely thick wall made of infinitely small particles which interact
with inner particles via the potential 
$45\epsilon(\sigma/r)^{12}/(4\pi\rho_{\mathrm{wall}}\sigma^3)$
where $\rho_{\mathrm{wall}}$ denotes the density of wall particles. The sum 
over the wall particles then yields $\epsilon (\sigma/z)^{9}$.
All simulation results are given in Lennard-Jones (LJ) units.
All lengths and energies are measured respectively in units of $\sigma$ and $\epsilon$,
temperature in units of $\epsilon/\kB$ ($\kB \myeq 1$) and time in units 
of $(m\sigma^2/\epsilon)^{1/2}$.

\begin{figure}
\begin{minipage}[h]{90mm}
\unitlength=1mm
\begin{picture}(90,90)
\put(-7,8){
\epsfysize=90mm 
\epsffile{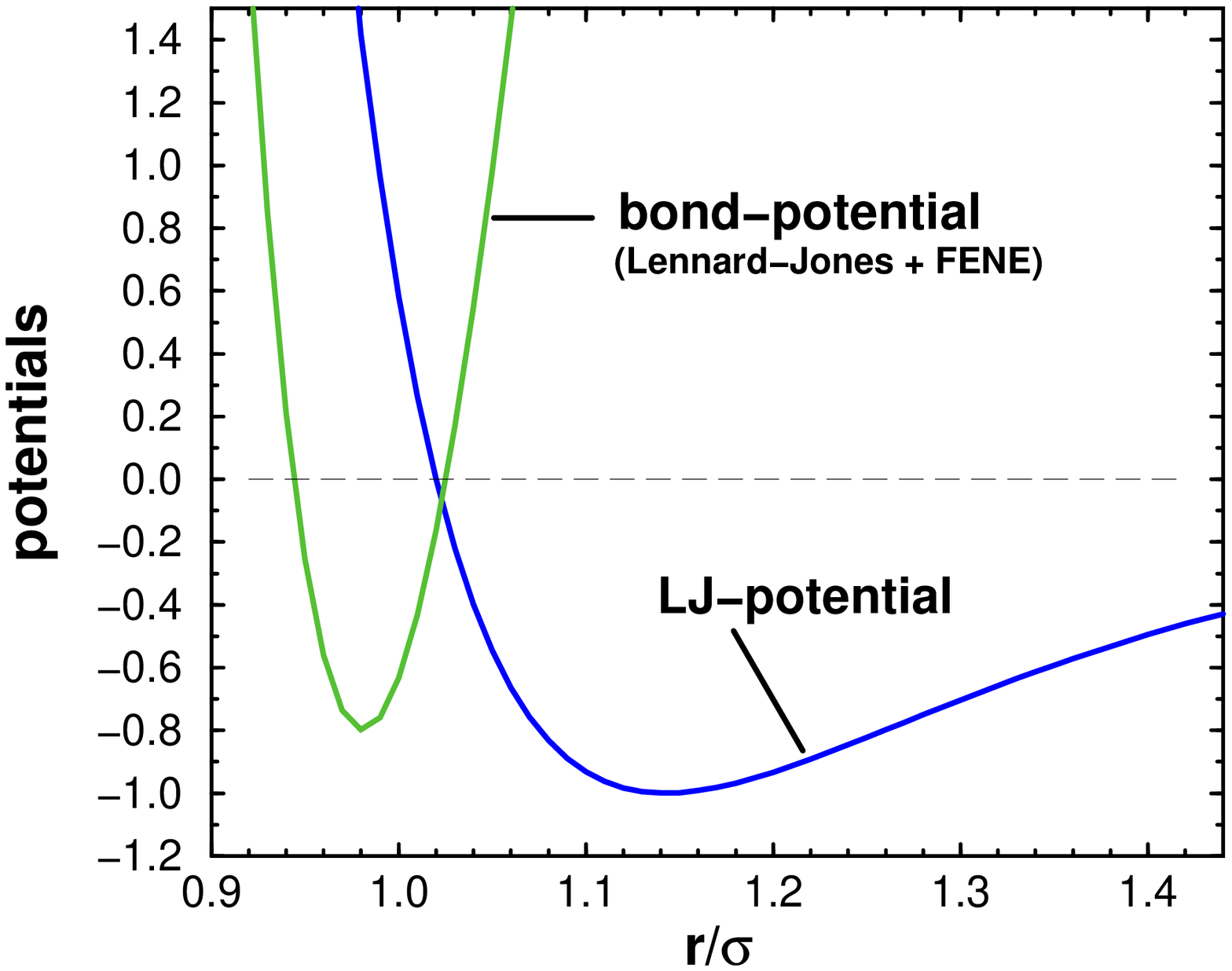}
}
\end{picture}
\end{minipage}
\hfill
\begin{minipage}[t]{55mm}
\begin{picture}(55,55)
\unitlength=1mm
\put(-5, -15){
\epsfysize=52mm 
\epsffile{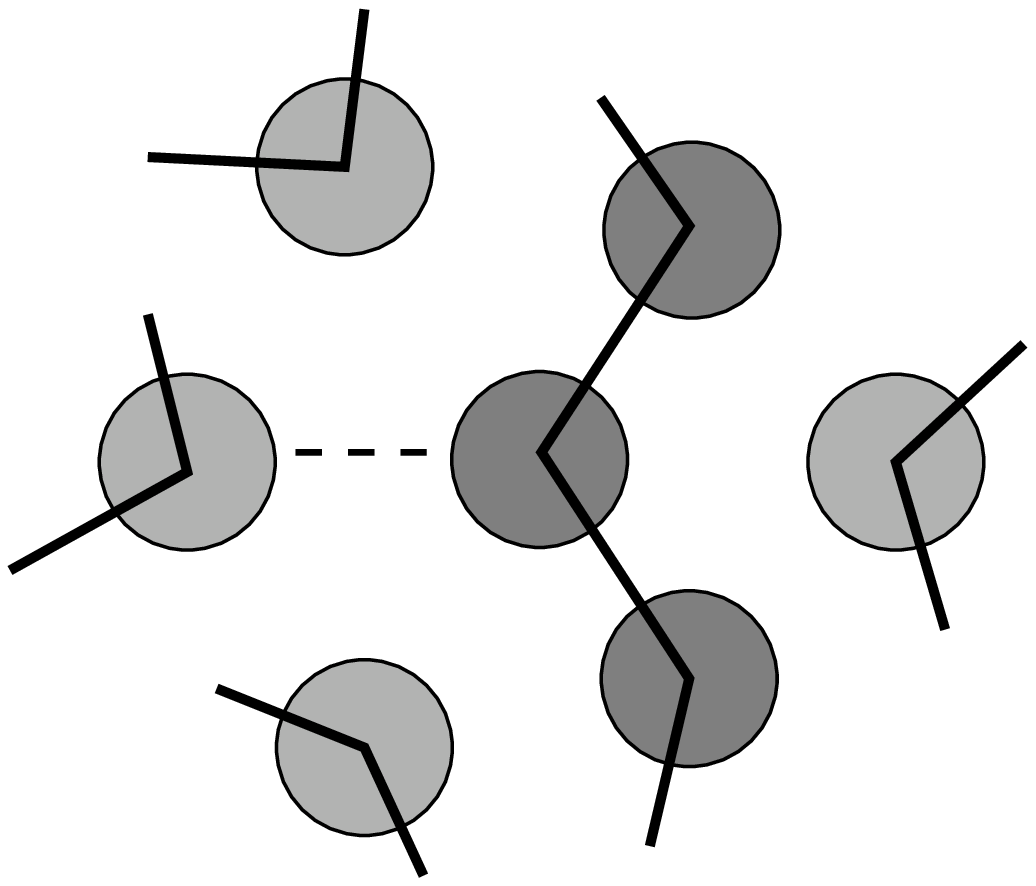}
\put(-41,28){\makebox(0,0)[lb]{{\small \sf LJ}}}
\put(-35,16){\makebox(0,0)[lb]{{\small \sf bond}}}
}
\end{picture}
\end{minipage}
\vspace{-5mm}
\caption[]{Left panel: Lennard-Jones potential (LJ) versus the bond 
potential, the sum of the LJ and FENE potentials. The bond potential is
shifted by -20 to lower values for the sake of comparison with the 
LJ potential.  The minimum position of the 
bond-potential is smaller than that of the LJ-potential:
The bond potential has its minimum at $r \simeq 0.96\sigma$,
whereas that of a pure LJ-potential lies at $r\!=\!\sqrt[6]{2}\sigma$.
Due to the incompatibility of these length scales and due to the flexibility of our
model (no bond angle or torsion potentials), one expects that cooling the
system would not lead to crystallization, but maintain the amorphous structure
typical of the liquid phase. This expectation is confirmed by the behavior of static
structure factor [see lower panel of Fig.~\ref{fig:fig2}].
Right panel: Schematic visualization of the local distortion of the structure due
to the two intrinsic length scales of the model.
}
\label{fig:fig1}
\end{figure}

The left panel of Fig.~\ref{fig:fig1} compares the bond potential,
i.e.\ the sum of LJ- and FENE-potentials, with the LJ-potential.
It shows that the bonded monomers prefer shorter distances than the non-bonded ones. 
Thus, our model contains two intrinsic length scales (see right 
panel of Fig.~\ref{fig:fig1}
for a schematic illustration). Since these length scales
are chosen to be incompatible with a (fcc or bcc) crystalline structure and since
our chains are flexible (no bond angle or torsion potentials),
one could expect that the system does not crystallize at
low temperatures, but remains amorphous \cite{MeyerMuellerPlathe::JCP::2001}. 

\begin{figure}
\epsfxsize=105mm 
\hspace*{15mm}\epsffile{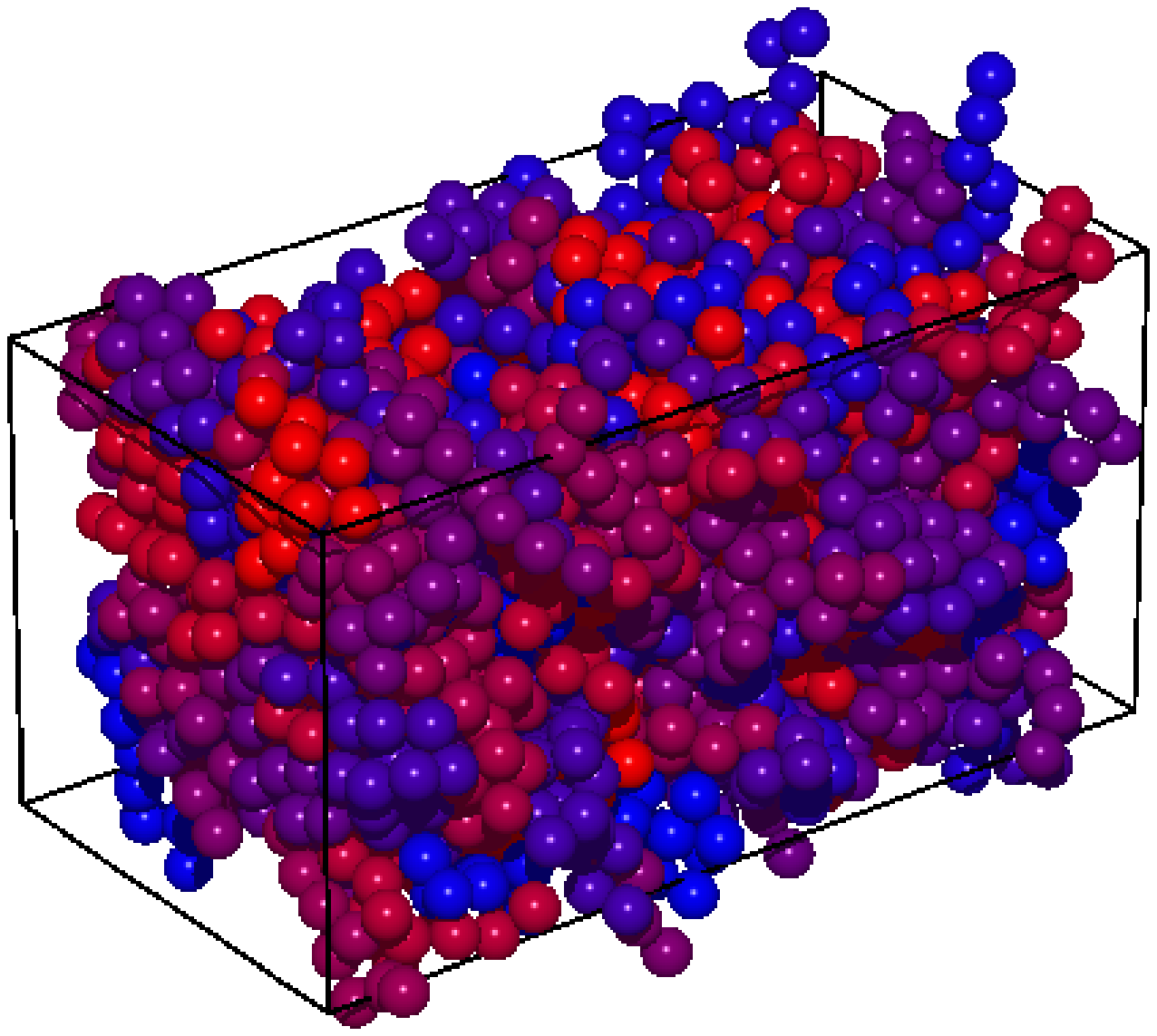}
\epsfxsize=100mm
\hspace*{25mm}\epsffile{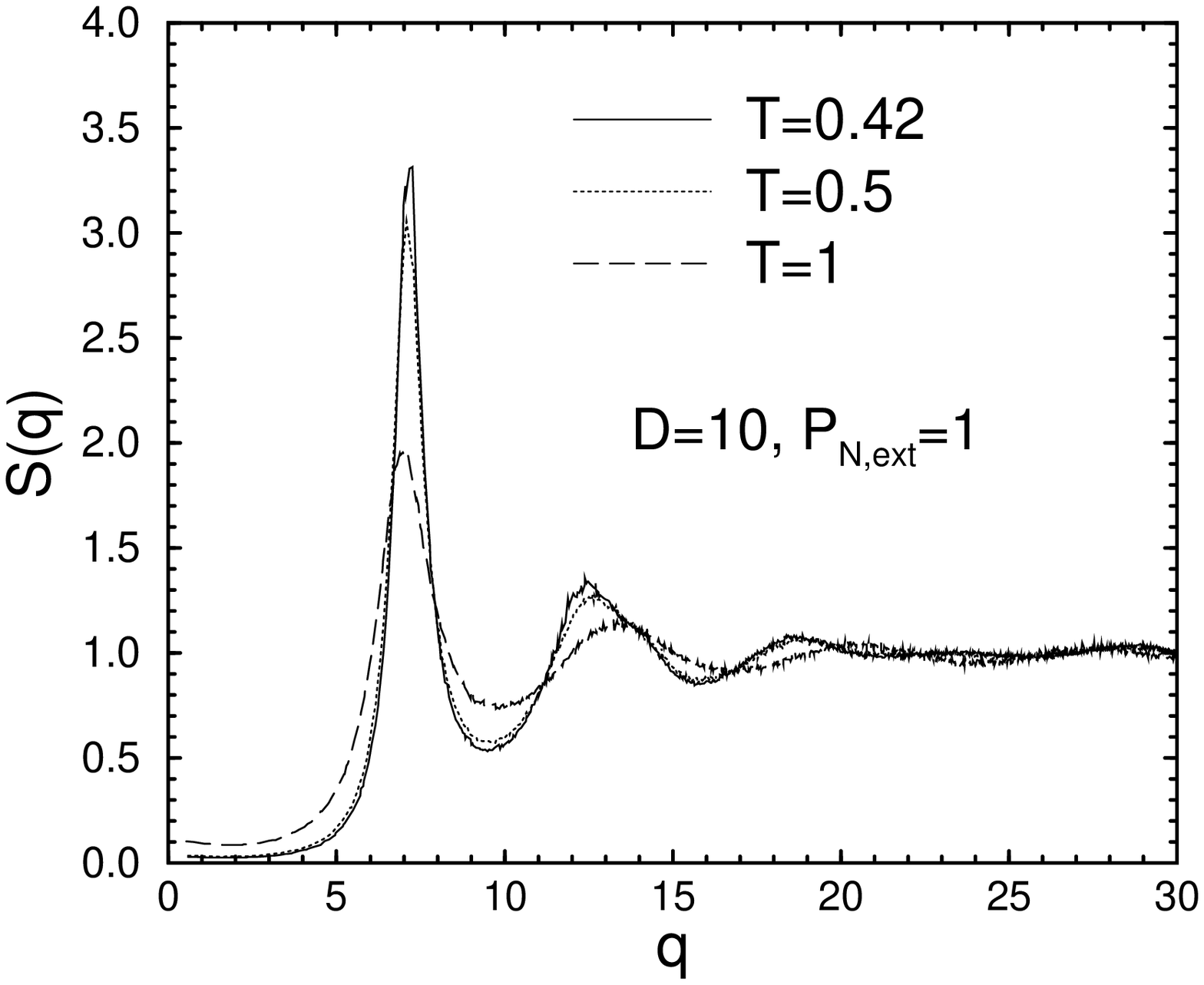}
\caption[]{
Upper panel: 
A snapshot of the simulation 
box for a film of thickness $D\myeq 20$ at $T\myeq 0.44$. Note that the mode 
coupling critical temperature of the system at this film thickness 
is $\Tc(D\myeq 20) \myeq 0.415$
(see text and also~\cite{Varnik-Baschnagel-Binder::PRE::2001}).
A temperature of $T\myeq 0.44$ thus corresponds to the supercooled 
regime at this film thickness. To visualize the chain structure, 
a continuous color code is used, monomers of a given chain having the same color. 
Lower panel: 
Structure factor $S(q)$ versus the modulus, $q$, of the wave vector for a 
film of thickness $D\myeq 10$ at three temperatures: 
$T\myeq 0.42,\; 0.5$, and $T\myeq 1$ 
(the critical temperature of mode-coupling theory for this film thickness is
$\Tc(D\myeq 10) \simeq 0.39 \pm 0.005$, see Table~\ref{tab1}).
At low temperatures, the peak of the structure factor is more pronounced
and its position is slightly shifted towards larger $q$. This is 
consistent with the fact that, at lower temperatures, the average 
interparticle distance decreases. Otherwise, $S(q)$ does not develop 
sharp peaks for larger $q$ at lower temperatures, thus indicating that 
the film remains amorphous.
}
\label{fig:fig2}
\end{figure}

This expectation is well borne out, as Fig.~\ref{fig:fig2} illustrates. The upper part of the
figure shows a snapshot of a film of thickness $D\myeq 20$ and at temperature $T\myeq 0.44$
which corresponds to the supercooled state. The visual inspection of this configuration
suggests that the structure is disordered.  This is corroborated by an analysis of the static
structure factor $S(q)$. 

The lower part of Fig.~\ref{fig:fig2} shows $S(q)$ for a film of  thickness $D\myeq 10$ at 
temperatures corresponding to the normal liquid 
state and to the supercooled state. $S(q)$ is calculated parallel to 
the walls [i.e.\ $q\myeq |\mbf{q}|,\; \mbf{q}\myeq (q_x,q_y)$] 
by averaging over all monomers in the system. 
Figure~\ref{fig:fig2} shows that the maximum value of the $S(q)$ increases at lower temperature
and that its position is slightly shifted towards larger $q$-values. Thus, the average interparticle 
distance decreases with decreasing temperature, since the density of the film increases in our
simulations at constant pressure \cite{Varnik-Baschnagel-Binder::JCP113::2000}.
However, these structural changes are rather
small. Even at a very low temperature of $T\myeq 0.42$ 
which is quite close to the critical temperature of the system at this 
film thickness ($\Tc(D \myeq10) \simeq 0.39$),
{\em no qualitative difference} is observed between the structure
factors at low and high temperatures. The packing
of the system thus remains liquid-like (i.e.\ amorphous)
at all studied temperatures.

All simulations have been carried out at constant normal pressure $\PNext\myeq p \myeq  1$ 
\cite{Varnik-Baschnagel-Binder::JCP113::2000}. However, to adjust the 
normal pressure, we do \emph{not} vary the wall-to-wall separation, 
$D$, but the surface area. For each temperature, the 
average surface area is calculated by an iterative 
approach~\cite{Varnik-Baschnagel-Binder::JCompPhys::2001}.
The system is then propagated until the instantaneous 
surface area reaches the computed average value. 
At this point the surface area (and thus the volume)
is fixed and a production run is started in $NVT$-ensemble, where
the temperature is adjusted using the Nos{\'e}-Hoover 
thermostat~\cite{Nose::JCP81,Hoover::PhysRevA31}.
This thermostat slows down or accelerates all particles
depending on the sign of the difference between the instantaneous
kinetic energy of the system and the desired value imposed, 
i.e.\  $3N\kB T/2$ ($N$ is the number of 
particles)~\cite{Nose::JCP81,%
Hoover::PhysRevA31,%
Nose-Klein::MolPhys50,%
Hoover::PhysRevLett48,%
Hoover::PhysRevA34,%
Nose::MolPhys52}.

One may therefore ask how reliable the resulting dynamics is when 
compared to pure Newtonian dynamics in the microcanonical ($NVE$) ensemble.
This question was already examined for the bulk system 
in Ref.~\cite{Bennemann-Paul-Binder-Duenweg::PRE57} and for 
our confined model
in Refs.~\cite{Varnik-Baschnagel-Binder::PRE::2001,Varnik::Dissertation::Mainz2000}.
In both cases, the results obtained from constant 
energy ($NVE$) simulations and from 
the Nos\'e-Hoover thermostat are identical.
More details about the applied simulation techniques can be found 
in~\cite{Varnik-Baschnagel-Binder::JCP113::2000,%
Varnik-Baschnagel-Binder::JCompPhys::2001,%
Varnik::Dissertation::Mainz2000}.
\section{Static Properties}
\label{section:statics}
In this section, we want to discuss the influence of 
confinement on both the chain conformation and the static 
properties of the melt. When computed within thin layers 
parallel to the wall, the structure of a chain varies with 
the distance from the wall. On the other hand, the chain structure 
averaged over the whole film does not depend much on film 
thickness. Contrary to this insensibility of the average chain 
conformation to the confinement, the dense packing of the melt exhibits 
a pronounced dependence on $D$.
\subsection{Effects of the Confinement on the Structure of a Chain}
\label{subsection:static::chain}
Let us first look at the $T$-dependence of the single chain structure 
factor $\Sc(q)$. Similarly to $S(q)$, $\Sc(q)$ is also calculated for 
$q$-vectors in direction parallel to the walls and by averaging over 
all chains in the system.

\begin{figure}
\epsfxsize=95mm 
\centering{\epsffile{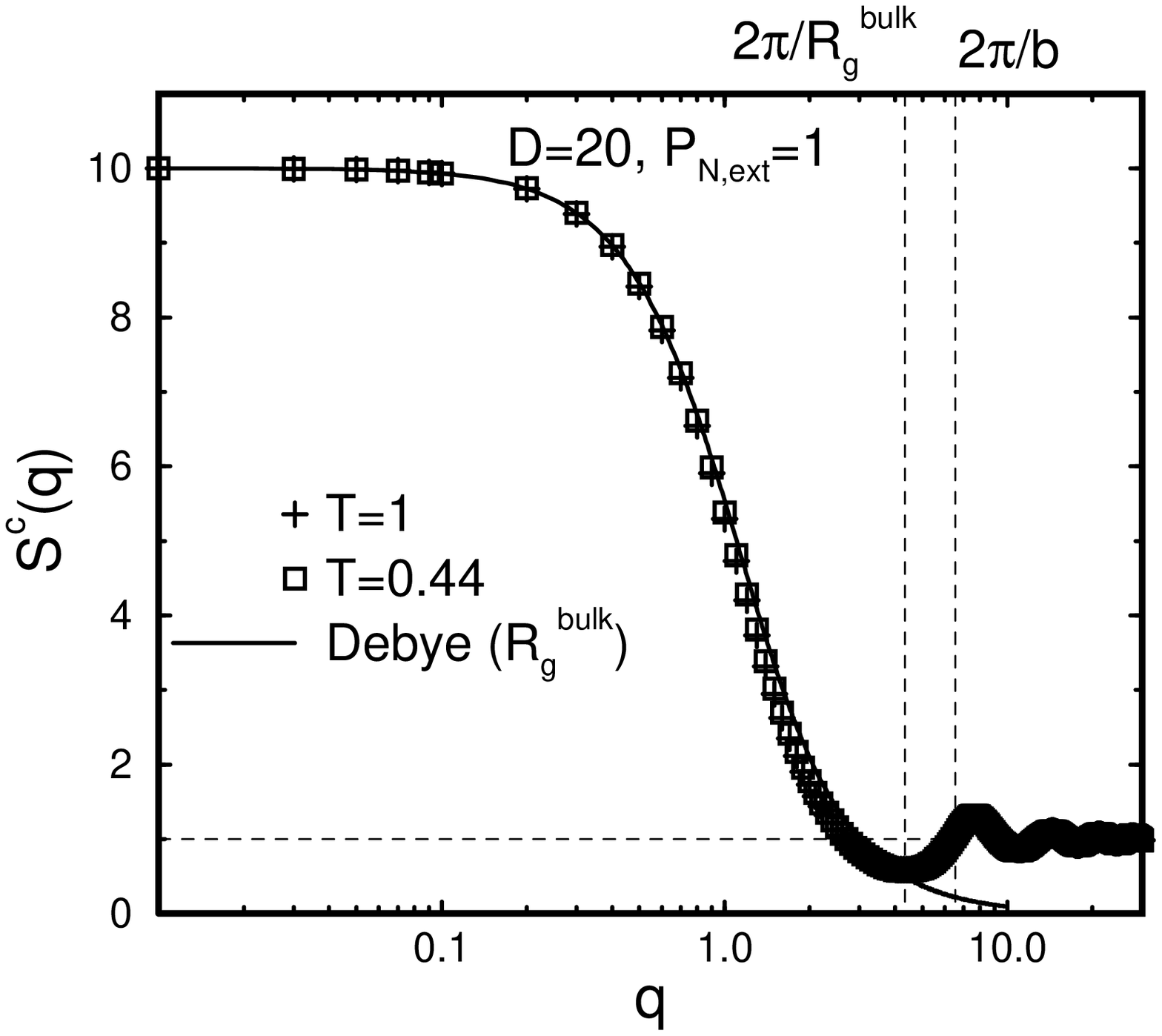}}
\epsfxsize=95mm 
\centering{\epsffile{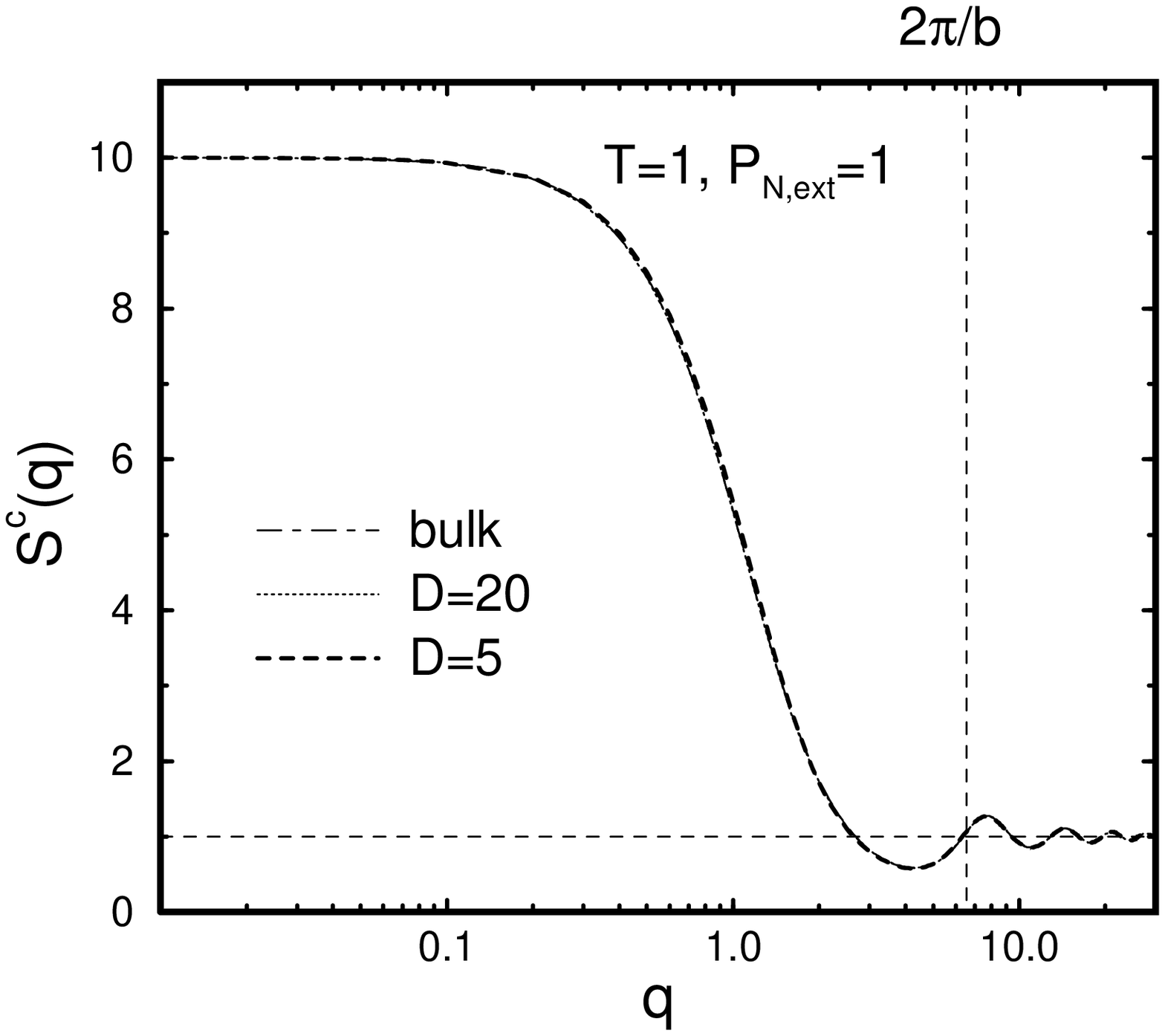}}
\vspace{-1mm}
\caption[]{
Upper panel:
Chain structure factor, $\Sc(q)$, for a film of thickness
$D\myeq 20$ averaged over the whole system. Two representative
temperatures are shown: $T\myeq 1$ (normal liquid state)
and $T\myeq 0.44$ (supercooled state) showing that $\Sc(q)$
is insensitive to a temperature variation.
$\Sc(q)$ in the film is computed in the direction parallel to 
the walls, i.e.\ $\mbf{q}\myeq (q_x,q_y),\; q \myeq |\mbf{q}|$,
whereas $\mbf{q}\myeq (q_x,q_y, q_z)$ in the bulk. 
The solid line indicates the Debye function 
with $\RgDebyesquare \myeq \Rgbulksquare$ 
[see Eq.~(\ref{eq:Debye::function})].
The vertical dashed lines mark the $q$-values $2\pi/b$
($b\myeq$ bond length) and $2\pi/\Rgbulk$. The horizontal dashed 
line indicates the large $q$-limit of $\Sc(q)$.
Lower panel: 
Same as in the upper panel, but now evaluated for the 
bulk and for films of various thicknesses at  
$T\myeq 1$ (normal liquid state). Apparently, $\Sc(q)$ is hardly 
affected by the confinement.
}
\label{fig:fig3}
\end{figure}

The upper panel of Fig.~\ref{fig:fig3} shows $\Sc(q)$ for
a film of thickness $D\myeq 20$ at two representative 
temperatures: a relative high temperature corresponding to
the normal liquid state ($T\myeq 1$) and a low temperature
representative of the supercooled state ($T\myeq 0.44$).
Contrary to the structure factor of the melt (see Fig.~\ref{fig:fig2}), 
$\Sc(q)$ is practically independent of the temperature, not only on 
scales larger than the radius of gyration, but also on the local scale of the 
intermonomer distance. This may be rationalized as follows: The interactions
of the monomers along the backbone of a chain do not include specific potentials
for the bond or torsional angles, which could make the chain expand with decreasing
temperature.  This would lead to a much stronger temperature dependence than
that resulting from potentials used.  Since the bond potential is very steep around 
the minimum and is in general very large, the bond length is 
essentially independent of $T$ ($b \myeq \sqrt{\langle \mbf{b} \cdot \mbf{b} 
\rangle} \myeq 0.966$ at $T\myeq 1$, $b \myeq \sqrt{\langle \mbf{b} \cdot \mbf{b} 
\rangle} \myeq 0.961$ at $T\myeq 0.46$. Here, $\mbf{b}$ is the bond vector).
The main effect is that the overall size of a chain slightly shrinks with 
decreasing temperature because the density increases. This also leads to a weak
increase of the peak of $\Sc(q)$ at $q \myeq  7.6$, which is, however, not
visible on the scale of Fig.~\ref{fig:fig3} and negligible compared to 
that of $S(q)$ in Fig.~\ref{fig:fig2}.

Furthermore, Fig.~\ref{fig:fig3} shows that, for $q \le 2 \pi / \Rgbulk$, 
the $q$-dependence of $\Sc(q)$ can be described well by a Debye-function,
\begin{equation}
S^{\rm Debye}(q)=\myfrac{2\Np}{(q^2\RgDebyesquare)^2}
\Big(\exp(-q^2\RgDebyesquare)-1+q^2\RgDebyesquare\Big) \;,
\label{eq:Debye::function}
\end{equation}
with $\RgDebyesquare \myeq \Rgbulksquare$, where $\Rgbulksquare$ was calculated
independently and inserted in Eq.~(\ref{eq:Debye::function}) [$\Np$ is the number of the 
monomers of a chain (degree of polymerization)]. The same observation has already been 
made for the model in the bulk~\cite{Bennemann-Paul-Binder-Duenweg::PRE57}. 
This agreement 
may be expected because the scattering for $q \le 2 \pi / \Rgbulk$ is determined by the 
overall size of a chain. On this
scale, the difference between the freely-jointed chain model utilized to derive the Debye 
function and the actual model of the simulation does not matter.

\begin{figure}
\epsfxsize=118mm 
\centering{\epsffile{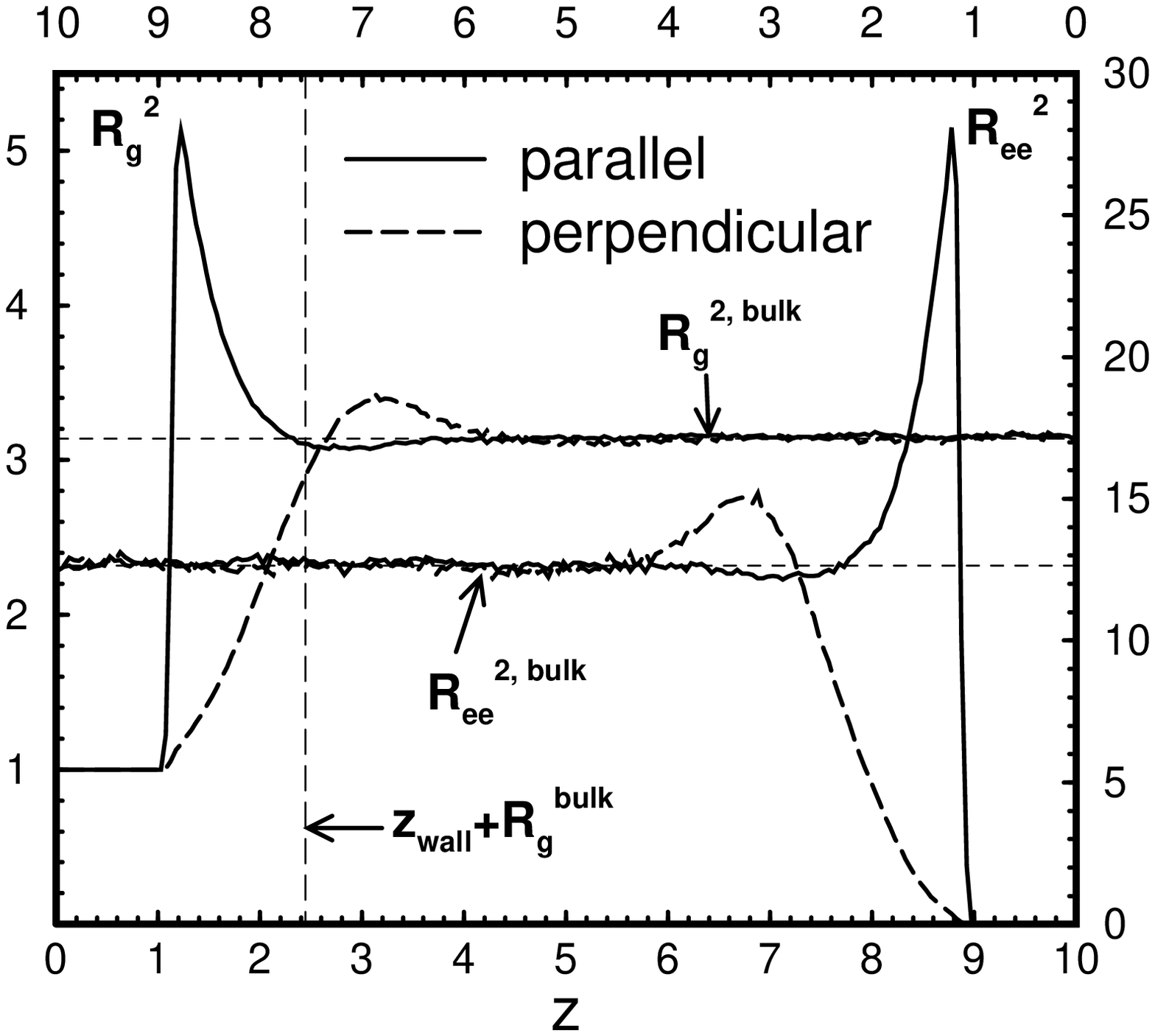}}
\epsfxsize=120mm 
\centering{\hspace*{-18mm}\epsffile{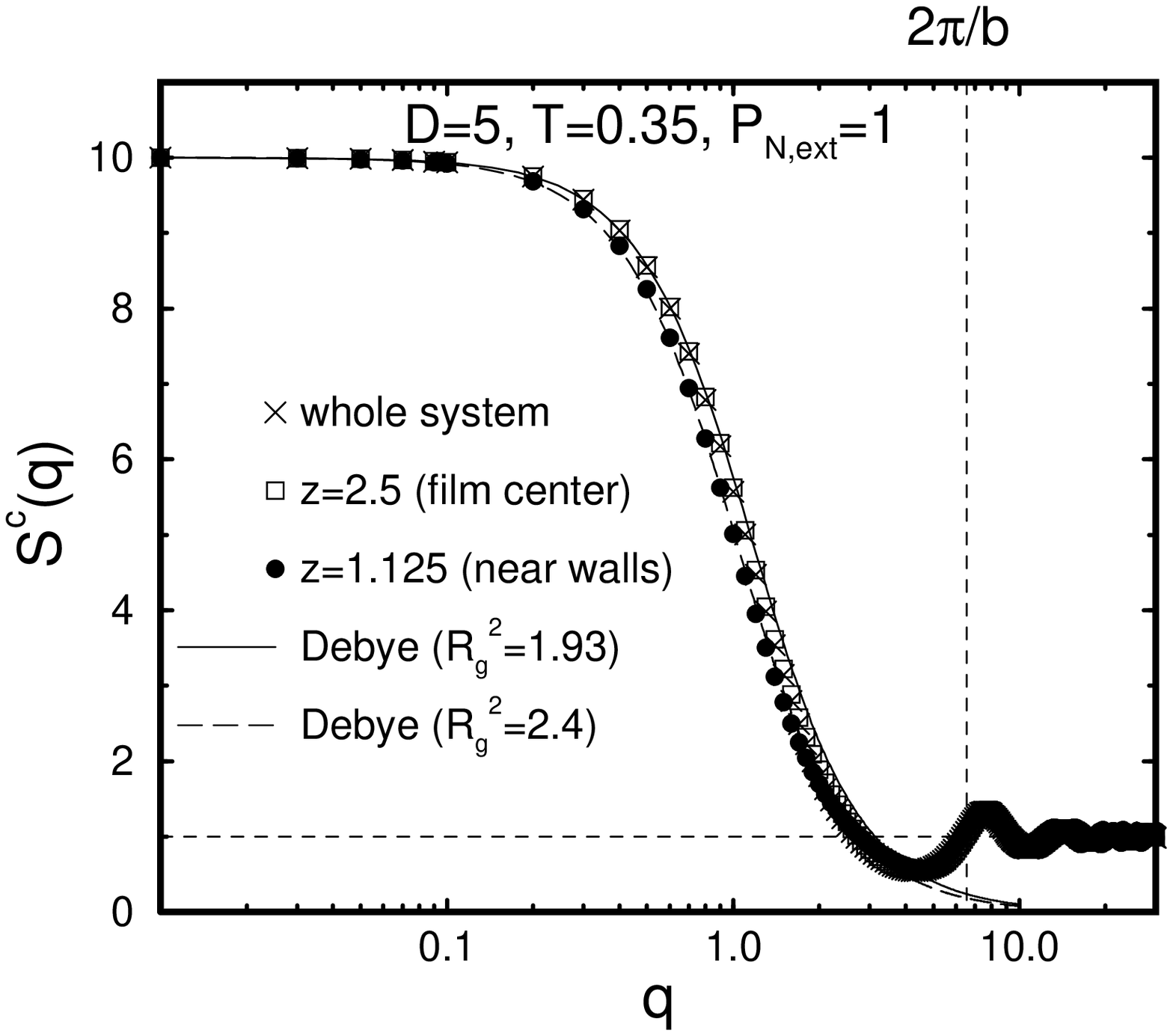}}
\caption[]{
Upper panel:
Components of the radius of gyration and 
of the end-to-end distance in directions parallel 
($\Rgpar$ and $\Reepar$) and perpendicular ($\Rgper$ and $\Reeper$)
to the wall versus the distance, $z$, from the wall.
$\Rgpar(z)$ and $\Reepar(z)$ behave 
qualitatively similar. They develop a maximum 
close to the wall and then converge towards a constant
(bulk) value in the film center (indicated by horizontal dashed lines).
Both the bulk radius of gyration and the end-to-end distance are 
known from previous simulations: $\Rgbulksquare \myeq 2.14$, and
$\Reebulksquare \myeq 12.6$ at
$T \myeq 1$ and pressure
$p \myeq 1$~\cite{Bennemann-Paul-Binder-Duenweg::PRE57}.
The parallel component is multiplied by a factor of $3/2$
and the perpendicular one  by $3$.
These factors take into account
that compared to three independent directions in the bulk, 
there are only two independent coordinates
in the parallel and one in the perpendicular directions in a planar system.
Lower panel:
Chain structure factor, $\Sc(q)$, evaluated within layers 
of thickness of $1/4$ monomer diameter for a film of thickness $D\myeq 5$ and 
at the low temperature $T\myeq 0.35$ (supercooled state).
Note that the mode-coupling critical temperature for this film 
thickness is $\Tc(D\myeq 5) \myeq 0.305 \pm 0.005$ 
[see Table~\ref{tab1}].
A layer close to the wall ($\bullet$) and 
the layer at the film center ($\Box$) are compared to the
average of  $\Sc(q)$ over the whole system ($\times$).
The chain structure in the film center coincides 
with the average behavior of the film. Thus, the deviations of the chain 
structure in the vicinity of the walls compared to the 
film center do not crucially influence the average over the 
whole system. The solid line gives the Debye function with 
$\RgDebyesquare \myeq 1.93$, whereas the long dashed line 
corresponds to $\RgDebyesquare \myeq 2.4$
[see Eq.~(\ref{eq:Debye::function})]. The values of $\RgDebyesquare$ are 
the average radii of gyration calculated over all chains whose center of 
mass lies in the interval of thickness $1/4$ around $z\myeq 1.125$ and 
$z\myeq 2.375$, respectively. The vertical dashed lines mark the $q$-values $2\pi/b$
($b\myeq$ bond length). The horizontal dashed line indicates
the large-$q$ behavior of $\Sc(q)$ (i.e., $\Sc(q\rightarrow \infty) \myeq  1$).
}
\label{fig:fig4}
\end{figure}

The lower panel of Fig.~\ref{fig:fig3} shows $\Sc(q)$ for films of various thicknesses $D$ 
at $T\myeq 1$. Obviously, the chain conformation averaged over the whole film does not depend 
much on $D$.  Contrary to the small effect of temperature on $\Sc(q)$, the independence of  
film thickness is less intuitive.  Figure~\ref{fig:fig4} illustrates this point.  In the upper panel 
we compare the parallel and perpendicular components of the radius of gyration 
and of the end-to-end distance, $\Rgpar$, $\Rgper$ and $\Reepar$, $\Reeper$, for
$D\myeq 20$ and $T\myeq 1$ (normal liquid phase).  The components are
plotted versus the distance, $z$, from the (left) wall, where $z$ denotes the position of 
chain's center of mass.  So, $\Rgpar(z)$, for instance, is the radius of gyration parallel 
to the wall, which is averaged over all chains whose centers of mass are located at $z$. The 
figure shows that both the radius of gyration and the end-to-end distance agree with the bulk
value if $z > z_{\rm w} \!+\! 2 \Rgbulk$. Here, $z_{\rm w} \myapprox 1$ denotes the 
physical position of the wall, 
i.e., the smallest distance between a monomer and $z_{\rm wall}\myeq D/2$.
As the chain's center of mass approaches the wall, $\Rgpar$ and $\Reepar$
first develop a shallow minimum and then increase to about twice the bulk value 
followed by a sharp decrease to zero in the very vicinity of the wall where practically 
no chain is present. On the other hand, the perpendicular components,  
$\Rgper$ and $\Reeper$, first pass through a maximum before decreasing 
to almost 0 at the wall.
This behavior has been observed in several other simulations 
(see \cite{BaschnagelBinderMilchev2000} and references therein), also for larger 
chain length than that used here \cite{BitsanisHadziioannou1990,WangBinder1991}. 

These results suggest that a spatially resolved version of the chain structure factor, $\Sc(q,z)$,
should depend upon the position of the chain with respect to the wall.  To test this idea
we divide the system into layers of thickness $1/4$ (in units of $\sigma$ which roughly corresponds
to the monomer diameter) and evaluate $\Sc(q, z)$ by taking into 
account all monomer pairs within the layer which belong 
to the same chain.  Here, $z$ is the distance of the 
middle of such a layer from the wall.
The lower panel of Fig.~\ref{fig:fig4} shows $\Sc(q, z)$
for an extremely thin film of thickness $D\myeq 5$ at $T\myeq 0.35$ and
for two typical layers: a layer in the film center ($z\myeq 2.375$) and a layer 
close to the wall ($z\myeq 1.125$)
[note that all layers with a distance smaller than $z\myeq 1.125$ to 
the wall are practically empty. Note also that there is no layer whose middle lies
exactly in the film center $z\myeq 2.5$. Only the boundary of the 
central layer ``touches'' the film center so that its middle is 
closer to the wall than $D/2\myeq 2.5$]. In the center of the film
$\Sc(q, z)$ can be described by a Debye function with $\RgDebyesquare\myeq 1.93$, whereas 
a larger radius ($\RgDebyesquare\myeq 2.4$) must be chosen close to the walls.
These values for $\RgDebyesquare$ are taken from the profile of the radius of 
gyration for $D\myeq 5$
at $T\myeq 0.35$, which yields $\Rgsquare\myeq 2.4$ and $\Rgsquare\myeq 1.93$ 
when averaging over the intervals 
$1 \leq z \leq 1.25$ and $2.25 \leq z \leq 2.5$, respectively.
Furthermore, Fig.~\ref{fig:fig4} shows that $\Sc(q, z)$ for the film center coincides
with $\Sc(q)$ obtained by averaging over the whole system.
Thus, the local variations of $\Sc(q, z)$ close to the wall disappear 
when the whole system is considered.  Qualitatively, this point can
be understood by noting that there are only few chains close to the wall,
where $\Rgpar(z)$ is larger than the bulk value, while most of chains are in the inner
part of the film, where $\Rgpar(z)$ is close to $R_{\rm g}^{2~{\rm bulk}}$.
Averaging over the whole system thus tends to cancel the effects of the walls on the
chain conformation.

\begin{figure}
\epsfxsize=120mm 
\centering{\epsffile{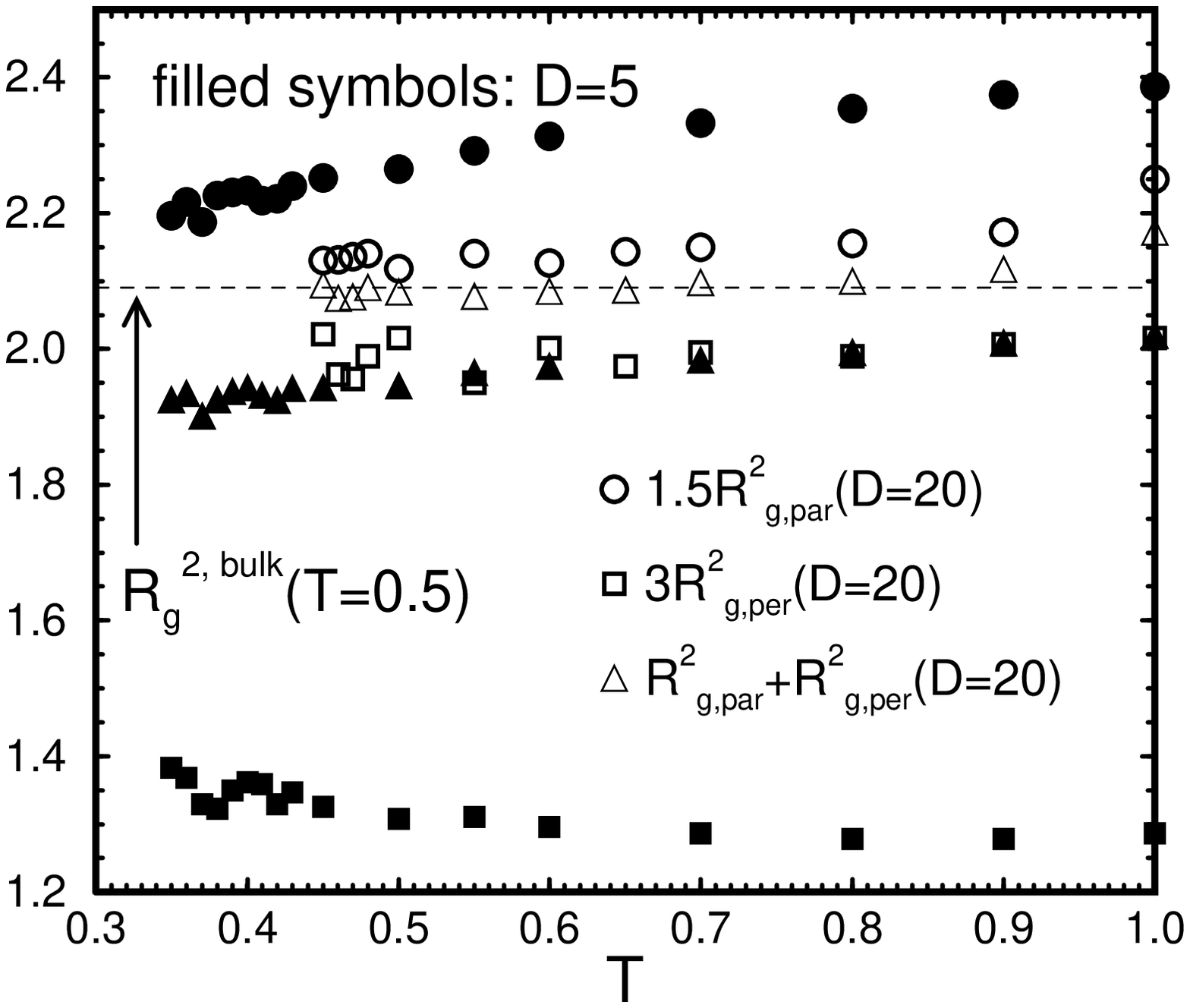}}
\epsfxsize=120mm 
\centering{\epsffile{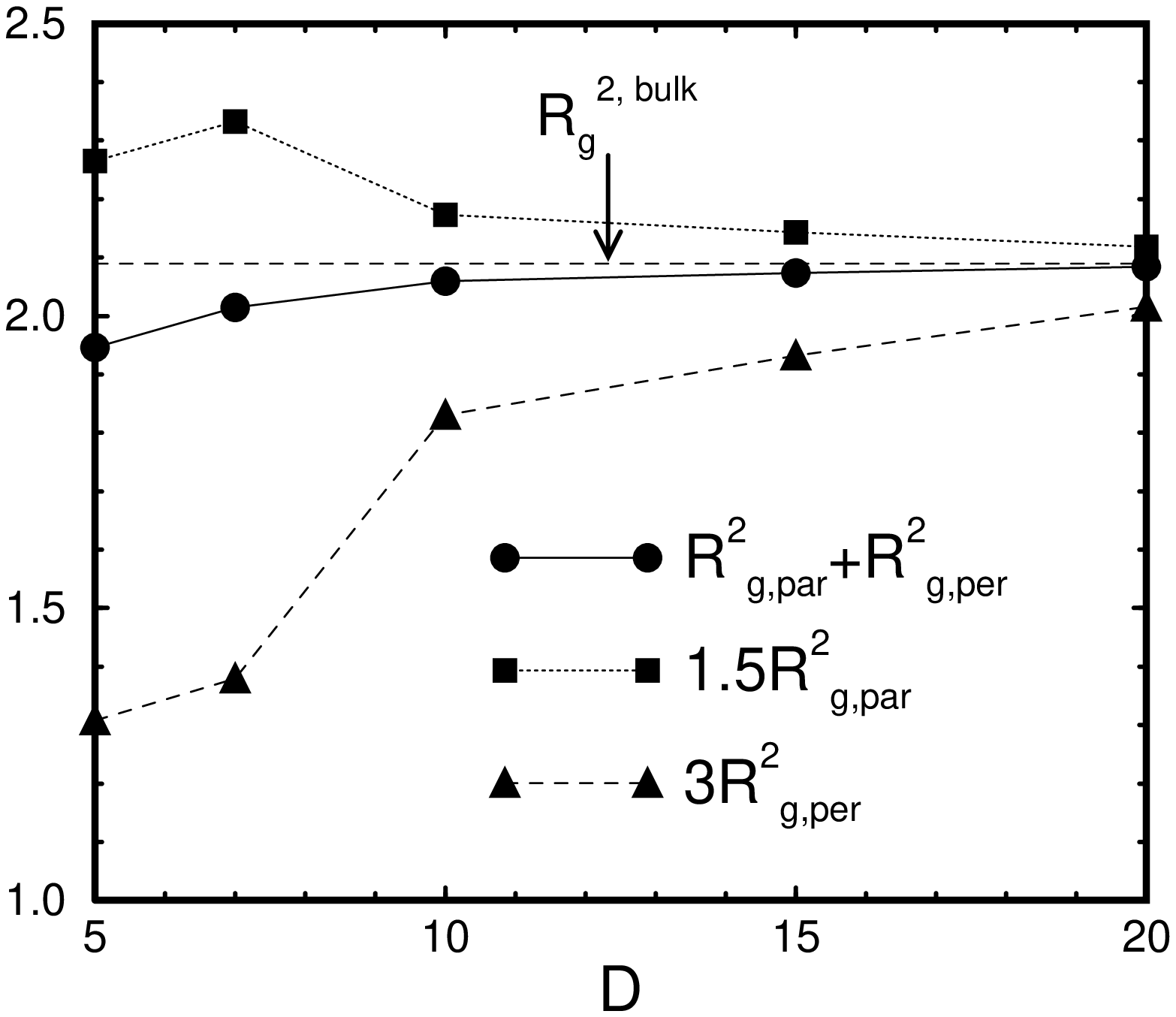}}
\caption[]{
Upper panel:
Parallel and perpendicular components of the radius of 
gyration (1.5 $\Rgpar$ and $3\Rgper$) and their 
sum, $\Rgsquare \myeq \Rgpar +\Rgper$, versus 
temperature. $\Rgpar$ and $\Rgper$ have been 
computed as (time) averages over all chains in the system. 
Data are shown for two film thicknesses: 
$D\myeq 20$ (open symbols) and $D\myeq 5$
(the corresponding filled symbols).
The factors $1.5$ and $3$ are introduced to 
simplify the comparison with the bulk value.
They account for the different number
of spatial components in the film for 
directions parallel ($2$) and perpendicular 
($1$) to the walls compared
to the bulk where there are $3$ such components.
Both in the case of $D\myeq 20$ and of $D\myeq 5$ 
chains prefer an alignment parallel to the walls.
Due to the opposite changes of $\Rgpar$ and $\Rgper$ with respect 
to the bulk value, the effect of the confinement on the chain's 
radius of gyration,  $\Rgsquare \myeq \Rgpar +\Rgper$, 
is less strong and practically vanishes for $D\myeq 20$.
The horizontal dashed line indicates the bulk radius of gyration,
$\Rgbulksquare \myeq 2.09$ (at $T\myeq 0.5$).
Lower panel:
Same quantities as in the upper panel, now versus film thickness, $D$,
at a temperature of $T\myeq 0.5$.
($\Rgsquare$: connected circles, $\Rgpar$: connected 
squares and $\Rgper$: connected triangles).
Again, the parallel orientation is favored by the walls. The 
overall radius of gyration, however, depends only slightly 
on film thickness.}
\label{fig:fig5}
\end{figure}

This point is further examined in Fig.~\ref{fig:fig5}.
The upper panel of the figure shows $1.5 \Rgpar$, $3\Rgper$, averaged over the whole film, and 
$ \Rgsquare \myeq \Rgpar +\Rgper$ versus temperature for two film thicknesses $D\myeq 5$ and $D\myeq 20$.
For all temperatures the parallel component is larger than the bulk value, whereas the perpendicular one
is smaller.  The disparity between the two components becomes more pronounced for small film thickness.
Nevertheless, the sum of these quantities, $\Rgsquare$, is fairly close to the bulk value for both
film thicknesses.  It slightly shrinks with decreasing temperature, since the density of the films 
increases.  Similar trends are observed if $D$ is varied and $T$ is kept constant.
The lower panel of Fig.~\ref{fig:fig5} depicts $1.5 \Rgpar$, $3\Rgper$ and $ \Rgsquare$ as a function of 
film thickness at $T\myeq 0.5$. $\Rgsquare$ depends only weakly on film thickness and is close to the
bulk value.  This explains why the chain structure factors of Fig.~\ref{fig:fig3} are essentially 
independent of $D$.
\subsection{The Packing Structure of the Melt}
\label{subsection:static::melt}
\begin{figure}
\epsfxsize=100mm 
\centering{\epsffile{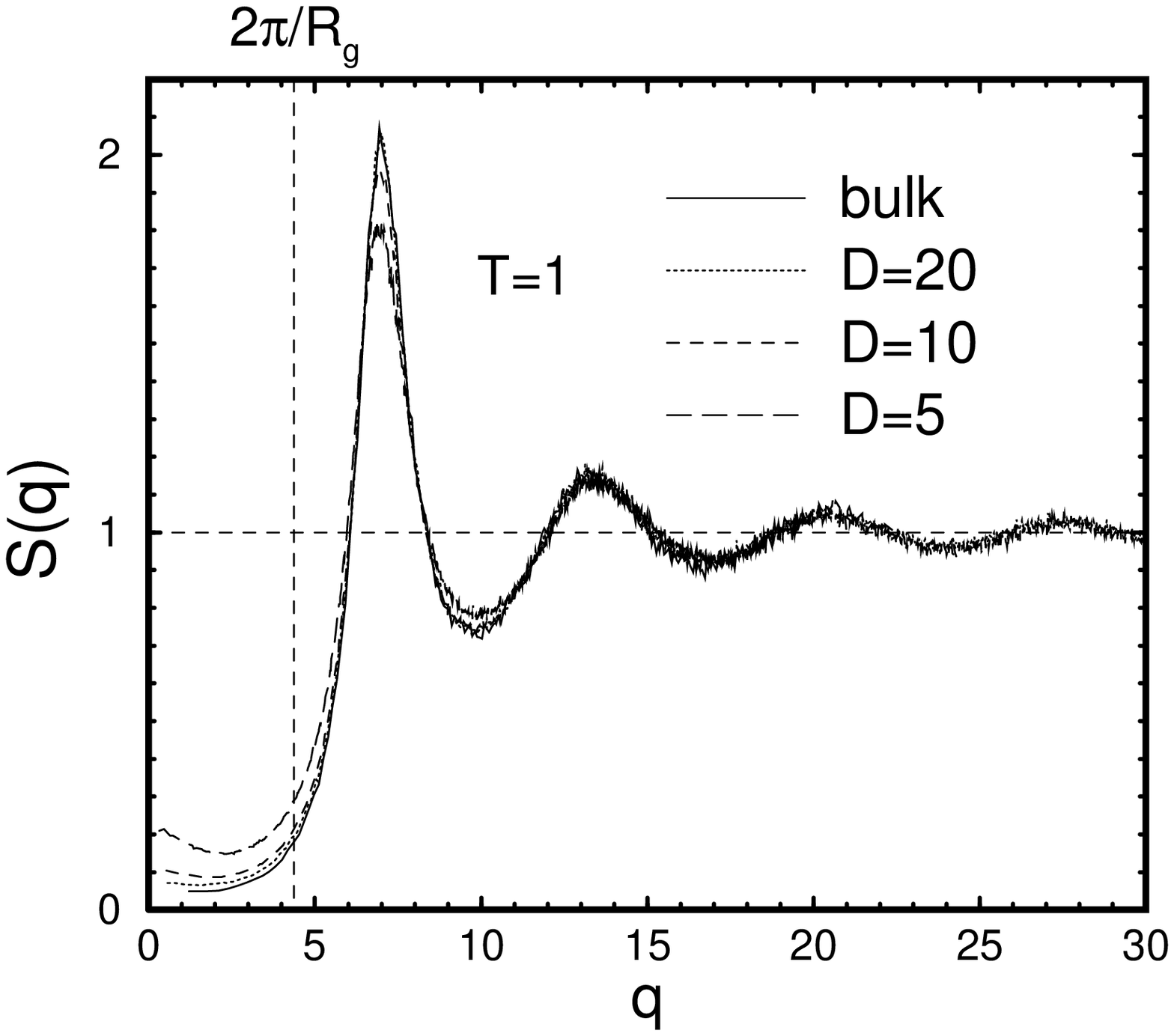}}
\epsfxsize=100mm 
\centering{\hspace*{0mm}\epsffile{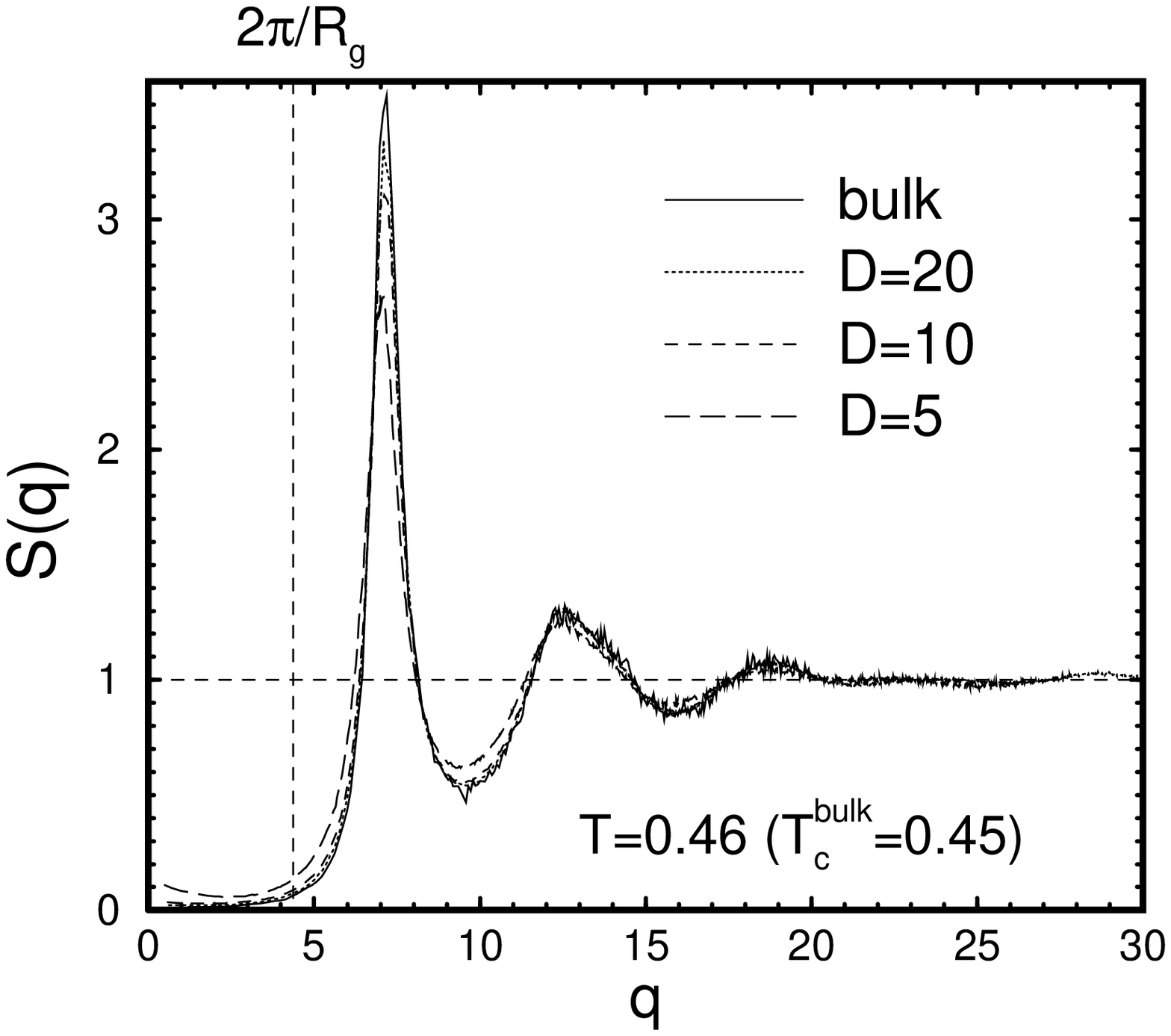}}
\caption[]{
Upper panel:
Structure factor, $S(q)$, of the melt for the bulk and films of various thicknesses
ranging from $D\myeq 5\, (\myapprox 3.5\Rg)$ to  $D\myeq 20\, (\myapprox 14\Rg)$
at a temperature of $T\myeq 1$ (normal liquid state). Note that the 
mode-coupling critical temperature of the system in the 
bulk is $\Tcbulk\myeq 0.45$~\cite{Bennemann-Paul-Binder-Duenweg::PRE57,%
Bennemann-Baschnagel-Paul::EPJB10} (see also Table~\ref{tab1}).
$S(q)$ in the film is computed by taking into account the coordinates parallel 
to the walls only, i.e.\ $\mbf{q}\myeq (q_x,q_y),\; q\myeq |\mbf{q}|$.
At $T\myeq 1$, the structure factor of the thickest 
film ($D\myeq 20$) coincides with that of the bulk.
For film thicknesses $D \le 10$, the influence of the walls 
on $S(q)$ is visible even at this high temperature.
The peak value of $S(q)$ decreases with stronger 
confinement (smaller $D$) and the position of the peak is slightly 
shifted towards smaller $q$, reflecting an increase of the 
(average) interparticle  distance. Instead of decreasing the
film thickness, these changes in $S(q)$ can also
be achieved by increasing the temperature only. So, the inverse film 
thickness plays a qualitatively similar role to that of the temperature.
For the largest film thickness shown here, however,
the structure factor of the film is identical to that of the 
bulk over a large $q$-range.
Lower panel: The same as in the upper panel, now for a lower 
temperature of  $T\myeq 0.46$ (supercooled state). Contrary to $T\myeq 1$, 
the structure factor of the film with $D\myeq 20$ no longer
coincides with $S(q)$ of the bulk. Thus, at lower temperatures,
the confinement has a stronger impact on the packing structure of the melt.
}
\label{fig:fig8}
\end{figure}

Figure~\ref{fig:fig8} compares the structure factor of the melt,
$S(q)$, in the bulk with that of films of various thicknesses at
$T\myeq 1$ (normal liquid state; upper panel of the figure) and 
at $T\myeq 0.46$ (supercooled state; lower panel of the figure).
Qualitatively, the behavior of the films and the bulk is identical.
The structure factor is small at small $q$-values, reflecting the 
low compressibility of the system. Then, it increases
and develops a peak at $q_{\rm max}$ which corresponds to
the local packing of monomers ($2\pi/q_{\rm max}\myapprox 1$)
before it decreases again and begins oscillating 
around $1$, the large-$q$ limit of $S(q)$. This behavior is 
characteristic of dense amorphous packing.

However, there are quantitative differences which become more pronounced 
at low temperature: While $S(q)$ of the film of thickness $D\myeq 20$  
(almost) coincides with the bulk data at $T\myeq 1$, deviations are clearly 
visible at $T\myeq 46$. Quite generally, 
the most prominent differences between the bulk and the film 
are found for small $q$ and for $q_{\rm max}$. The compressibility
of the film is higher, the value of $q_{\rm max}$ is shifted to
slightly lower $q$ and the magnitude of $S(q_{\rm max})$ is smaller
than in the bulk. Keeping the film thickness fixed, one can observe 
similar changes of $S(q)$ as the temperature increases (see Fig.~\ref{fig:fig2}).
Therefore, the local packing of the monomers in the
films seems to resemble that of the bulk at some higher temperature.
Since the local structure of the melt has an important
influence on its dynamic behavior in the supercooled
state~\cite{Goetze::LesHouches::1989}, 
Fig.~\ref{fig:fig8} suggests that the film relaxes 
more easily than the bulk at the same temperature. Indeed, we will 
see later that the dynamics of the system is much faster in the 
film than in the bulk when compared at the same temperature.
\section{Dynamics}
\label{section:Dynamics}
This section discusses the dynamics of the films at low temperatures 
and for various thicknesses, $D$, ranging from about 3 to about 14 times the 
bulk radius of gyration. To this end, the incoherent intermediate scattering 
function and various mean-square displacements were calculated. We will see 
that, despite the geometric confinement, the films exhibit several dynamic features 
which are in agreement with predictions of mode-coupling theory (MCT) 
\cite{Goetze::LesHouches::1989,Goetze-Sjoegren::TransportTheoryStatPhys,%
Goetze-Sjoegren::RepProgPhys55,Goetze::JPCM::1999}.  In this respect, the films
behave as the bulk \cite{Bennemann-Baschnagel-Paul::EPJB10,%
Bennemann-Paul-Baschnagel-Binder::JPCM11,%
Bennemann-Baschnagel-Paul-Binder,%
Aichele-Baschnagel::EurPhysJE::I::D,%
Aichele-Baschnagel::EurPhysJE::II::D}.  
However, the onset of MCT effects is shifted to lower temperatures compared to the
bulk.  The presence of the smooth walls accelerates the dynamics and this influence
of confinement is the stronger, the smaller $D$.
\subsection{Confinement leads to Faster Dynamics}
\label{subsection:confinement::faster::dynamics}
An interesting dynamic correlation function is the incoherent intermediate
scattering function $\phis(t)$.  It measures density fluctuations on various
length scales, which are caused by the displacement of individual particles.
For a planar system, we define $\phis(t)$ for $q$-vectors parallel to the wall, 
i.e.,
\begin{eqnarray}
\phis(t) = \Big< \myfrac{1}{N} \sum_{i=1}^{N}
\exp \Big({\rm i} \qpar\, [ \ripar(t)- \ripar(0)] \Big) \Big> \;.
\end{eqnarray}
Here, $N$ is the total number of monomers in the system,
$\qpar\myeq(q_x, q_y)$, $q \myeq |\qpar| \myeq \sqrt{q_x^2+q_y^2}$
and $\ripar \myeq (x_i, y_i)$.

\begin{figure}
\epsfxsize=140mm 
\centering{\epsffile{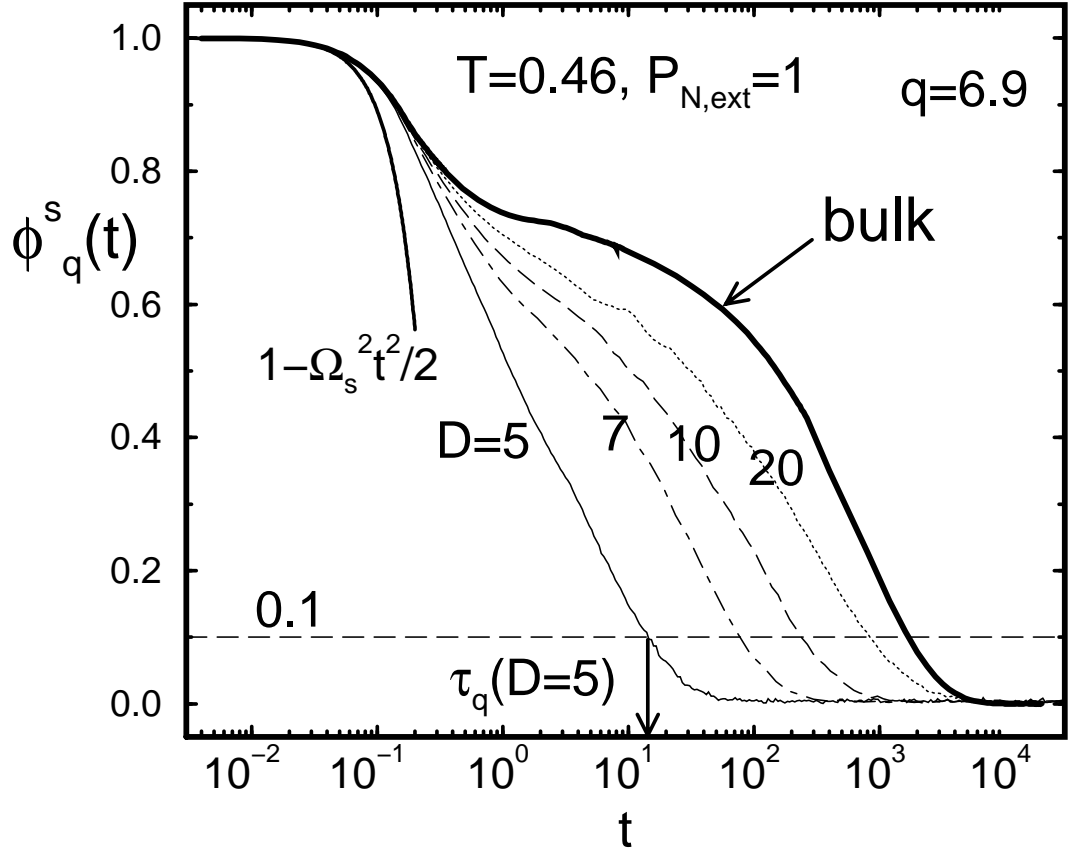}}
\caption[]{
Comparison of the incoherent intermediate scattering function, $\phis(t)$, at 
$T\myeq 0.46$ ($\Tcbulk\myeq 0.45$) for the bulk and films of various 
thicknesses ranging from $D\myeq 5\myapprox 3.5 \Rg$ to $D \myeq 20 \myapprox 14 
\Rg$. $P_{\rm N,ext}\myeq 1$ denotes the pressure of the simulation. The scattering
functions are calculated at $q\myeq 6.9$ (maximum of $S(q)$, see Fig.~\ref{fig:fig2}).
The initial decay of $\phis(t)$ is determined by free ballistic motion, i.e.,
$\phis(t)\myeq 1-(\Omegas t)^2/2$ with $\Omegas \myeq q \sqrt{\kB T}$. The dashed 
horizontal line at $\phis(t)\myeq 0.1$ and the vertical arrow on the curve for $D\myeq 5$
indicate how the $\alpha$-relaxation time, $\tau_q$, is defined. It is
the time value when the scattering function has decayed to 0.1.
}
\label{fig:fig9}
\end{figure}

Figure~\ref{fig:fig9} compares the relaxation of $\phis(t)$ at the maximum
of the static structure factor ($q\myeq 6.9$) in the bulk with that in films of various 
thicknesses ranging from $D\myeq 5\myapprox 3.5 \Rg$ to $D \myeq 20 \myapprox 
14 \Rg$.  The temperature studied, $T\myeq 0.46$, is slightly above $\Tcbulk$ 
($\myeq 0.45$) \cite{Bennemann-Baschnagel-Paul::EPJB10,%
Bennemann-Paul-Baschnagel-Binder::JPCM11,%
Bennemann-Baschnagel-Paul-Binder,%
Aichele-Baschnagel::EurPhysJE::I::D,%
Aichele-Baschnagel::EurPhysJE::II::D}. In the bulk, a two-step relaxation is observed:
At very short times, $\phis(t)$ can be described by
$\phis(t)\myeq 1-(\Omegas t)^2/2$ with the microscopic frequency 
$\Omegas \myeq q \sqrt{\kB T}$~\cite{Goetze::LesHouches::1989}.
This corresponds to free particle motion. At later times, the relaxation of
$\phis(t)$ is strongly protracted.  There is an intermediate time window
($\beta$-relaxation regime of MCT~\cite{Goetze::LesHouches::1989}),
where the correlator changes rather slowly with time  before the 
final structural relaxation ($\alpha$-relaxation) sets in at long times.
This two-step decay is a characteristic feature of $\phis(t)$ for
temperatures close to $\Tc$ and reflects the temporary ``arrest'' of 
a monomer in its local environment (``cage effect''~\cite{Goetze::LesHouches::1989}).
It has been analyzed in detail in Refs.~\cite{Bennemann-Paul-Binder-Duenweg::PRE57,%
Bennemann-Baschnagel-Paul::EPJB10,%
Aichele-Baschnagel::EurPhysJE::I,%
Aichele-Baschnagel::EurPhysJE::II}.

Compared to the bulk, the film data relax faster. If the film thickness
decreases, this acceleration is enhanced and the two-step decay gradually 
disappears. At $D\myeq 5$, no intermediate $\beta$-relaxation is observed.
The same changes also occur in the bulk if the temperature increases~\cite{Bennemann-Baschnagel-Paul::EPJB10,Bennemann-Baschnagel-Paul-Binder,Aichele-Baschnagel::EurPhysJE::I}.
This suggests that the inverse film thickness qualitatively plays a 
similar role as the temperature.

The acceleration of the dynamics in the film compared to the bulk
is not limited to the main peak of $S(q)$. It is also found for
other $q$-values, for instance, for very small wave vectors. Since the time dependence of
$\phis(t)$ is directly related to the monomer mean-square displacement (MSD) in 
the low-$q$ limit, it is instructive to illustrate the acceleration of the dynamics 
by an investigation of the MSD. For a polymer system, various kinds of 
MSD's may be defined. An important 
example is the mean-square displacement of the innermost monomer,
\begin{eqnarray}
g_{1} (t) &=& \frac{3}{2M} \sum^{M}_{i=1} \left \langle
\big[ \mbf{r}^{\rm inner}_{i, \parallel} (t) - \mbf{r}^{\rm inner}_{i, \parallel} (0)
\big]^{2}    \right \rangle \; .
\label{eq:g1a::def}
\end{eqnarray}
Here, $M$ is the total number of chains in the melt and
$\mbf{r}^{\rm inner}_{i, \parallel} \myeq (x^{\rm inner}_{i},\; y^{\rm inner}_{i})$
is the projection of the position vector of the innermost monomer of the $i$-th chain
onto the $xy$-plane (which is parallel to the walls).
The factor $3/2$ is introduced to simplify the comparison
with bulk results. It takes into account that only two independent directions 
($x$ and $y$) contribute to the MSD in the film,
whereas all three directions are considered in the bulk\footnote{As most of the comparisons with bulk results are done using the film data in parallel direction, we drop
the index ``$\parallel$'' to simplify the notation.
To avoid ambiguities, when film data in perpendicular 
direction are discussed, an index ``$\bot$'' will then be used.}.
Similarly, one defines the mean-square displacements of 
the chain's center of mass,
\begin{eqnarray}
g_3 (t) &=& \frac{3}{2M} \sum^{M}_{i=1}
\left \langle
\big[ \mbf{R}^{\rm cm}_{i, \parallel}(t)-\mbf{R}^{\rm cm}_{i, \parallel} \big]^{2}
\right \rangle \;,
\label{eq:g3a::def}
\end{eqnarray}
where $\mbf{R}^{\rm cm}_{i, \parallel}$ is the projection of vector to 
the $i$-th chain's center of mass onto a plane parallel 
to the wall. Again, the factor $3/2$ accounts for the difference in the number of 
independent components, as above.

\begin{figure}
\epsfxsize=120mm 
\centering{\epsffile{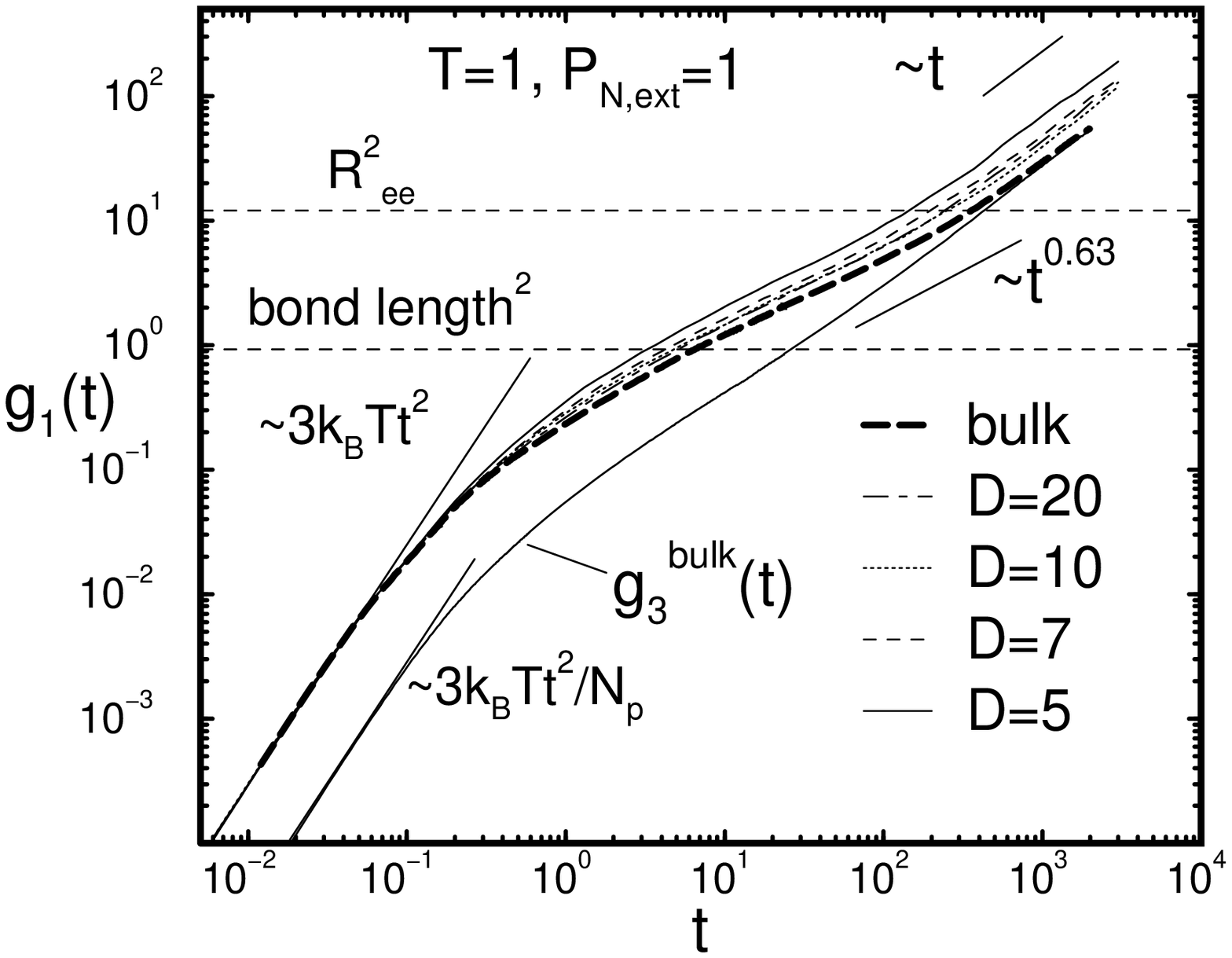}}
\epsfxsize=120mm 
\centering{\epsffile{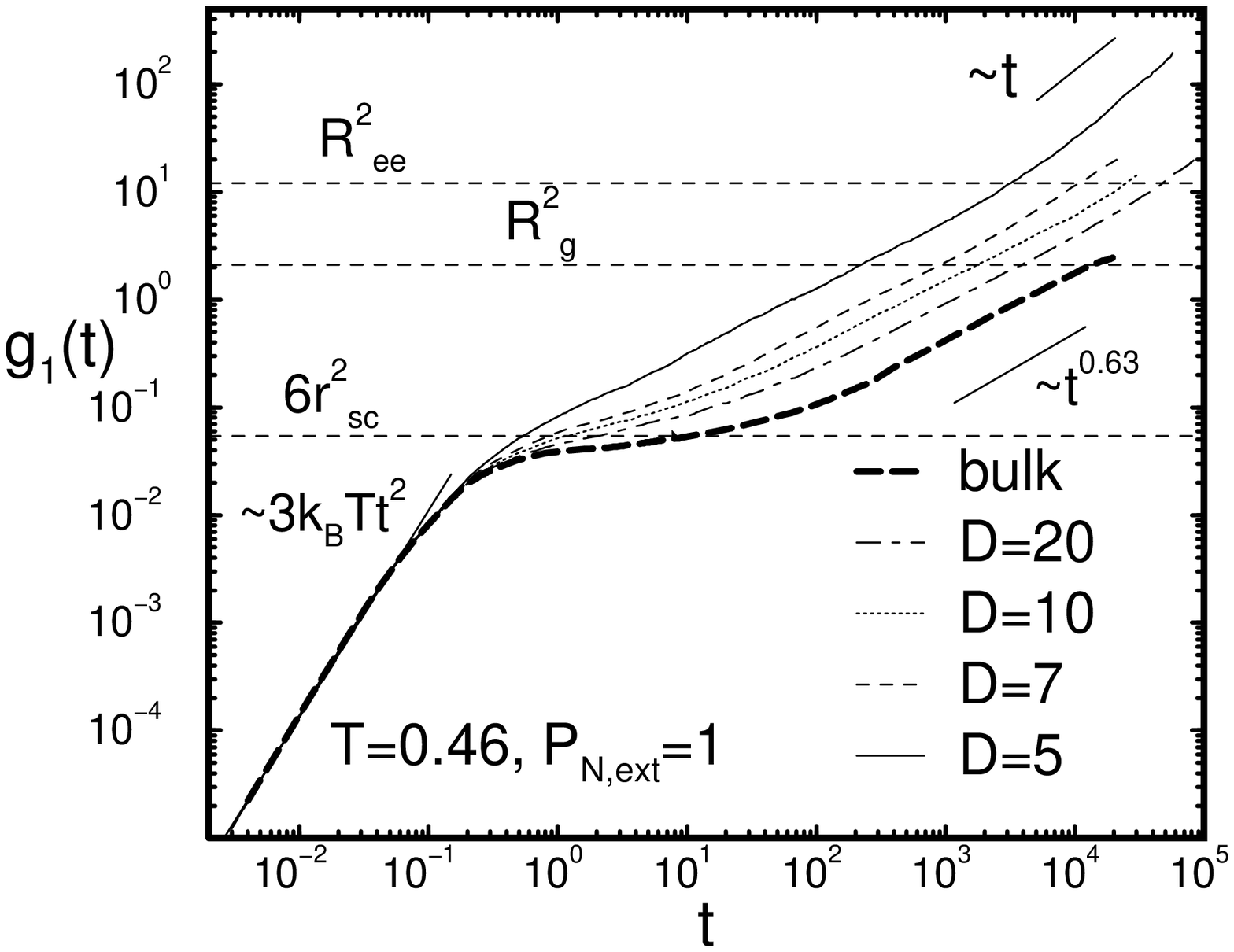}}
\caption[]{
Upper panel:
Log-log plot of the mean-square displacement of innermost
monomers, $g_1(t)$, at $T\myeq 1$
(normal liquid state far above $\Tcbulk \myeq 0.45$).
The figure compares the bulk
data with the displacements measured parallel to the walls in films 
of different thicknesses ranging from $D\myeq 5$ ($\approx \! 3.5 R_\mr{g}$)
to  $D\myeq 20$ ($\approx \! 14 R_\mr{g}$).
The film data were multiplied by $3/2$ to account for the 
different number of spatial directions
used to calculate $g_1(t)$ (i.e., 3 directions for the bulk, but only 2 for the
films). The bulk end-to-end distance $\Reesquare$ ($\myeq 12.3$) 
and the average bond length $b^2\myeq 0.922$ are indicated as
horizontal dashed horizontal lines.
The solid lines represent the behavior of $g_1(t)$
expected in different time regimes: ballistic at short times ($g_1(t)\sim t^2$),
diffusive at late times ($g_1(t) \sim t$),
and dominated by chain connectivity for times where $g_1(t) > b^2$
($g_1(t)\sim t^x$; $x \myeq  0.63 \myeq $ effective exponent in the bulk).
For comparison, the MSD of chain's center of mass, $g_3(t)$,
is also depicted: At short times, $g_3(t) \myeq 3 \kB T t^2/\Np$,
where $\Np \myeq 10$ is the degree of polymerization 
(number of the monomers of a chain).
Contrary to the innermost monomers, the motion of the 
chain's center of mass is not affected by chain connectivity. 
Therefore, $g_3(t)$ does not show a $t^{0.63}$-behavior,
but continuously crosses over to the diffusive regime 
which is reached for $g_3(t) \ge \Rgsquare$ $(\myeq 2.14\; {\rm at}\; T\myeq 1)$.
The effects of confinement are rather weak at this high temperature:
$g_1(t)$ for $D\myeq 10$ and  $D \myeq 20$ are practically indistinguishable 
from that of the bulk. For strong confinement ($D\myeq 5$ and $D \myeq 7$), 
however, a dependence on film thickness is observed.
Lower panel:
Same as in the upper panel, now at a low temperature of $T\myeq 0.46$ 
(supercooled state close to $T_\mr{c}^\mr{bulk}\myeq 0.45$).
The lowest horizontal line shows the plateau value $6r_{\rm sc}^2$ of 
a MCT-analysis ($\simeq 0.054$).
Again, the solid lines represent the behavior of $g_1(t)$
expected in different time regimes: ballistic at short times ($g_1(t)\sim t^2$),
diffusive at late times ($g_1(t)\sim t$), and dominated by chain connectivity
for times where $g_1(t) > b^2$ ($g_1(t)\sim t^x$; 
$x \myeq  0.63 \myeq $ effective exponent in the bulk).
At this low temperatuure, the effect of the walls is  
much more pronounced compared to $T\myeq 1$ (see the upper panel).
In both panels $P_{\rm N, ext}\myeq 1$ denotes the pressure of the simulation. 
}
\label{fig:fig10}
\end{figure}

Figure~\ref{fig:fig10} depicts $g_1(t)$ for two typical temperatures: $T\myeq 1$ (upper panel), 
which corresponds to the normal liquid state, and $T\myeq 0.46$ (lower panel), which belongs
to the supercooled state in the bulk.
Both panels compare $g_1(t)$ for the bulk and for films of 
various thicknesses. The influence of the walls is rather small
at $T\myeq 1$ so that $g_1(t)$ of the bulk almost overlaps 
with that of the film if $D \!\ge\! 10$.
However, the  lower panel of Fig.~\ref{fig:fig10} shows that
the effect of the walls on the mobility becomes significant 
at all studied thicknesses with progressive supercooling.
Outside the initial ballistic regime ($g_1(t) \myeq 3T t^2$), the motion resembles that of the
bulk, but is the faster, the smaller the film thickness. For the bulk and $D \gtrsim 7$,
$g_1(t)$ exhibits several regimes. In agreement with the predictions of the mode-coupling 
theory~\cite{Goetze::LesHouches::1989,%
Goetze-Sjoegren::TransportTheoryStatPhys,%
Goetze-Sjoegren::RepProgPhys55},
a plateau regime emerges after the ballistic motion.
At low temperature, the tagged particle remains temporarily in the ``cage''
formed by its neighbors. However, contrary to simple (atomic) liquids, where a 
direct crossover from the plateau to the diffusive regime occurs,
an intermediate subdiffusive regime emerges due to the connectivity 
of the monomers~\cite{Rouse::JChemPhys21:1953}.
In this regime, which is present for all $D$, the center of mass already crosses over to the
asymptotic diffusive motion, $g_3(t) \simeq t$, whereas the 
motion of the innermost monomer is described by a power 
law $g_1(t) \sim t^x$ with an effective exponent 
$x \simeq 0.63$. The innermost  monomer reaches the center of mass only 
if $g_1(t)$ is larger than the end-to-end distance 
of a chain. In this limit, the diffusive motion is 
dominated by the motion of the chain's center of mass
and $g_1(t)$ coincides with $g_3(t)$.

\begin{figure}
\epsfxsize=105mm 
\centering{\epsffile{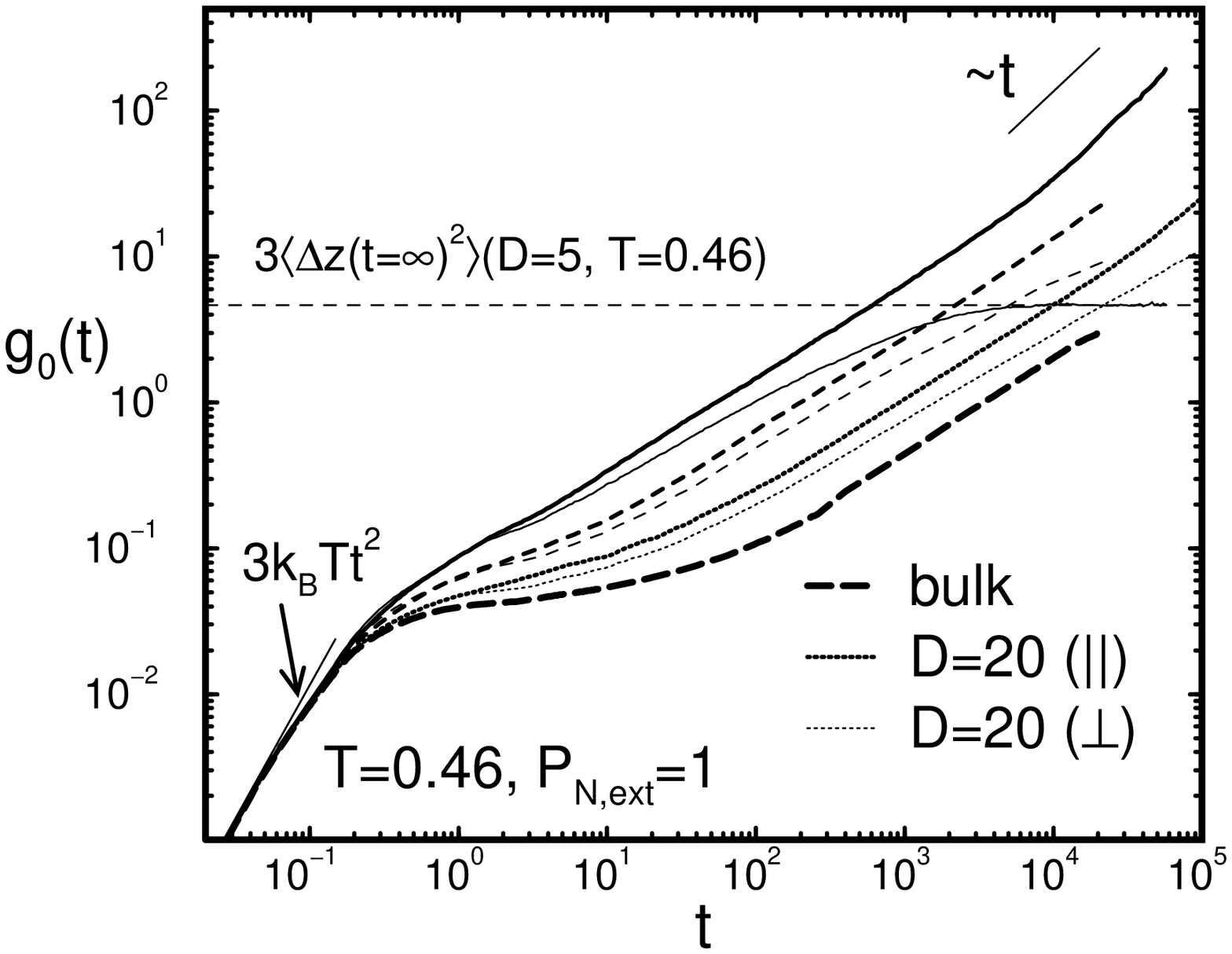}}
\epsfxsize=105mm 
\centering{\epsffile{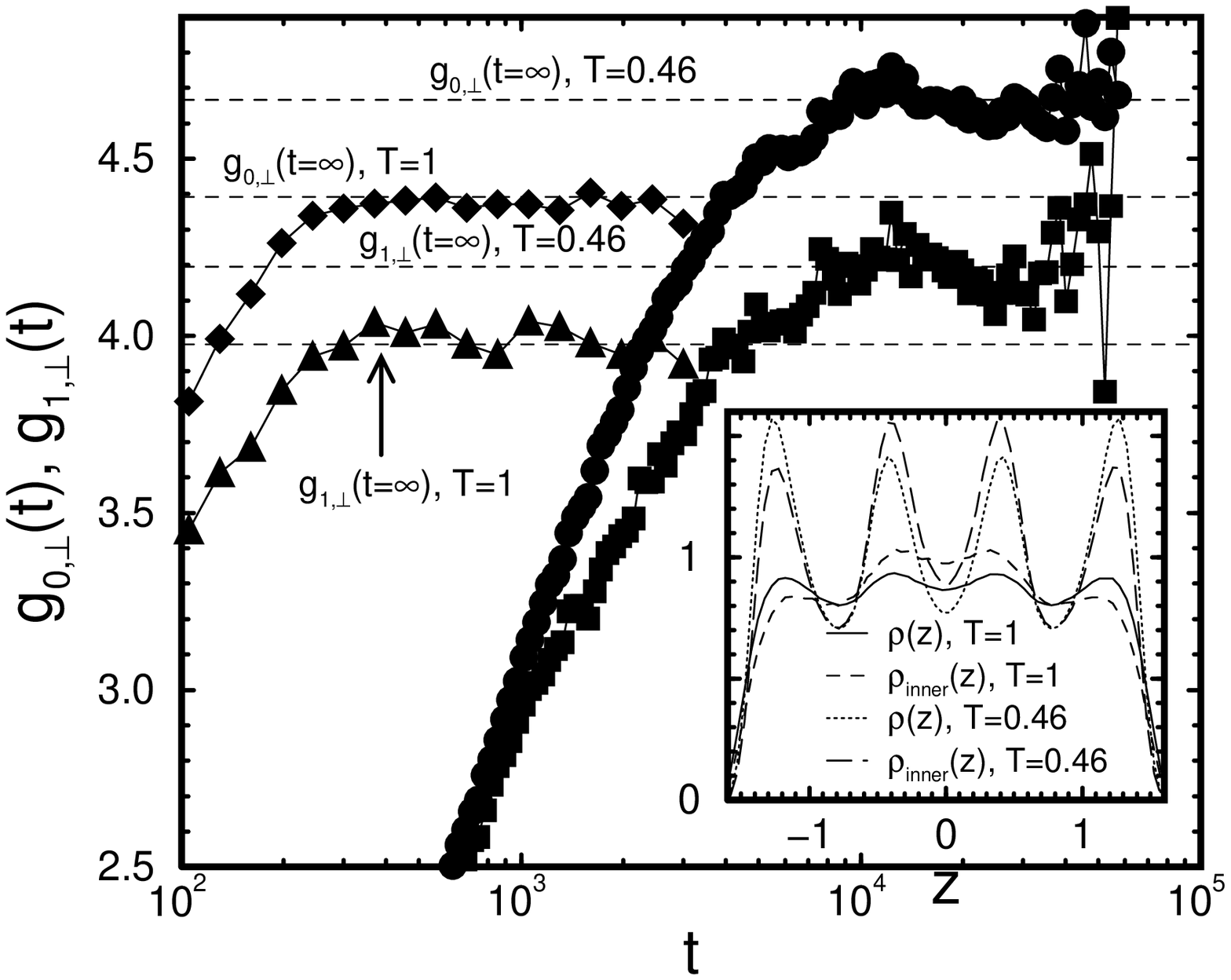}}
\caption[]{
Upper panel:
Log-log plot of the mean-square displacement of all
monomers, $g_0(t)$, at $T\myeq 0.46$ 
(supercooled state). The figure compares the bulk
data with the displacements measured parallel ($\parallel$)
and perpendicular ($\bot$) to the walls in films  of different
thicknesses: $D\myeq 5$ ($\parallel$: thick solid line,
$\perp$: thin solid line), $D\myeq 7$ ($\parallel$: thick dashed
line, $\perp$: thin dashed line) and $D\myeq 20$ ($\parallel$: thick
dotted line, $\perp$: thin dotted line).
The film data were multiplied by $3/2$ for the 
parallel direction and by $3$ for the perpendicular one
to account for the  different number of 
spatial directions used to calculate $g_0(t)$ (i.e., 3 
directions for the bulk, 2 for the parallel direction 
and 1 for the perpendicular direction in the film).
The dashed horizontal line indicates the large-$t$ limit 
of $g_{0, \bot}(t)$ expected from Eq.~(\ref{eq:Deltaz::D5::large_t::limit})
for a film of thickness $D\myeq 5$.
Lower panel:
Test of Eq.~(\ref{eq:Deltaz::D5::large_t::limit}) for
$g_{0, \bot}(t\myeq \infty)$
and $g_{1, \bot}(t\myeq \infty)$ at $T\myeq 1$ and $T\myeq 0.46$
[$g_{0,\bot}(t\; ; T\myeq 0.46)\myeq$ connected circles,
$g_{1,\bot}(t\; ; T\myeq 0.46)\myeq$ connected squares,
$g_{0,\bot}(t\; ; T\myeq 1)\myeq$ connected diamonds and
$g_{1,\bot}(t\; ; T\myeq 1)\myeq$ connected triangles]. The film
thickness is again $D\myeq 5$.
The density profiles of all monomers and 
of the innermost monomer are depicted in the inset.
At the lower temperature of $T\myeq 0.46$,
both density profiles develop pronounced peaks
close to the walls, whereas the density in the film center
is rather decreased compared to $T\myeq 1$. Larger transversal 
distances thus contribute more to the MSD in the $z$-direction.
Therefore, we expect an increase of $g_{0, \bot}(t\myeq \infty)$ and 
$g_{1, \bot}(t\myeq \infty)$ at lower  $T$ [see also the text].
As seen in this panel, this expectation is indeed observed in 
the long time part of $g_{0, \bot}(t)$ 
and $g_{1, \bot}(t)$, respectively.
}
\label{fig:fig11}
\end{figure}

In addition to the parallel displacements the analysis of the motion perpendicular to the wall 
is also interesting because the oscillations of the monomer density profile, which can propagate through the whole film for small $D$ and low $T$, could possibly suppress perpendicular motion substantially.  To check that, Fig.~\ref{fig:fig11} shows the MSD of all monomers, 
$g_0(t)$, computed in direction parallel to wall, 
\begin{eqnarray}
g_{0} (t) &=& \frac{3}{2N} \sum^{N}_{i=1}
\left \langle 
\big[ \mbf{r}_{i, \parallel}(t)-\mbf{r}_{i, \parallel} \big]^{2}
\right \rangle \;,
\label{eq:g0a::def}
\end{eqnarray}
and perpendicular to it
\begin{eqnarray}
g_{0, \bot} (t) &=& \frac{3}{N} \sum^{N}_{i=1}
\left \langle 
\big[ z_i(t) - z_i(0) \big]^{2}
\right \rangle \;.
\label{eq:g0a::per::def}
\end{eqnarray}
Here, $N$ is again the total number of monomers and $\mbf{r}_{i, \parallel} \myeq(x_i,\; y_i)$. 
Furthermore, $z_i$ denotes the $z$ position of $i$-th monomer.
The factors $3/2$ for $g_0(t)$ and $3$ for $g_{0,\bot}(t)$ account for the difference of
independent components: two in parallel and one perpendicular direction compared to $3$ in a 
bulk system.

The upper panel of Fig.~\ref{fig:fig11} depicts $g_0(t)$ at $T\myeq 0.46$ 
(supercooled state). It compares the bulk data with the displacements parallel
and perpendicular to the walls in films of thicknesses $D\myeq 5,\; 7$ 
and $20$.  The comparison of $g_{0,\bot}(t)$
for the films and of $g_0(t)$ for the bulk reveals 
that the confinement does not only accelerate the dynamics in 
parallel, but also in perpendicular direction if $g_{0, \bot}(t, D)$ is sufficiently
smaller than the film thickness.  For all $D$, one then finds $g_{0, \bot}(t, D) > 
g_{0}^{\rm bulk}(t)$ and $g_{0, \bot}(t, D_1) > g_{0, \bot}(t, D_2)$ if $D_1 < D_2$.
However, this increase of the mobility is less pronounced than that of the dynamics in 
parallel direction, so that, for a given film thickness,
$g_{0, \parallel}(t, D)>g_{0, \bot}(t, D)$.

Of course, the perpendicular displacement cannot grow infinitely. It must be limited by the
film thickness. This is illustrated by $g_{0, \bot}(t, D \myeq 5)$ which crosses over to a 
constant of approximately 5 at late times.  In fact, using the density profile, $\rho(z)$, 
one can compute the large time limit of $g_{0,\bot}(t)$ by
\begin{equation}
\lim_{t \to \infty} g_{0, \bot}(t) = 3 \, \myfrac{ \int^{+D/2}_{-D/2} \D z 
\int^{+D/2}_{-D/2} \D z' \rho(z) \rho(z') (z-z')^2 \, }
{\Big(\int_{-D/2}^{+D/2} \rho(z) \D z \Big)^2}\;.
\label{eq:Deltaz::D5::large_t::limit}
\end{equation}
Equation~(\ref{eq:Deltaz::D5::large_t::limit}) can be 
understood as follows:
The contribution of two points $z$ and $z'$ to the long time
limit of $g_{0,\bot}(t)$ is equal to $(z-z')^2$ multiplied by
the probability $p^{(2)}(z,t;\, z', t')\, \D z\, \D z'$ of finding a 
tagged particle at a given time, $t'$, in the interval $[z', z'+\D z']$ 
provided that it was at a previous time, $t$ in $[z, z + \D z]$.
As $|t-t'|$ grows, this probability becomes independent of 
the initial position of the particle, i.e. 
$p^{(2)}(z,t;\, z', t') \myeq p(z, t)\, p(z', t')$, where
$p(z, t)\, \D z$ is the probability of finding a particle
at time $t$ in $[z, z+ \D z]$. Obviously, this probability 
does not depend on the choice of the time origin, i.e.
$p(z, t) \myeq p(z) \myeq \rho(z) / \int \rho(z')\D z'$.
Putting all this together and adding a factor of $3$ for the 
sake of comparison with bulk results, we obtain 
$3(z-z')^2 \rho(z)\,\rho(z')\, \D z \D z'/(\int \rho(z)\D z)^2$
for the contribution of the pair $(z,z')$ to 
$\lim_{t \to \infty} g_{0, \bot}(t)$.
Integration over both variables $z$ and $z'$ yields
Eq.~(\ref{eq:Deltaz::D5::large_t::limit}).

For a film of thickness $D \myeq 5$ at a temperature of $T \myeq 0.46$
insertion of the density profile (obtained from the simulation)
in Eq.~(\ref{eq:Deltaz::D5::large_t::limit}) yields
$\lim_{t \to \infty} g_{0, \bot}(t)\myeq 4.667$. This is
the value to which $g_{0,\bot}(t, D\myeq 5)$ converges for large $t$
(Fig.~\ref{fig:fig11}). 
Note that by replacing the monomer density profile
$\rho(z)$ in Eq.~(\ref{eq:Deltaz::D5::large_t::limit})
by the density profile of the innermost monomer, one 
can obtain the long time limit of $g_{1,\bot}(t)$, 
i.e. $g_{1,\bot}(t\myeq \infty)$.
In a similar way, Eq.~(\ref{eq:Deltaz::D5::large_t::limit})
can be adapted to compute $g_{4, \bot}(t\myeq \infty)$
by using the density profile of end monomers
and/or  $g_{3,\bot}(t\myeq \infty)$ by replacing $\rho(z)$
by the profile of the density of the chain's center of mass,
$\rho_{\rm cm}(z)$.

The general validity of Eq.~(\ref{eq:Deltaz::D5::large_t::limit}) is 
demonstrated in the lower panel of Fig.~\ref{fig:fig11}. Here, $g_{0,\bot}(t)$
and  $g_{1,\bot}(t)$ are plotted for two temperatures,
$T\myeq 0.46$ and $T\myeq 1$.
The corresponding long time limits have been computed as discussed above.
At both temperatures the results obtained by
Eq.~(\ref{eq:Deltaz::D5::large_t::limit}) are in 
good agreement with long time behavior of $g_{0, \bot}(t)$ 
and $g_{1, \bot}(t)$. The inset of the lower panel of 
Fig.~\ref{fig:fig11} depicts the density profiles of all 
monomers and  of the innermost monomer at both investigated 
temperatures. As seen from this inset, the density close
to the walls increases appreciably at lower temperatures.
Furthermore, the formation of strong density peaks close to 
the walls is accompanied by a slight decrease of the density
in the center of the film. As a consequence, the relative weight of
larger transversal distances in Eq.~(\ref{eq:Deltaz::D5::large_t::limit})
increases compared to that of smaller $(z-z')$.
One can therefore expect an increase of $g_{0, \bot}(t\myeq \infty)$ 
at lower $T$. As the density of the innermost monomer changes
in a similar way with temperature, we also expect an increase of 
$g_{1, \bot}(t\myeq \infty)$ when decreasing $T$.
This expectation is nicely born out in the lower panel of 
Fig.~\ref{fig:fig11}.
\subsection{Incoherent intermediate scattering function}
\label{subsection:incoherent::scattering::function}
The discussion of the previous sections suggested that mode-coupling theory (MCT) could
also be a relevant theoretical framework to describe the dynamics of the supercooled 
polymer films.  In this section we want to test this suggestion by an analysis of 
the incoherent intermediate scattering function $\phi_q^{\rm s}(t)$.  

A quantitative application of MCT to the simulation data requires an intricate fit procedure which must simultaneously optimize several parameters subject to various theoretical constraints (some of them have to be independent of temperature, others independent of the wave vector).  Before attempting this analysis simple tests should be carried out to check whether the approach is worthwhile at all.  Two such tests can be performed.  

Mode-coupling theory predicts that there is an intermediate time window in which the scattering function remains close the time- and temperature independent non-ergodicity parameter $f_q^{\rm sc}$.  This time window is called $\beta$-relaxation regime \cite{Goetze::LesHouches::1989,Goetze::JPCM::1999}.  In this regime the scattering function can be written as
\begin{equation}
\phi_q^{\rm s}(t) = f_q^{\rm sc} + h_q^{\rm s}G(t) \; ,
\label{eq:FacTheo}
\end{equation}
where $G(t)$ and $h_q^{\rm s}$ are the $\beta$-correlator and the critical amplitude, respectively \cite{Goetze::LesHouches::1989,Goetze::JPCM::1999}.  Equation~(\ref{eq:FacTheo}) shows that the time-dependent corrections to $f_q^{\rm sc}$ have an important property.  The space- and the time dependences factorize from one another.  

This ``factorization theorem'' \cite{Goetze::LesHouches::1989,Goetze::JPCM::1999} suggests a simple test \cite{Aichele-Baschnagel::EurPhysJE::I,Aichele-Baschnagel::EurPhysJE::II,KobHorbachBinder1999,GleimKob2000} which uses the simulation data directly without invoking any fit procedure.  If $t'$ and $t''$ denote two times belonging to the $\beta$-regime,  then the ratio
\begin{equation}
R^{\rm s}_q(t) = \frac{\phi^{\rm s}_q(t)
- \phi^{\rm s}_q(t')}{\phi^{\rm s}_q(t'') -
\phi^{\rm s}_q(t')} = \frac{G(t) - G(t')}{G(t'') - G(t')}
= R(t) 
\label{eq:DefRofT}
\end{equation}
should only depend on temperature and time, but not on $q$, 
provided Eq.~(\ref{eq:FacTheo}) holds.

Note that $\phis(t)$ varies slowly for times around the plateau.
Therefore, the denominator of Eq.~(\ref{eq:DefRofT})
is fairly small and the accuracy of the test is predicated upon
an appropriate choice of the parameters $t''$ and $t'$.
To obtain a satisfactory signal-to-noise ratio, one would like to
make $t''$ and $t'$ as different as possible. However, one has to 
be careful not to take $t''$ and $t'$ outside the $\beta$-relaxation regime, where
Eq.~(\ref{eq:DefRofT}) is no longer valid. 
We find that $t''\myeq 1$ and $t'\myeq 50$ is a reasonable compromise
for all studied film thicknesses $D\myeq 5, \; 10$ and $D\myeq 20$.

With this choice two observations can be made from Fig.~\ref{fig:fig5}: First, 
there is indeed an intermediate time window for all $D$ where the correlators, 
measured at different $q$, collapse, whereas they splay out at both shorter and 
longer times.  This is a qualitative evidence for the factorization theorem.
However, the best agreement with Eq.~(\ref{eq:DefRofT}) is obtained for $D\myeq 10$.
In the other two cases the superposition of the scattering functions for $t''<t<t'$ 
is not as good.  A possible reason for this difference could be that, for $D\myeq 10$, 
Eq.~(\ref{eq:DefRofT}) is tested at a distance of 
$T-\Tc(D\myeq 10)\myeq 0.42-0.39 \myeq 0.03$ from the critical
temperature of this film thickness. The tests for $D\myeq 20$
and $D\myeq 5$, however, correspond to 
$T-\Tc(D\myeq 20)\myeq 0.46-0.415 \myeq 0.045$ and 
$T-\Tc(D\myeq 5)\myeq 0.35-0.305 \myeq 0.045$, respectively. 
As Eq.~(\ref{eq:DefRofT}) is an asymtotic relation which 
is expected to hold the better, the closer $T$ is to $\Tc$,
Fig.~\ref{fig:fig5} still suggests that the 
factorization theorem is not only satisfied in the bulk 
\cite{Bennemann-Baschnagel-Paul::EPJB10,Aichele-Baschnagel::EurPhysJE::I}, 
but also in the polymer films.

The second observation concerns the order of the $q$-values before and after the 
$\beta$-regime.  This order is preserved.  The top curve, $\phi_{q\myeq 1}^{\rm s}(t)$, 
at short times is also the top curve in the $\alpha$-regime.  Similarly, the bottom 
curve, $\phi_{q\myeq 18.5}^{\rm s}(t)$, before the $\beta$-regime also remains below all 
other $q$-values when leaving the $\beta$-regime again.  This behavior reflects the 
theoretical prediction that the short- and long-time corrections to Eq.~(\ref{eq:FacTheo}) 
exhibit the same $q$-dependence.  It was pointed out in simulations of a binary 
Lennard-Jones mixture \cite{GleimKob2000} and also found for our model in the bulk 
\cite{Aichele-Baschnagel::EurPhysJE::I}.

The second test of the applicability of MCT deals with the late-time relaxation of 
$\phi_q^{\rm s}(t)$.  An important prediction of the theory for the $\alpha$-process 
is the time-temperature superposition principle (TTSP) 
\cite{Goetze::LesHouches::1989,Goetze::JPCM::1999}.  
This means that $\phi_q^{\rm s}(t)$ 
is not a function of time and temperature separately, but only of the scaled 
time $t/\tau_q(T)$, where $\tau_q(T)$ is the $\alpha$-relaxation time.  So, we have
\begin{equation}
\phi_q^{\rm s}(t) = F_q(t/\tau_q) \;.
\label{eq:DefTTSP}
\end{equation} 
The function $F_q(t/\tau_q)$ can often be well approximated by a 
Kohlrausch-Williams-Watts (KWW) function
\begin{equation}
F_q(t/\tau_q^{\rm K})= f_q^{\rm K} \exp \Big [ (t/\tau^{\rm K}_q)^{\beta_q^{\rm K}} \Big ] \;.
\label{eq:DefKWW}
\end{equation}
For time-temperature superposition to hold the amplitude $f_q^{\rm K}$ and the 
stretching exponent $\beta_q^{\rm K}$ must be independent of temperature.  
Only the Kohlrausch relaxation time $\tau_q^{\rm K}$ is a function of $T$.  
MCT predicts that $\tau_q^{\rm K}$ is proportional to the $\alpha$-relaxation 
time and increases upon cooling as  \cite{Goetze::LesHouches::1989,Goetze::JPCM::1999}
\begin{equation}
\tau_q^{\rm K} \propto \tau_q \sim (T - \Tc)^{-\gamma} \;.
\label{eq:tauKvsT}
\end{equation}
Equation~(\ref{eq:tauKvsT}) also implies that any time from the window of the 
$\alpha$-process should exhibit the same temperature dependence and can thus be 
used to test the TTSP.  This means that it should be possible to collapse the 
late time decay of the scattering functions, measured at different $T$, onto a 
common $T$-independent master curve by plotting $\phi_q^{\rm s}(t)$ versus $t/\tau_q$.  
A convenient definition of $\tau_q$ is to simply read off the time when $\phi_q^{\rm s}(t)$ 
has decay to a certain value.  Again, such a test has the advantage that no 
complicated fit procedure is involved.  It works directly with the simulation data.  
A possible choice is $\phi_q^{\rm s}(\tau_q) \myeq 0.1$ 
\cite{Aichele-Baschnagel::EurPhysJE::II}.  This low value warrants that the 
scattering function has decayed sufficiently so that possible perturbations 
from the $\beta$-relaxation are completely negligible. 

The resulting master curves are shown in Fig.~\ref{fig:fig13} for $D\myeq 5,10,20$.
For all film thicknesses, even for the extreme case of $D \myeq 5$, which corresponds 
to three atomic layers only, the TTSP is borne out by the simulation data 
and extends to shorter rescaled times with decreasing temperature, as 
predicted by MCT for homogeneous systems.  In this respect, the films 
behave identical to the bulk 
\cite{Bennemann-Baschnagel-Paul::EPJB10,Aichele-Baschnagel::EurPhysJE::II}.  
However, the results for $D\myeq 10$ suggest that there could also be a qualitative 
difference.  They include $T\myeq 0.4$, which is very close to the estimated
$\Tc$, i.e. $T-\Tc \myeq 0.01$.
Contrary to the bulk \cite{Bennemann-Baschnagel-Paul::EPJB10,Aichele-Baschnagel::EurPhysJE::II}, 
the film data at this $T-\Tc$ exhibit no apparent violation of the TTSP.
Whether this is a general property of the confined systems or just a special 
feature of $D\myeq 10$ is not clear at present.

While the scattering functions exhibit time-temperature superposition for all 
thicknesses studied, they do not superimpose if different thicknesses are 
compared.  This is illustrated in Fig.~\ref{fig:fig14}.  The figure  
shows $\phi_q^{\rm s}$ versus $t/\tau_q$ for $D\myeq 5,20$ and the bulk at 
comparable distances to the corresponding critical temperature (i.e., 
$T-\Tc \simeq 0.045$).  Obviously, the shape of the late-time relaxation 
depends on $D$. Note that, compared to $\betaK_q(D\myeq 20)$,  
the stretching exponent of the {\em thinner} film, i.e., $\betaK_q(D\myeq 5)$
is closer to that of the bulk. A possible explanation could be as follows:
Due to the presence of the walls, there is a distribution of 
the relaxation times along the transversal direction. Regions closer
to the wall decay faster and thus exhibit a smaller relaxation time.
On the other hand, as the  temperature decreases, the influence of the walls 
is ``felt'' throughout the film for all thicknesses studied. There is practically 
no region of (bulk-like) constant relaxation time in the inner part of the film. 
Now, if we divide a film into layers in which the relaxation time is calculated,
the number of layers is smaller in a thin than in a thick film. The simplest
assumption then is that the larger number of layers in the thick film leads to a 
broader distribution of relaxation times.
Of course, if $D$ increases further, there will be a region of
bulk-like behavior which grows in the middle of the film and starts to 
dominate properties of the system for very large $D$.
The film thicknesses studied here, however, 
are far from this limit.

\section{Determination of $\mbf{\Tc(D)}$}
\label{section:Determination::of::Tc}
The analysis of the preceeding section showed that the relaxation of the films
is qualitatively compatible with predictions of mode-coupling theory and that 
it speeds up with increasing confinement.  To some extent, the dynamics of the films
corresponds to the behavior of the bulk obtained at a higher temperature.
This suggests that confinement
reduces the characteristic temperatures of the film compared to those of the bulk.
In this section, we want to quantify this reduction by determining the critical
temperature $\Tc(D)$ and the Vogel-Fulcher-Tammann temperature $T_0(D)$.

Mode-coupling theory~\cite{Goetze::LesHouches::1989,Goetze-Sjoegren::TransportTheoryStatPhys,Goetze-Sjoegren::RepProgPhys55} predicts that the relaxation times of any correlation function, which  couples to  density fluctuations, should exhibit the power-law temperature dependence of Eq.~(\ref{eq:tauKvsT}) if $T$ is close to $\Tc$.  Indeed, the $\alpha$-relaxation times of the incoherent and coherent scattering functions in the bulk may be described by Eq.~(\ref{eq:tauKvsT}) in some $q$-dependent temperature interval~\cite{Bennemann-Baschnagel-Paul::EPJB10,Aichele-Baschnagel::EurPhysJE::II}.  Deviations are found both for temperatures too close and to far away from $\Tcbulk$.  These deviations are theoretically expected. Since Eq.~(\ref{eq:tauKvsT}) is an asymtotic result of the idealized MCT, it can be violated if $T$ is very close to $\Tc$ due to relaxation channels which are not treated by the idealized theory, and very far from $\Tc$ where the asymtotic formula is no longer applicable.  Similarly, the $q$-dependence of the temperature interval, where Eq.~(\ref{eq:tauKvsT}) is valid, has been rationalized by calculating the leading-order corrections to the asymptotic behavior within the framework of idealized MCT \cite{Franosch-Fuchs-Goetze-Mayr-Singh::PhysRevE55::1997,Fuchs-Goetze-Mayr::PhysRevE58::1998}.  Due to our findings in the bulk and due to the 
results of the previous section we expect to obtain similar results when analyzing the films.

To examine this expectation, we define relaxation times as the time 
needed by a given mean-square displacement, like $g_1$, $g_3$ or 
$g_0$ [see Eqs.~(\ref{eq:g1a::def}),~(\ref{eq:g3a::def})
and~(\ref{eq:g0a::def})], to reach the monomer size
\begin{eqnarray}
g_{i} (t \myeq \tau) &:=& 1 \mbox{~(defining equation for $\tau$)}\;.
\label{eq:tau_gi::def}
\end{eqnarray}
This is a reasonable choice because the bulk analysis revealed that a monomer, which has moved across its own diameter, contributes to the $\alpha$-relaxation \cite{Aichele-Baschnagel::EurPhysJE::I}.  Using Eq.~(\ref{eq:tau_gi::def}) we computed $\tau(g_{i} \myeq 1)$ as a function of temperature for various film thicknesses, where, in addition to $g_0$, $g_1$ and $g_3$, the MSD of the end monomers,
\begin{eqnarray}
g_{4} (t) &=& \frac{3}{2M} \sum^{M}_{i=1}
\left \langle 
\big[ \mbf{r}^{\rm end}_{i, \parallel}(t)-\mbf{r}^{\rm end}_{i, \parallel} \big]^{2}
\right \rangle \;,
\label{eq:g4a::def}
\end{eqnarray}
has also been used.  The resulting relaxation times were then fitted by Eq.~(\ref{eq:tauKvsT}) as follows: First, all (three) parameters of Eq.~(\ref{eq:tauKvsT}) were adjusted.  The results for $\gamma$ obtained from $\tau(g_{0}\myeq 1)$, $\tau(g_{1}\myeq 1)$, $\tau(g_{3}\myeq 1)$ and $\tau(g_{4}\myeq 1)$ agreed well within the error bars. Therefore, we fixed $\gamma$ at the average value for the given film thickness and repeated the fits to optimize the remaining parameters.

Table~\ref{tab1} compiles $\Tc(D)$ and 
$\gamma_{\rm MSD}(D)$ obtained in this way.
Figure~\ref{fig:fig15} shows a representative
example of this analysis.  The upper panel depicts $\tau^{-1/\gamma}$
versus $T$ for a film of thickness $D\myeq 5$.
The intersection of the straight lines (MCT-fit results) with the 
$T$-axis yields the critical temperature at this film thickness: 
$\Tc(D \myeq 5) \myeq 0.305 \pm 0.005$.
Note that, despite the highly non-linear relationship between
the MSD's used to define the  various $\tau$'s, all 
fits yield the same $\Tc$.

To test this analysis the determined critical temperature can be 
used to linearize the relaxation time by plotting $\tau$ 
versus $T\!-\!\Tc$ on a  log-log scale.  The lower panel of 
Fig.~\ref{fig:fig15} shows that the power law~(\ref{eq:tauKvsT}) 
is a good approximation close to $\Tc$.  Analogous to the bulk, there are
deviations for large $T\!-\!\Tc$, where the asymptotic regime is left.  On the
other hand, no deviations are observed for small $T\!-\!\Tc$.  In the bulk
\cite{Bennemann-Baschnagel-Paul::EPJB10,Bennemann-Baschnagel-Paul-Binder}, they
were only found for $T\!-\!\Tc \lesssim 0.02$ and thus for temperatures
smaller than those simulated for $D\myeq 5$ up to now.

Furthermore, as indicated by the solid line in the lower panel,
$\tau(T)$ can also be described by a Vogel-Fulcher-Tammann (VFT) equation,
i.e., by
\begin{equation}
\tau(T) \propto  \exp \left [ \myfrac{c(D)}{T-\Tnull(D)} \right] \; ,
\label{eq:VFT::law::for::tau}
\end{equation}
where $c$ is a constant which can depend on film thickness.
The possibility of describing the same data by both a power law (MCT)
and a VFT-fit has also been observed for the bulk (see Fig.~10 
in~\cite{Bennemann-Paul-Binder-Duenweg::PRE57}).
We therefore use the VFT-formula as an independent approach to determine the variation
of $\Tnull$ with film thickness. Table~\ref{tab1} contains the results.
A plot of $\Tc(D)$ and $\Tnull(D)$ is shown in Fig.~\ref{fig:fig17}. Since we expect
$T_0 < \Tg < \Tc$, the figure suggests that also the glass transition temperature, 
$\Tg$, should be reduced for stronger confinement.  Qualitatively, this result parallels
those reported for experiments on supported~\cite{Keddie::Jones::Cory::EuroPhysLett28} and on
freely standing polystyrene films~\cite{Forrest::2000,Forrest::PRL77::page2002,Dalnoki-VeressMurray}
as well as for MD simulations of a polymer model similar to ours
in the case of weak monomer-substrate attraction~\cite{Torres-Nealey-dePablo::PRL85}.

The critical temperatures, $\Tc(D)$, were determined, for instance, 
from the mean-square displacement of all monomers [see Eq.~(\ref{eq:tau_gi::def})] 
and thus from a quantitiy which corresponds to the low-$q$ limit of the incoherent 
scattering function.  Figure~\ref{fig:fig18} shows that the same $\Tc(D)$ can 
also be used to linearize the relaxation times $\tau_q$ 
at maximum of the static structure factor. Here, $\tau_q$ was defined 
by $\phis(t \myeq \tau_q) \myeq 0.3$.  The upper panel depicts a log-log 
plot of $\tau_q$ versus $T-\Tc$ for $D\myeq 5, 7, 10,15, 20$ and for the bulk.  
It illustrates that a shift of the temperature scale by $\Tc(D)$ leads to a very 
good superposition of the film data for all $T$ studied and also of the film 
and the bulk if $T-\Tc > 0.1$.  
For smaller $T-\Tc$ deviations occur.
The relaxation time of the bulk increases less steeply than that of the films.  
This implies that the exponent $\gamma_{\phi}$ of Eq.~(\ref{eq:tauKvsT}) is  
smaller in the bulk than in the films.  
If one fits Eq.~(\ref{eq:tauKvsT}) to that $T$-interval, where the film data are 
linear, one obtains $\gamma_{\phi}$-values which range between 
$\gamma_{\phi}(D\myeq 10)\myeq 2.68$
and $\gamma_{\phi}(D\myeq 5)\myeq 3.24$ and which do not increase 
monotonically with decreasing $D$ [see table~\ref{tab1}].
The lower panel of Fig.~\ref{fig:fig18} illustrates 
the same data in a  slightly different way. 
Here, $\tau_q^{-1/\gamma_{\phi}}$ is depicted versus $T-\Tc(D)$.
The exponents, $\gamma_{\phi}$, used for this plot are
obtained from fits to  Eq.~(\ref{eq:tauKvsT}) using $\tau_{q}$
as input data and keeping $\Tc(D)$ constant.
As expected, curves at all film thicknesses converge towards the origin of coordinates.
In other words, the computed critical temperatures are consistent with  a
temperature  dependence of $\tau_{q}$ according to  Eq.~(\ref{eq:tauKvsT}).
On the other hand, the curves splay out at large $T-\Tc(D)$ 
indicating that the prefactor,  $B$, in $\tau_q^{-1/\gamma_{\phi}}\myeq B\, |T-\Tc(D)|$ 
depends on the film thickness, $D$.

As mentioned above, although not monotonically, the exponent 
$\gamma_{\phi}$ seems to increase with decreasing $D$ [see also table~\ref{tab1}]. 
To some extent, this finding is unexpected, given the variation of the 
Kohlrausch exponent $\beta_q^{\rm K}$ in Fig.~\ref{fig:fig14}.
For the bulk, mode-coupling theory predicts $\lim_{q\rightarrow \infty}\beta_q^{\rm K}\myeq b$ \cite{Fuchs::JNon-CrystSolids::1994}.  Here, $b$ is the von-Schweidler exponent which is related to 
$\gamma_{\phi}$ by $\gamma_{\phi} \myeq  1/2a + 1/2b$ \cite{Goetze::LesHouches::1989,Goetze-Sjoegren::TransportTheoryStatPhys,Goetze-Sjoegren::RepProgPhys55}.  The exponents $a$ (exponent of the ``critical decay'') and $b$ are correlated with one another.  A decrease or an increase of $b$ entails a decrease or an increase of $a$.  When applying these bulk predictions to the films the $D$-dependence of $\beta_q^{\rm K}$ in  Fig.~\ref{fig:fig14} [$\beta_q^{\rm K}(D\myeq 20) < \beta_q^{\rm K}(D\myeq 5) < \beta_q^{\rm K\, bulk}$] suggests $b(D\myeq 20) < b(D\myeq 5) < b_{\rm bulk}$ and thus $\gamma_{\phi}(D\myeq 20) > \gamma_{\phi}(D\myeq 5) > \gamma_{\phi, \;\rm bulk}$.  
While $\gamma_{\phi}(D) > \gamma_{\phi,\; \rm bulk}$ for all $D$, the results of table~\ref{tab1}
do not confirm the expected order between the different film thicknesses.

However, one should not conclude from the preceding discussion that a consistent application of MCT might not be possible.  On the one hand, it is not clear whether the $D$-dependence of the finite-$q$ result in Fig.~\ref{fig:fig14} ($\beta_q^{\rm K}$ at $q\myeq 6.9$) is the same as in the limit $q\rightarrow \infty$ (i.e., as for $b$).  The corrections to the asymptotic behavior could depend on both $q$ and $D$.  On the other hand,  the fitted value for $\gamma_{\phi}$ at $D\myeq 5$ is the least certain because the lowest temperature simulated corresponds to a rather large $T-\Tc(D\myeq 5) \simeq 0.045$ contrary to $T-\Tc \simeq 0.025$ for $D\myeq 20$.  Therefore, the simulations have to be extended to lower temperature to test whether there are clear deviations from the superposition of $\tau_q$ at low $T$ in Fig.~\ref{fig:fig18}.  The present quality of the fits would rather suggest that the 
$\gamma_{\phi}$-values for the different film thicknesses are very close to each other.

Within the framework of MCT, quantities, such as the exponents $a$, $b$ or $\gamma$ and 
the critical temperature, are determined by the thermodynamic properties of the 
glass former, in particular by the static structure factor 
\cite{Goetze::LesHouches::1989,Goetze-Sjoegren::TransportTheoryStatPhys,Goetze-Sjoegren::RepProgPhys55}.
The discussion of Fig.~\ref{fig:fig8} already suggested that $S(q)$ 
changes with decreasing $D$ in a quite similar fashion as the bulk $S(q)$ if 
the temperature is increased.  Therefore, the question arises of whether it is 
possible to superimpose the bulk and film results for $S(q)$ by comparing the data for the 
same $T-\Tc(D)$.  This would imply that the reduction of $\Tc$ in the films is 
closely related to the fact that the development of the local packing, characteristic 
of the supercooled bulk, is shifted to lower temperature by the presence of the 
smooth walls.  Figure~\ref{fig:fig19} shows that such as superpostions for the same $T-\Tc(D)$ is 
possible if the confinement is not too strong (here, for $D \ge 10$).  The upper 
panel compares $S(q)$ of the bulk with that of a film of thickness $D\myeq 10$ for 
$T-\Tc\myeq 0.01$.  With the exception of the (slightly) different amplitude of the 
first maximum, both structure factors are identical over the whole $q$-range.  The 
lower panel shows the same comparison for the bulk and films of thicknesses $D\myeq 5, 10$ 
and $20$ at a larger distance from $\Tc$, i.e., for $T-\Tc\myeq 0.05$.  While $S(q)$ of 
the bulk and film still coincide for $D \ge 10$, this is no longer the case for the thinnest 
film studied ($D \myeq 5$).  The influence of the confinement on the packing  
structure of the system at this film thickness cannot be explained by a mere shift of 
the temperature axis.
\section{Summary}
\label{section:conclusion}
Results of extensive MD simulations of thin (non-entangled) polymer films
are presented, which focus on the influence of confinement
on the sluggish dynamics of the system and in particular on the glass transition
temperature.  The film geometry is realized by introducing two perfectly 
smooth and purely repulsive walls. All simulations are carried out at 
constant normal pressure $\PNext \myeq 1$.

The static properties of the system show 
that chains close to the walls prefer a parallel alignment.
However, when averaged over the whole film and all directions, 
the influence of the walls on the chain conformations becomes very weak.
In particular, we find that the chains' radius of gyration, $\Rgsquare$, 
does not depend much on film thickness. Even for the extreme case 
of $D\myeq 5$, $\Rgsquare$ lies only by $10\%$ below the corresponding bulk value.

On the level of the overall packing structure of the melt, we observe that the structure factor, $S(q)$, of a film of thickness $D$ resembles that of the bulk at a higher temperature.  If the confinement is not too strong ($D \ge 10 $), $S(q)$, measured for a particular $D$ at some $T'$, almost 
coincides with the bulk result for that temperature $T''$ which lies at the 
same distance to $\Tcbulk$ as $T'$ to $\Tc(D)$ [i.e. $T''-\Tcbulk \myeq T'-\Tc(D)$].
This indicates that $T-\Tc(D)$ is a relevant parameter for our confined system.

This static property of our model finds a counterpart in the dynamic behavior.  Our main findings for the dynamics can be summarized as follows:  (1) The relaxation of the supercooled films is accelerated compared to the bulk so that characteristic temperatures, such as the mode-coupling critical temperature, $\Tc(D)$, or the Vogel-Fulcher-Tammann temperature, $T_0(D)$, decrease with decreasing film thickness.  As we expect $T_0 \le \Tg \le \Tc$, our results suggest that also $\Tg(D)$ should decrease with decreasing film thickness.  (2) The films exhibit several features predicted by mode-coupling theory, such as the space-time factorization property in the intermediate time window of the $\beta$-process, time-temperature superposition of the $\alpha$-relaxation, and a power-law increase of the $\alpha$-relaxation time in a $T$-interval that is close, but not too close to $\Tc$.  This implies that the implications of the cage effect, whose approximate mathematical treatment leads to these predictions, also seems to be an important factor to understand the dynamics of the confined system in the supercooled state above $\Tc$.  (3) A comparison of the mean-square displacements in direction parallel and perpendicular to the walls shows that not only the parallel motion, but also the perpendicular motion is accelerated compared to the bulk if the displacement is sufficiently smaller than the film thickness.  However, for a given thickness, the parallel motion is always faster than that in transverse direction.  In other words, the enhancement of the dynamics in more pronounced when relaxation processes parallel to the walls are considered.  Due to the film geometry, it is clear that the long-time limit of the perpendicular mean-square displacements must be finite. We gave an expression which allows a computation of this limitting value from the density profile of the particle that corresponds to the MSD under consideration (i.e., inner monomer, center of mass, etc.).

\section*{Acknowledgement}
We thank J. Horbach for helpful discussions on various 
aspects of this work. We gratefully acknowledge the financial 
support by the ``Deutsche Forschungsgemeinschaft'' (DFG) under the project number 
SFB262 and by BMBF under the project number 03N6015.
We are also indebted to the European Science 
Foundation for financial support by the ESF Programme on ``Experimental and 
Theoretical Investigations of Complex Polymer Structures'' (SUPERNET).
Generous grants of simulation time by 
the computer center at the university of Mainz (ZDV), 
the NIC in J\"ulich and 
the RHRK in Kaiserslautern are also acknowledged.

\newpage
\begin{table}
\begin{tabular}{|c||c|c|c|c|c|c|}
$D$        & 5     & 7     & 10    & 15    & 20    & bulk \\
\hline
$\Tnull$ & 0.204 $\pm$ 0.007 & 0.253 $\pm$ 0.013 & 0.288 
$\pm$ 0.006 & 0.297 $\pm$ 0.007 
& 0.308 $\pm$ 0.004 & 0.328$\pm$ 0.008 \\
\hline
$T_\mr{c}$ & 0.305 $\pm$ 0.006 & 0.365 $\pm$ 0.007 & 0.390 
$\pm$ 0.005 & 0.405 $\pm$ 0.008 
& 0.415 $\pm$ 0.005 & 0.450 $\pm$ 0.005 \\
\hline
$\gamma_{\rm MSD}$ & 
2.5 $\pm$ 0.2 & 
2.4 $\pm$ 0.2 & 
2.1 $\pm$ 0.1 & 
2.2 $\pm$ 0.1 & 
2.1 $\pm$ 0.1 & 
1.84 $\pm$ 0.1 \\
\hline
$\gamma_{\phi}$ &
3.24$\pm$ 0.08 & 
3.15 $\pm$ 0.1 &
2.68  $\pm$ 0.08 & 
2.76  $\pm$ 0.1 & 
2.74 $\pm$ 0.1 & 
2.09 $\pm$ 0.07
\end{tabular}
\caption[]{Survey of the VFT-temperature, $\Tnull$, the mode-coupling 
critical temperature, $\Tc$, and the 
critical exponents, $\gamma_{\rm MSD}$ and $\gamma_{\phi}$, 
for different film thicknesses $D$ and for the bulk.
$\Tnull$ was determined via fits to Eq.~(\ref{eq:VFT::law::for::tau}) 
both for the film and for the bulk.
The critical temperature, $\Tc(D)$, was obtained from fits to 
Eq.~(\ref{eq:tauKvsT}). In both cases, fits were done for the relaxation time 
extracted from mean-square displacements [see Eq.~(\ref{eq:tau_gi::def})].
$\Tcbulk$ was known from previous analyses of 
incoherent and coherent scattering~\cite{Bennemann-Baschnagel-Paul::EPJB10,%
Aichele-Baschnagel::EurPhysJE::I,%
Aichele-Baschnagel::EurPhysJE::II}. 
The same result for $\Tcbulk$ is also obtained by applying 
Eq.~(\ref{eq:tauKvsT}) to the bulk-MSD's. 
The corresponding critical exponent $\gamma_{\rm MSD}$ 
is listed in the third row of
the table. The critical exponent, $\gamma_{\phi}$ (last row), 
is obtained from fits to Eq.~(\ref{eq:tauKvsT}), now using the 
$\alpha-$relaxation times, $\tau_{q\myeq 6.9}$, defined by
$\phis(t\myeq \tau_q)\myeq 0.3$. In this case, $\Tc(D)$ was kept
fixed and only the prefactor and $\gamma_{\phi}$ were varied.
}
\label{tab1}
\end{table}
%
%
\vspace*{-20mm}
\newpage
\drawxyzaxes{240}{330}
\newpage
\vspace*{-5mm}
\newpage
\newpage
\newpage
\newpage
\newpage
\newpage
\newpage
\begin{figure}
\epsfxsize=115mm
\centering{\epsffile{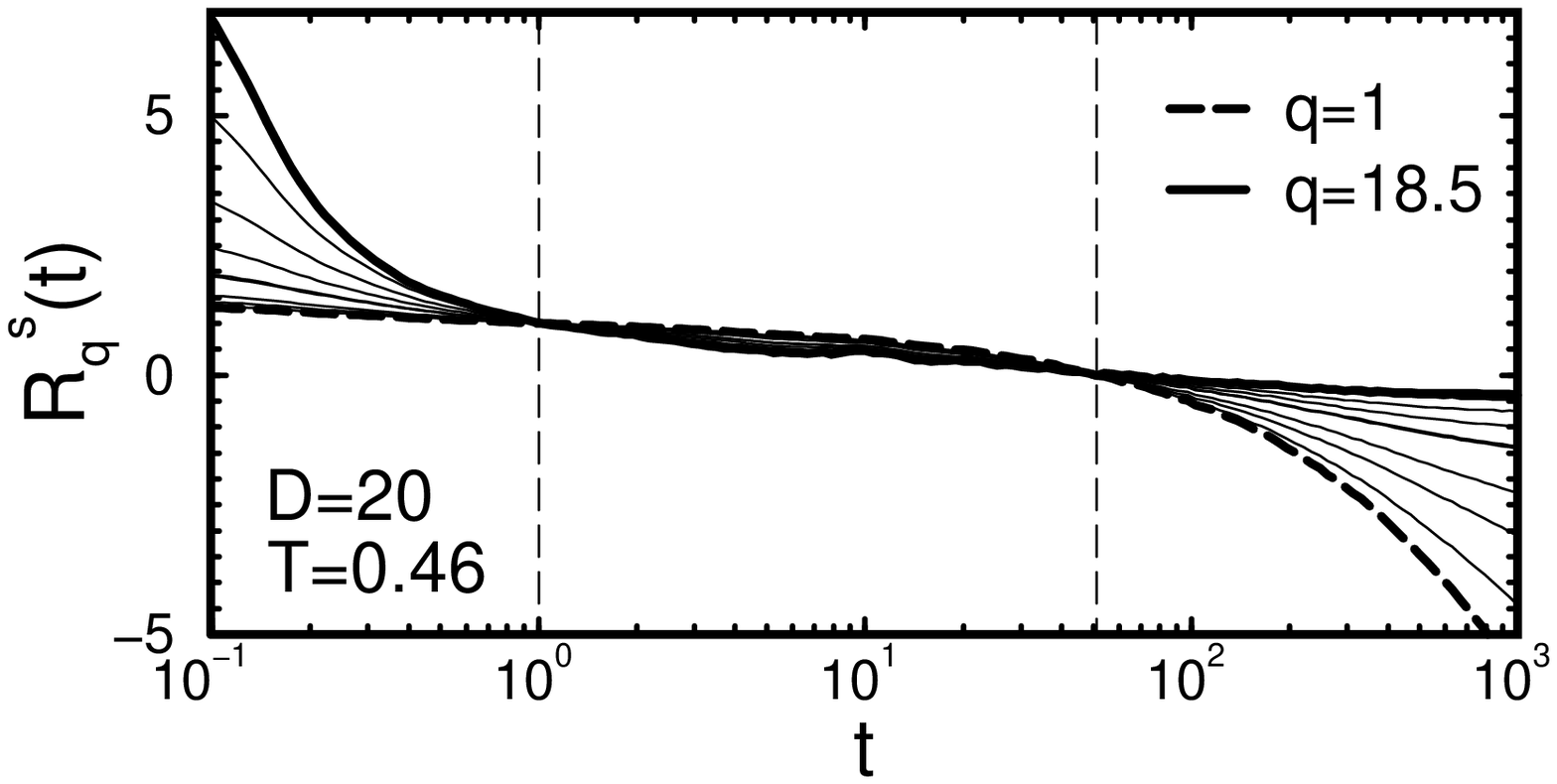}}
\epsfxsize=115mm
\centering{\epsffile{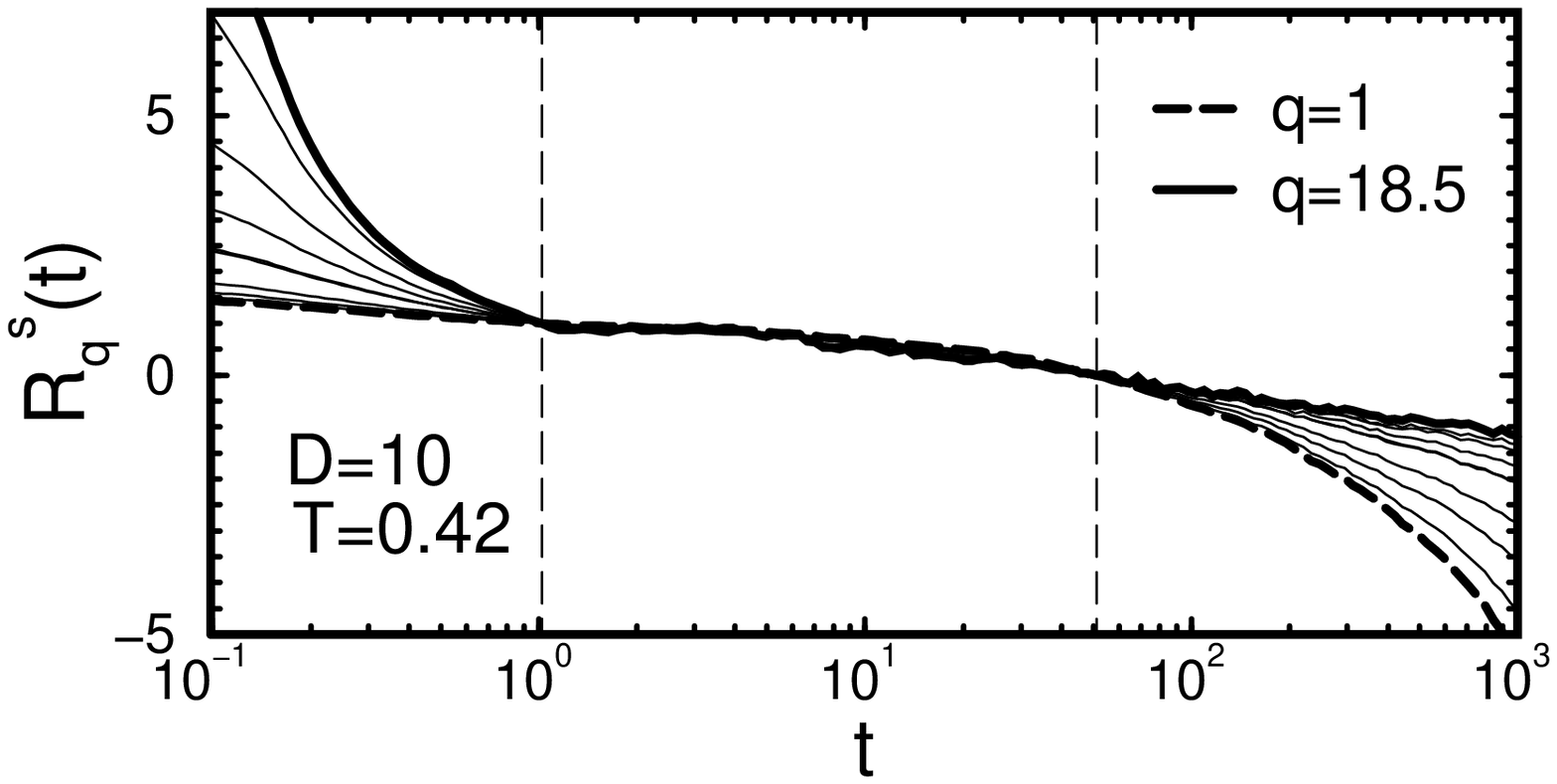}}
\epsfxsize=115mm
\centering{\epsffile{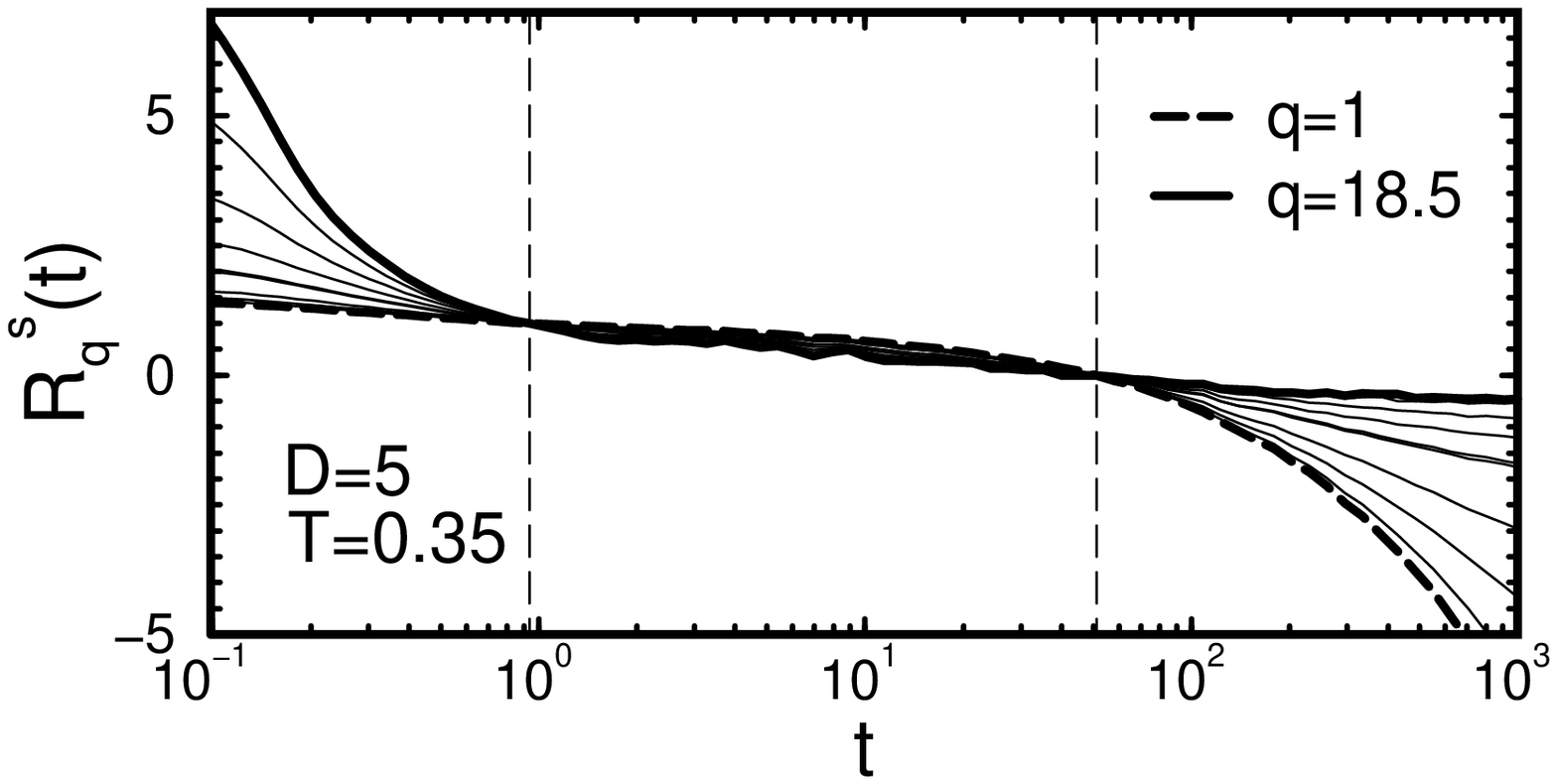}}
\caption[]{Test of the space-time factorization theorem [Eq.~(\ref{eq:FacTheo})] 
for the incoherent intermediate scattering function $\phi_q^{\rm s}(t)$ via the 
ratio $R_q^{\rm s}(t)$ defined by Eq.~(\ref{eq:DefRofT}).  The three panels show the 
results for different film thicknesses and temperatures: $D\myeq 20$, $T\myeq 0.46$ 
($\hat{\myeq }\; T-\Tc \myeq  0.045$), $D\myeq 10$, $T\myeq 0.42$ ($\hat{\myeq }\; T-\Tc \myeq  0.03$), 
$D\myeq 5$, $T\myeq 0.35$ ($\hat{\myeq } \;T-\Tc \myeq  0.045$).  The smallest ($q\myeq 1$) and the 
largest ($q\myeq 18.5$) wave vectors are highlighted by a thick dashed line and a 
thick solid line, respectively.  The thin solid lines in between refer to 
the following $q$-values (from bottom to top): $q\myeq 1.8$, 3, 4.3, 6.9, 7.1, 
9.5, 12.5, 16. The vertical dashed lines indicate the choices for $t''$ (left line)
and $t'$ (right line). By definition, $R_q^{\rm s}(t'')\myeq 1$ and $R_q^{\rm s}(t') 
\myeq 0$.
}
\label{fig:fig12}
\end{figure}
\newpage
\begin{figure}
\epsfxsize=81.75mm
\centering{\epsffile{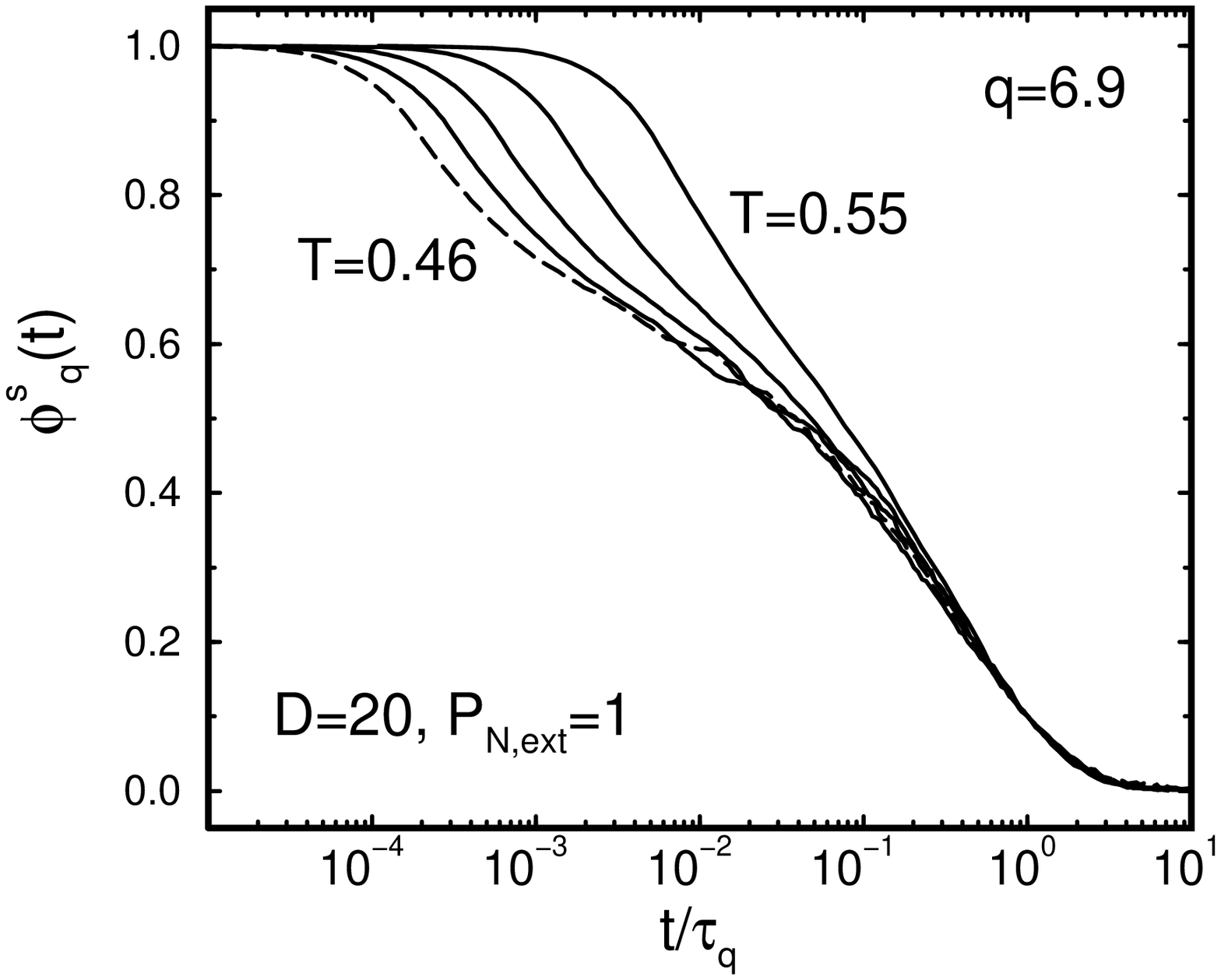}}
\epsfxsize=81.75mm
\centering{\epsffile{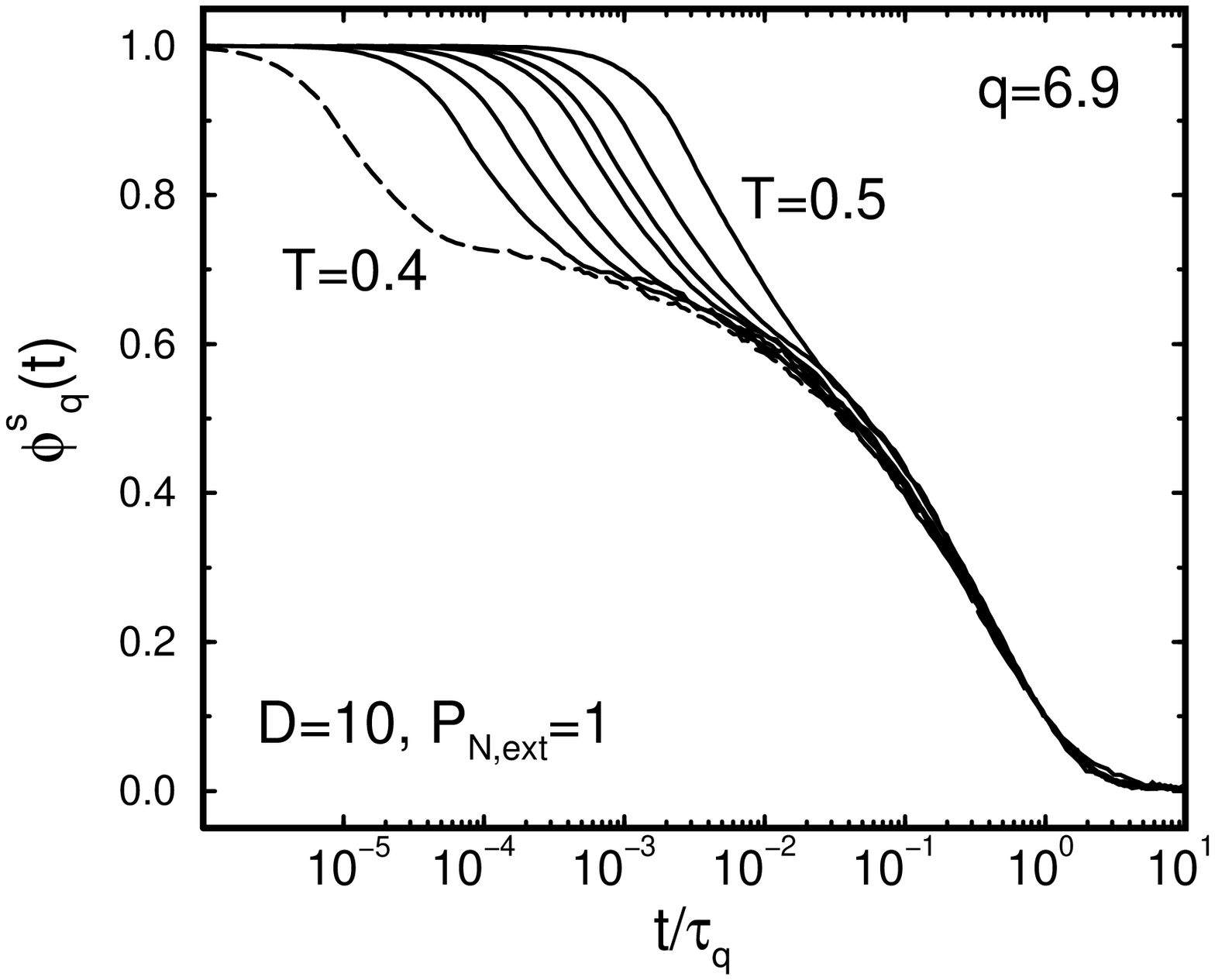}}
\epsfxsize=81.75mm
\centering{\epsffile{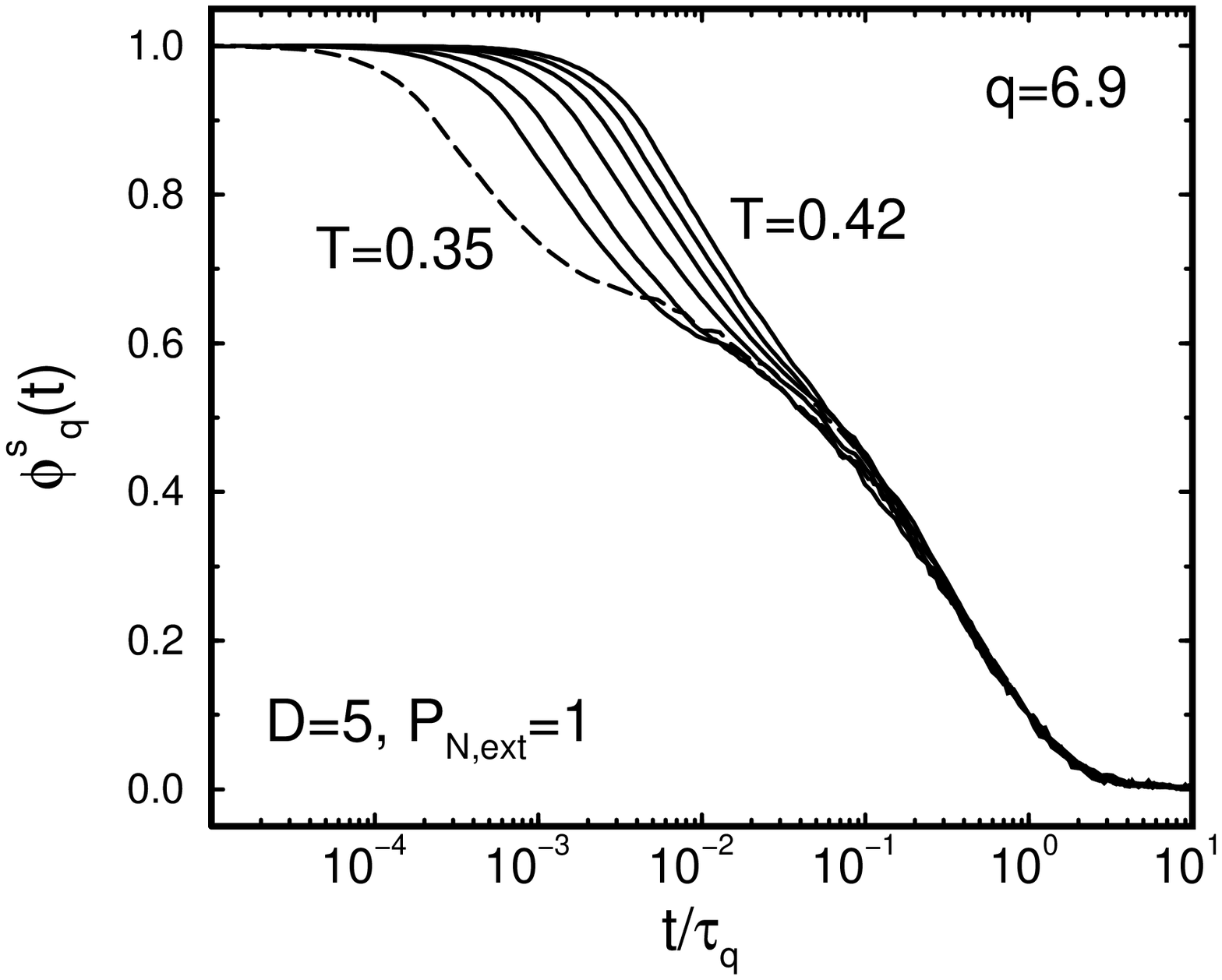}}
\caption[]{Test of the time-temperature superposition principle 
[Eq.~(\ref{eq:DefTTSP})] for three different film thicknesses. Top panel: $D\myeq 20$. 
Temperatures shown are from left to right: $T\myeq 0.46\; (\hat{\myeq }\; T-\Tc \myeq  0.045)$, 
0.47, 0.48, 0.48, 0.5, 0.55 .  Middle panel: $D\myeq 10$. Temperatures shown are from 
left to right: $T\myeq 0.4 \;(\hat{\myeq }\; T-\Tc \myeq  0.01)$, 0.42, 0.43, 0.44, 0.45, 
0.46, 0.47, 0.5.  Bottom panel: $D\myeq 5$. Temperatures shown are from left to 
right: $T\myeq 0.35 \; (\hat{\myeq }\; T-\Tc \myeq  0.045)$, 0.37, 0.38, 0.39, 0.4, 0.41, 0.42.  
In all three panels the time axis is rescaled by the $\alpha$-relaxation time, $\tau_q$, 
defined by $\phis(t\myeq \tau_q)\myeq 0.1$, and the lowest temperature is 
depicted by a dashed line.
}
\label{fig:fig13}
\end{figure}
\newpage
\begin{figure}
\epsfxsize=140mm
\centering{\epsffile{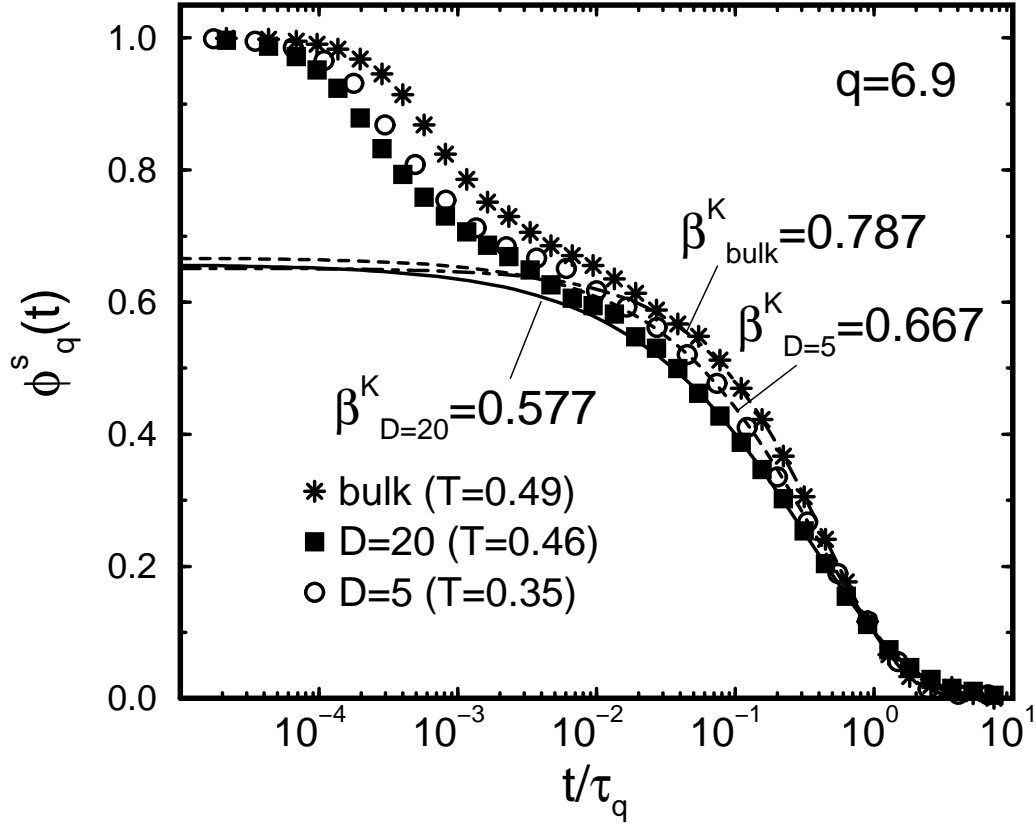}}
\caption[]{Incoherent intermediate scattering function $\phi_q^{\rm s}(t)$,
calculated at the maximum of $S(q)$ ($q\myeq 6.9$), 
versus rescaled time $t/\tau_q$ for $D\myeq 5,20$ and the bulk.  $\tau_q$ is 
defined by $\phis(t\myeq \tau_q)\myeq 0.1$.  The temperatures shown roughly 
correspond to the same distance to $\Tc$, i.e., to $T-\Tc \simeq 0.045$,
for both the films and the bulk.  In addition, the results of KWW-fits
[Eq.~(\ref{eq:DefKWW})] to the long-time behavior of $\phis(t)$ are 
depicted ($D\myeq 5$: dashed line, $D\myeq 20$: solid line, bulk: dot-dashed line).
The corresponding stretching exponents are indicated in the figure.
In all three cases, only data with $t/\tau_q\!>\!0.05$ were used 
for the KWW-Fit.
}
\label{fig:fig14}
\end{figure}
\newpage
\begin{figure}
\epsfxsize=105mm
\hspace*{15mm}%
\epsffile{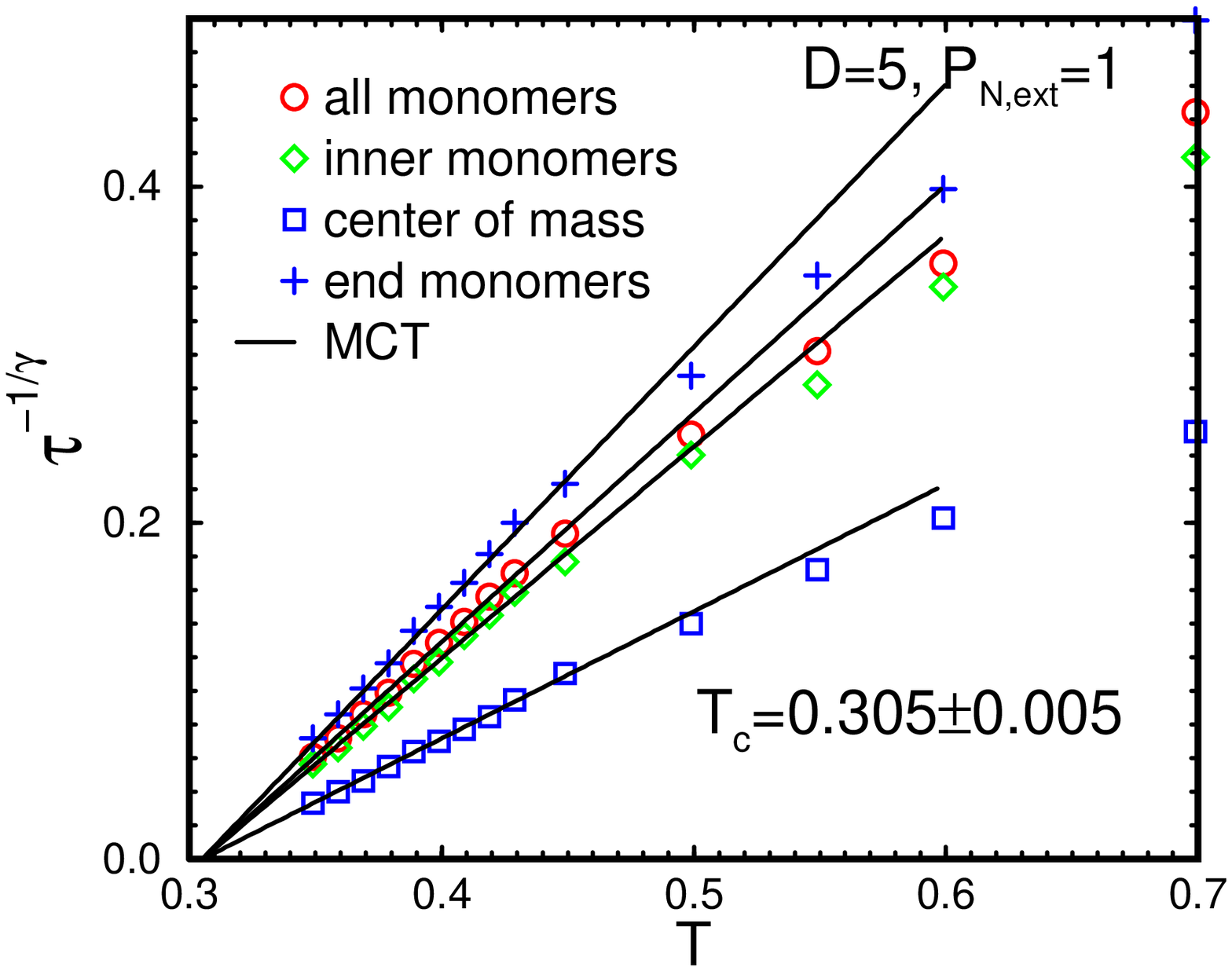}
\epsfxsize=105mm 
\hspace*{20mm}
\epsffile{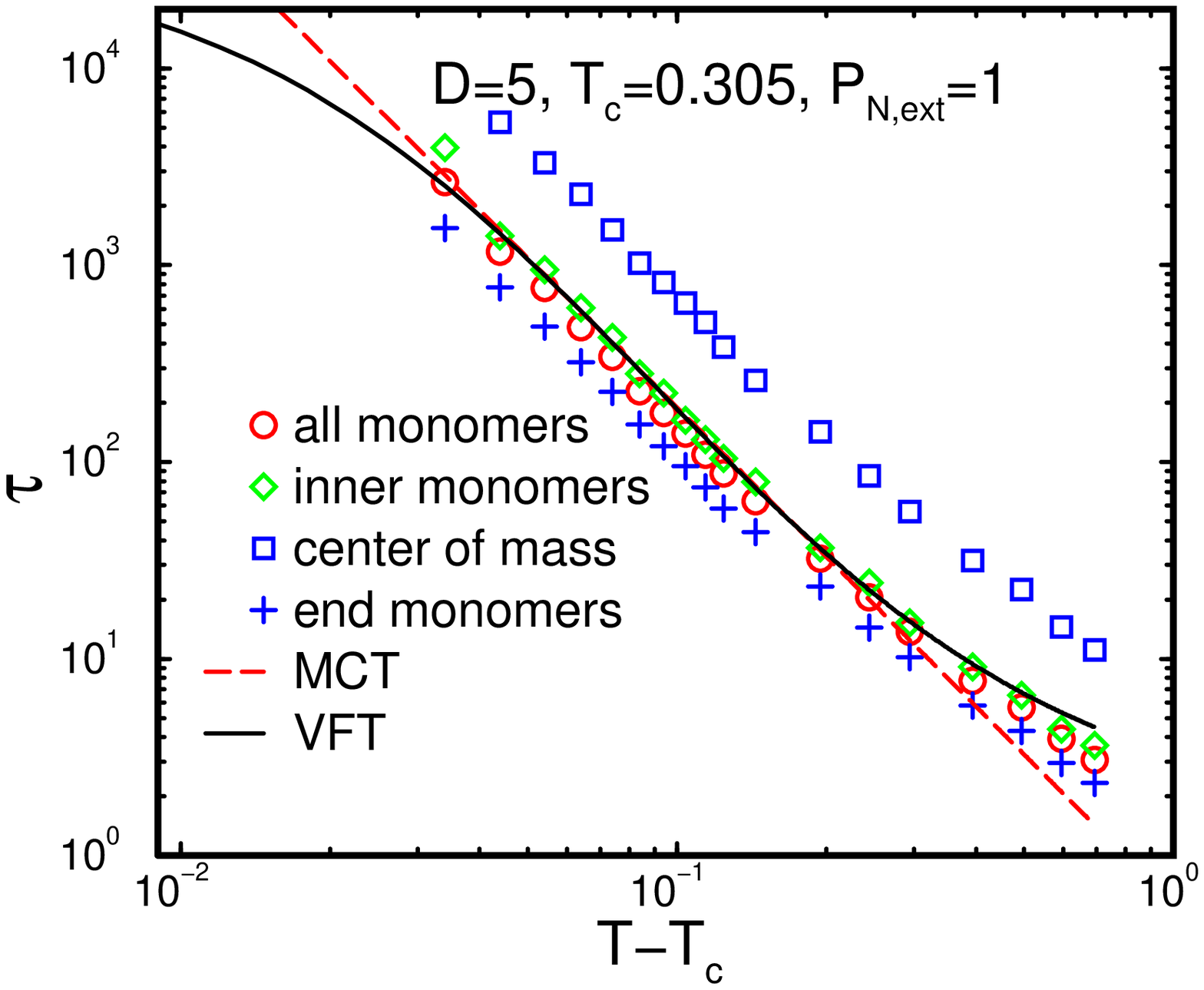}
\caption[]{Upper panel: 
Plot of $\tau^{-1/\gamma}$ versus $T$ for
$D\myeq 5$.  The relaxation time $\tau$ was determined by 
Eq.~(\ref{eq:tau_gi::def}) using the mean-square displacements (MSD) of
inner-, end- and all monomers and of the chain's center of mass.
The exponent $\gamma$ was first determined by fits to 
Eq.~(\ref{eq:tauKvsT}) with three open parameters.
Since the results for all MSD's agreed within the error bars ($\gamma \myeq 2.5\pm 0.2$), 
$\gamma \myeq 2.5$ was used in the plot. The fits to Eq.~(\ref{eq:tauKvsT})
are represented by straight lines. The intersection of these lines with the 
$T$-axis determines the critical temperature of the film ($\Tc(D\myeq 5) \myeq 0.305 
\pm 0.005$).
Lower panel: Different representation of the same data. Now, $\tau$ is plotted 
versus $T-\Tc$ ($\Tc(D\myeq 5) \myeq 0.305$).
The dashed line indicates the MCT-fit [Eq.~(\ref{eq:tauKvsT})] using $\gamma\myeq 2.5$,
whereas the solid line represents a fit to the VFT-equation [Eq.~(\ref{eq:VFT::law::for::tau})]
with $T_0(D\myeq 5) \myeq 0.204 \pm 0.026$. Both fits shown here were done for $g_1(t)$.
}
\label{fig:fig15}
\end{figure}
\newpage
\begin{figure}
\epsfxsize=140mm
\centering{\epsffile{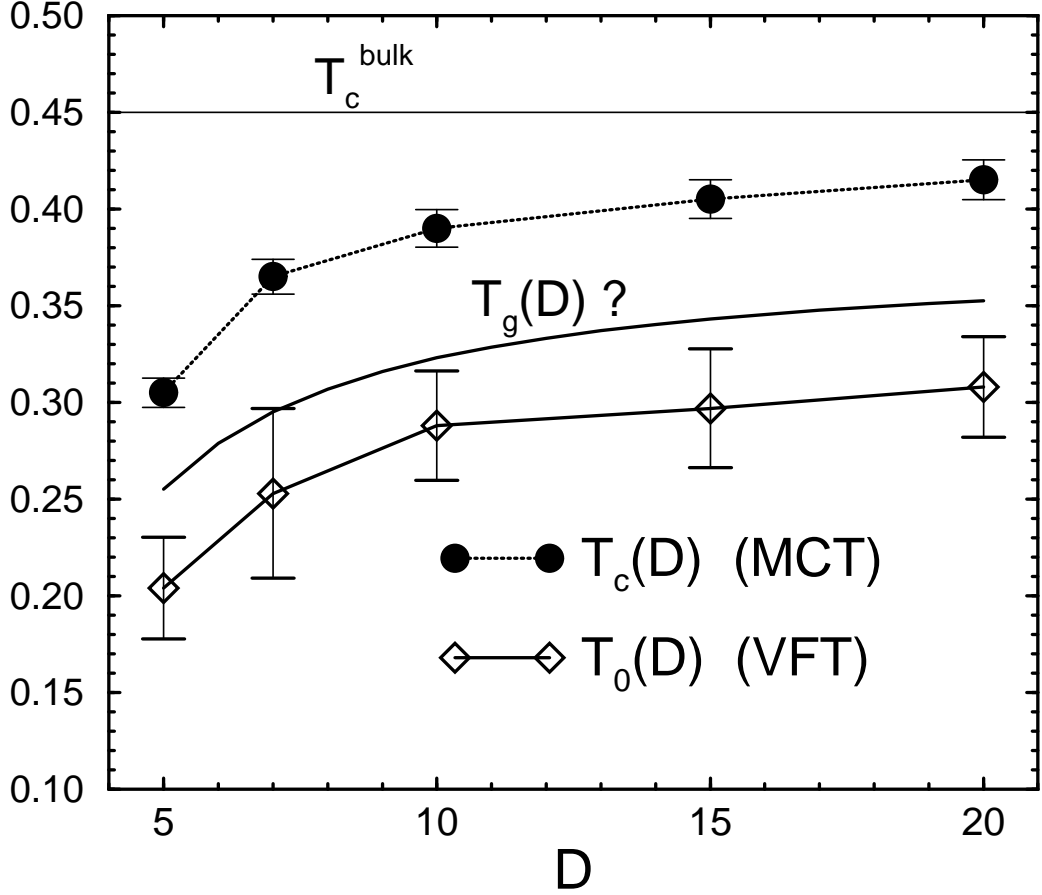}}
\caption[]{Mode-coupling critical temperature, $T_\mr{c}(D)$,
and the VFT-temperature $T_\mr{0}(D)$ versus film thickness $D$.
$T_\mr{c}(D)$ was obtained from fits to 
Eq.~(\ref{eq:tauKvsT}). Similarly, the  
VFT-temperatures, $\Tnull(D)$, and $T^{\rm bulk}_0$ 
are results of fits to  Eq.~(\ref{eq:VFT::law::for::tau}). 
Note that, the glass transition temperature $\Tg$ lies ${\em between}$
these two temperatures: $\Tc<\Tg<T_0$. Thus, $\Tg(D)$ 
lies somewhere between $T_0(D)$ and $\Tc(D)$. 
The solid line gives a suggestion for the possible 
form of $\Tg(D)$.
}
\label{fig:fig17}
\end{figure}
\newpage
\begin{figure}
\epsfxsize=110mm
\centering{\epsffile{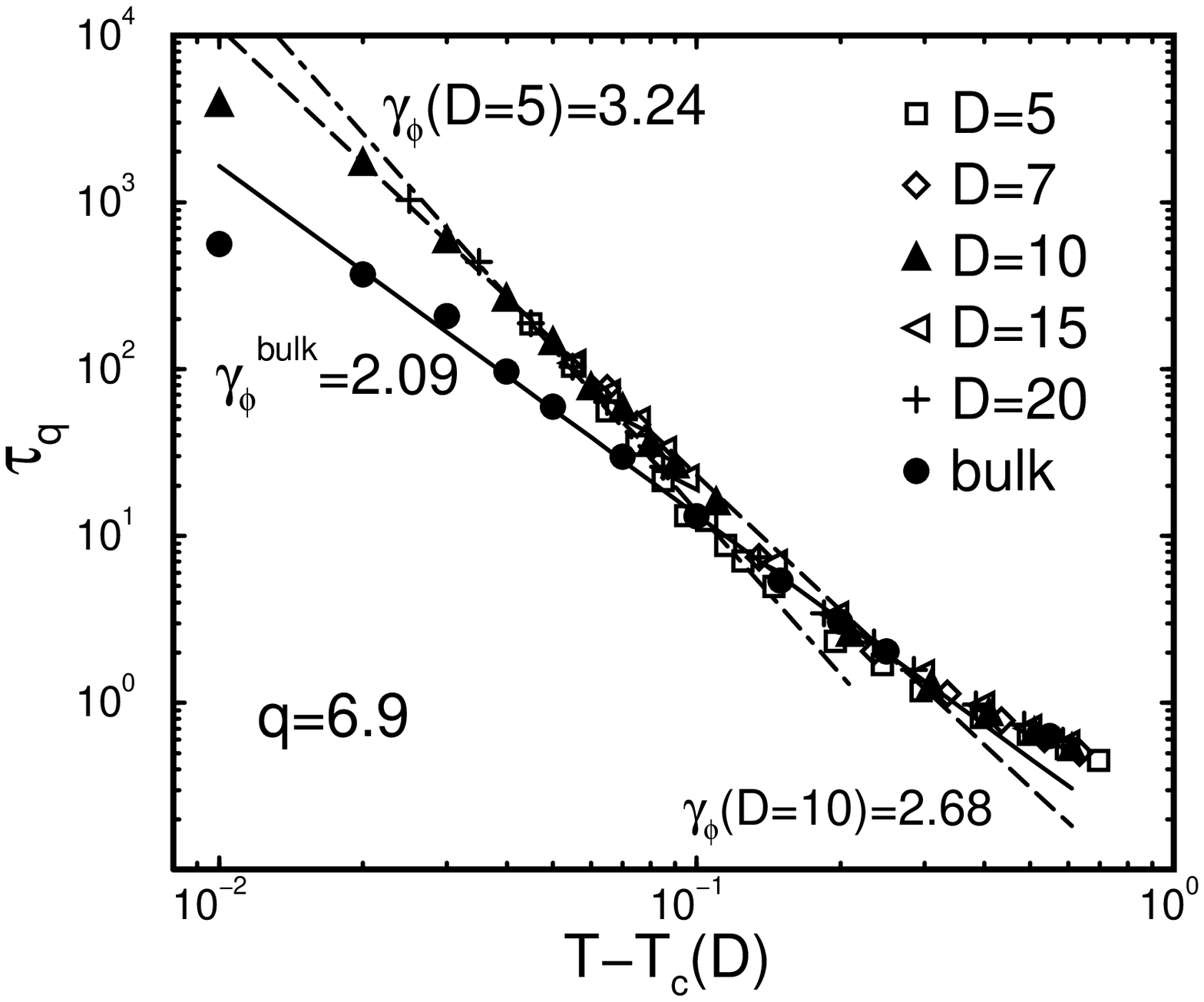}}
\epsfxsize=110mm
\centering{\epsffile{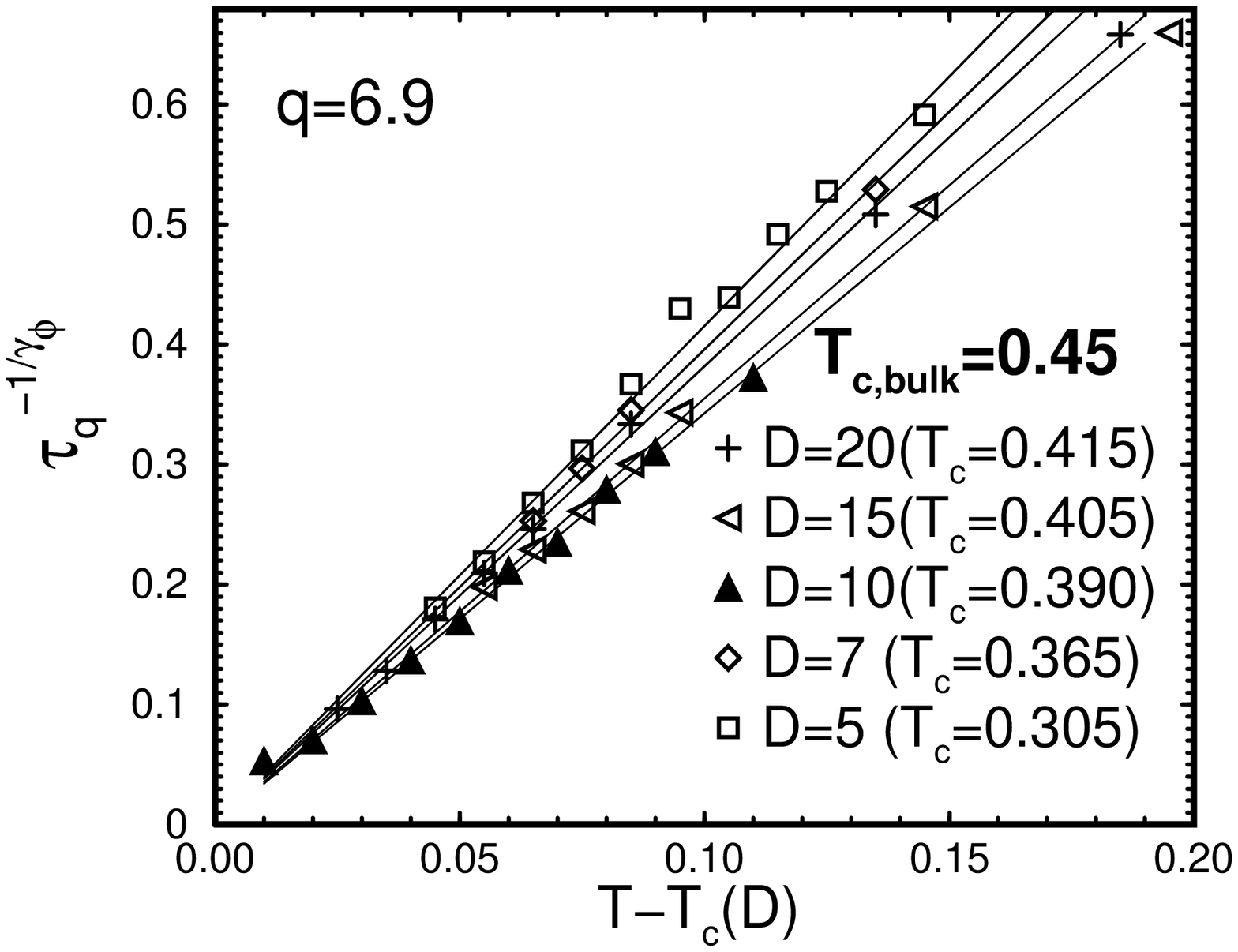}}
\caption[]{
Upper panel:
Relaxation times, $\tau_q$, at the maximum of $S(q)$ ($q\myeq 6.9)$ versus $T-\Tc$
for various film thicknesses ranging from
the smallest value studied ($D\myeq 5\myapprox 3.5\Rg$)
to the largest one ($D \myeq 20\myapprox 14 \Rg$).
$\tau_q$ was determined from the incoherent 
scattering function by requiring $\phis(t\myeq \tau_q)\myeq 0.3$,
whereas the critical temperatures, $\Tc(D)$,
were obtained from the analysis of the mean-square 
displacements, i.e.\ from power-law fits to 
$\tau$ by Eq.~(\ref{eq:tauKvsT}).  The lines represent fits of $\tau_q$ 
to Eq.~(\ref{eq:tauKvsT}) in the regime where the data are linear: $D\myeq 5$
(dash-dotted line), $D\myeq 10$ (dashed line), bulk (solid line).  The resulting
values for $\gamma_{\phi}(D)$ are indicated in the figure.

Upper panel: Using the critical exponents, 
$\gamma_{\phi}(D)$, obtained from the upper panel of this figure, 
$\tau_q^{-1/\gamma_{\phi}}$ is plotted versus $T-\Tc(D)$ for the film data.
The lines are again fits to Eq.~(\ref{eq:tauKvsT}).  The deviations from the
power law Eq.~(\ref{eq:tauKvsT}), visible for $D\myeq 10$ at $T-\Tc(D) \myeq 0.01$ in the 
upper panel, cannot be seen in this representation of the data. 
}
\label{fig:fig18}
\end{figure}
\newpage
\begin{figure}
\epsfxsize=100mm
\centering{\epsffile{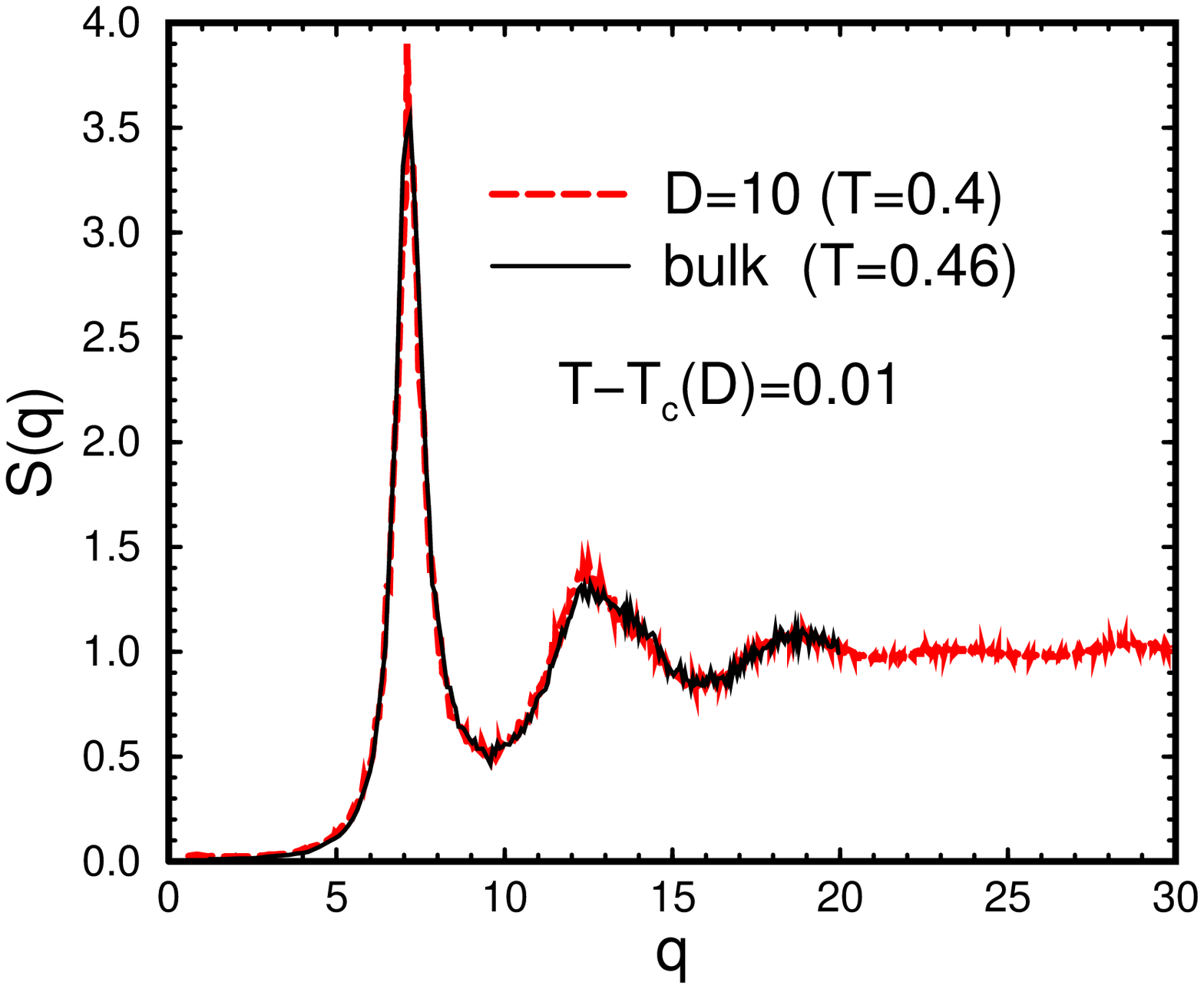}}
\epsfxsize=100mm
\centering{\epsffile{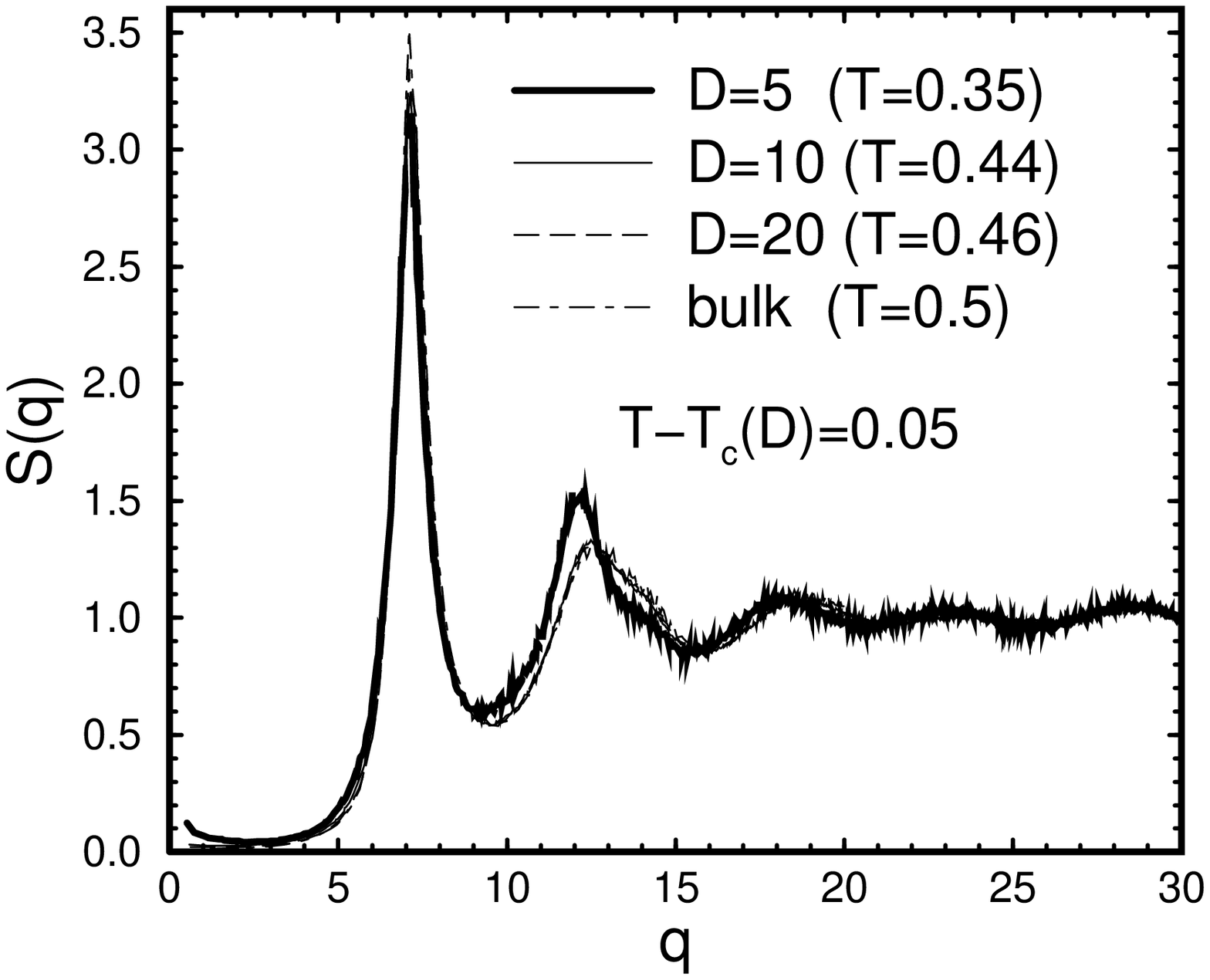}}
\caption[] {
Upper panel: Collective structure factor, $S(q)$, of the melt
for a film of thickness $D\myeq 10$
and for the bulk at temperatures $T\myeq 0.4$ 
and $T\myeq 0.46$, respectively.
Note that $\Tc(D\myeq 10)\myeq 0.39$ and
$\Tcbulk \myeq 0.45$. Thus, the chosen 
temperatures have the same distance from the 
corresponding critical temperature, i.e.\ 
$0.4 \!-\! \Tc(D\myeq 10) \myeq 0.46 \!-\!\Tcbulk\myeq 0.01$.
Except the (slightly) different amplitude of the main peak,
$S(q)$ of the film and the bulk are essentially identical when
compared for the same difference to the critical temperature.
Lower panel: Same as in the upper panel for a distance of $0.05$ 
from $\Tc(D)$, where data of many film thicknesses
are compared to that of the bulk.  Again, for film thicknesses 
$D\ge 10$, $S(q)$ of the film overlaps with that of the bulk. For 
the extreme case of $D\myeq 5$, however, deviations are not 
negligible.
}
\label{fig:fig19}
\end{figure}

\begin{thebibliography}{10}

\bibitem{Zhang::PolEngSci::1999}
X. Zhang and J. Bell, Polymer Engineering and Science {\bf 39},  119  (1999).

\bibitem{Rayss::JApplPolSci::1993}
J. Rayss, W. Podkoscielny, and J. Widomski, J. Appl. Pol. Sci. {\bf 49},  835
  (1993).

\bibitem{Armstrong::ElectrochemicaActa::1993}
R. Armstrong and D. Wright, Electrochimica Acta {\bf 38},  1799  (1993).

\bibitem{Garbassi-Morra-Occhiello::POLYMER::SURFACES}
F. Garbassi, M. Morra, and E. Ochiello, {\em Polymer Surfaces: From Physics to
  Technology} (John Wiley \& Sons, New York, 1998).

\bibitem{Laetsch-Hiraoka-Baron::MRS::385::1995}
S. L\"atsch, H. Hiraoka, and J. Bargon,  in {\em Polymer/Inorganic Interfaces
  II}, edited by L. Drzal, R. Opila, N. Peppas, and C. Schutte (Material
  Research Society, Pittsburgh, 1995), Vol.~385, p.\ 239.

\bibitem{Adam-Gibbs::JCP::43}
G. Adam and J. Gibbs, J. Chem. Phys. {\bf 43},  139  (1965).

\bibitem{Forrest::2000}
J. Forrest and J. Mattsson, Phys. Rev. E {\bf 61},  R53  (2000).

\bibitem{Forrest::PRL77::page2002}
J. Forrest, K. Dalnoki-Veress, J. Stevens, and J. Dutcher, Phys. Rev. Lett.
  {\bf 77},  2002  (1996).

\bibitem{Keddie::Jones::Cory::EuroPhysLett28}
J. Keddie, R.~L. Jones, and R. Cory, Europhys. Lett. {\bf 27},  59  (1994).

\bibitem{Keddie::FaradayDiscuss98}
J.~L. Keddie, R.~A.~L. Jones, and R.~A. Cory, Faraday Discuss. {\bf 98},  219
  (1994).

\bibitem{Forrest::PRL77::page4108}
J. Forrest, K. Dalnoki-Veress, J. Stevens, and J. Dutcher, Phys. Rev. Lett.
  {\bf 77},  4108  (1996).

\bibitem{ForrestJones2000}
J. Forrest and R. Jones,  in {\em Polymer Surfaces, Interfaces and Thin Films},
  edited by A. Karim and S. Kumar (World Scientific, Singapore, 2000), pp.\
  251--294.

\bibitem{Forrest::PRE56}
J. Forrest, K. Dalnoki-Veress, and J. Dutcher, Phys. Rev. E {\bf 56},  5705
  (1997).

\bibitem{Torres-Nealey-dePablo::PRL85}
J. Torres, P. Nealey, and J. de~Pablo, Phys. Rev. Lett. {\bf 85},  3221
  (2000).

\bibitem{deGennses::EurPhysJE2000}
P. de~Gennes, Eur. Phys. J. E {\bf 2},  201  (2000).

\bibitem{Bennemann-Paul-Binder-Duenweg::PRE57}
C. Bennemann, W. Paul, K. Binder, and B. D\"unweg, Phys.~Rev.~E {\bf 57},  843
  (1998).

\bibitem{Bennemann-Baschnagel-Paul::EPJB10}
C. Bennemann, J. Baschnagel, and W. Paul, Eur.~Phys.~J~B {\bf 10},  323
  (1999).

\bibitem{Cohen-Turnbull::JCP::31}
M.~H. Cohen and D. Turnbull, J. Chem. Phys. {\bf 31},  1164  (1959).

\bibitem{Cohen-Turnbull::JCP::34}
M.~H. Cohen and D. Turnbull, J. Chem. Phys. {\bf 34},  120  (1961).

\bibitem{Cohen-Turnbull::JCP::52}
M.~H. Cohen and D. Turnbull, J. Chem. Phys. {\bf 52},  3038  (1970).

\bibitem{Cohen-Grest::PhysRevE1979}
M. Cohen and G. Grest, Phys. Rev. B {\bf 20},  1077  (1979).

\bibitem{Goetze::LesHouches::1989}
W. G\"otze,  in {\em Les Houches 1989, Session LI}, edited by J. Hansen, D.
  Levesque, and J. Zinn-Justin (North-Holland, Amsterdam, 1989), pp.\ 287--503.

\bibitem{Goetze-Sjoegren::TransportTheoryStatPhys}
W. G\"otze and L. Sj\"ogren, Transport Theory and Statistical Physics {\bf 24},
   801  (1995).

\bibitem{Goetze-Sjoegren::RepProgPhys55}
W. G\"otze and L. Sj\"ogren, Rep. Prog. Phys. {\bf 55},  241  (1992).

\bibitem{Goetze::JPCM::1999}
W. G\"otze, J. Phys.: Condens.\ Matter {\bf 11},  A1  (1999).

\bibitem{GoetzeVoigtmann::PRE2000}
W. G\"otze and T. Voigtmann, Phys. Rev. E {\bf 61},  4133  (2000).

\bibitem{Kob::JPhysCondensedMatt::1999}
W. Kob, J. Phys.: Condens. Matter {\bf 11},  R85  (1999).

\bibitem{Bennemann-Paul-Baschnagel-Binder::JPCM11}
C. Bennemann, W. Paul, J. Baschnagel, and K. Binder, J. Phys.: Condens. Matter
  {\bf 11},  2179  (1999).

\bibitem{Bennemann-Baschnagel-Paul-Binder}
C. Bennemann, J. Baschnagel, W. Paul, and K. Binder, Comp.\ Theo.\ Poly.\ Sci.
  {\bf 9},  217  (1999).

\bibitem{Aichele-Baschnagel::EurPhysJE::I::D}
M. Aichele and J. Baschnagel, Eur. Phys. J. E {\bf 5},  229  (2001).

\bibitem{Aichele-Baschnagel::EurPhysJE::II::D}
M. Aichele and J. Baschnagel, Eur. Phys. J. E {\bf 5},  245  (2001).

\bibitem{Kremer-Grest::JCP92}
K. Kremer and G. Grest, J. Chem. Phys. {\bf 92},  5057  (1990).

\bibitem{Varnik-Baschnagel-Binder::JPhysIV10::ConfProc}
F. Varnik, J. Baschnagel, and K. Binder, J. Phys. IV {\bf 10},  239  (2000).

\bibitem{Varnik-Baschnagel-Binder::JCP113::2000}
F. Varnik, J. Baschnagel, and K. Binder, J. Chem. Phys. {\bf 113},  4444
  (2000).

\bibitem{MeyerMuellerPlathe::JCP::2001}
H. Meyer and F. M\"uller-Plate, J. Chem. Phys.  (to be published).

\bibitem{Varnik-Baschnagel-Binder::JCompPhys::2001}
F. Varnik, J. Baschnagel, and K. Binder, An Iterative Method for MD or MC
  Simulations of Planar Systems at Constant Normal Pressure and Constant Plate
  Separation, preprint.

\bibitem{Nose::JCP81}
S. Nos\'e, J. Chem. Phys. {\bf 81},  511  (1984).

\bibitem{Hoover::PhysRevA31}
W. Hoover, Phys. Rev. A {\bf 31},  1695  (1985).

\bibitem{Nose-Klein::MolPhys50}
S. Nos\'e and M. Klein, Mol. Phys. {\bf 50},  1055  (1983).

\bibitem{Hoover::PhysRevLett48}
W. Hoover, A. Ladd, and B. Moran, Phys. Rev. Lett. {\bf 48},  1818  (1982).

\bibitem{Hoover::PhysRevA34}
W. Hoover, Phys. Rev. A {\bf 34},  2499  (1986).

\bibitem{Nose::MolPhys52}
S. Nos\'e, Mol. Phys. {\bf 52},  255  (1984).

\bibitem{Varnik-Baschnagel-Binder::PRE::2001}
F.~Varnik and J.~Baschnagel, and K.~Binder, Reduction of the Glass Transition
  Temperature in Polymer Films: A Molecular-Dynamics Study, submitted to Phys.
  Rev. E (2001).

\bibitem{Varnik::Dissertation::Mainz2000}
F. Varnik, Ph.D. thesis, University of Mainz, 2000, available from
  http://ArchiMeD.uni-mainz.de/pub/2001/0007/.

\bibitem{BaschnagelBinderMilchev2000}
J. Baschnagel, K. Binder, and A. Milchev,  in {\em Polymer Surfaces, Interfaces
  and Thin Films}, edited by A. Karim and S. Kumar (World Scientific,
  Singapore, 2000), pp.\ 1--49.

\bibitem{BitsanisHadziioannou1990}
I.~A. Bitsanis and G. Hadziioannou, J. Chem. Phys. {\bf 92},  3827  (1990).

\bibitem{WangBinder1991}
J.-S. Wang and K. Binder, J. Phys. I France {\bf 1},  1583  (1991).

\bibitem{Aichele-Baschnagel::EurPhysJE::I}
M. Aichele and J. Baschnagel, Eur. Phys. J. E {\bf 5},  229  (2001).

\bibitem{Aichele-Baschnagel::EurPhysJE::II}
M. Aichele and J. Baschnagel, Eur. Phys. J. E {\bf 5},  245  (2001).

\bibitem{Rouse::JChemPhys21:1953}
P. Rouse, J. Chem. Phys. {\bf 21},  1272  (1953).

\bibitem{KobHorbachBinder1999}
W. Kob, J. Horbach, and K. Binder,  in {\em Slow dynamics in complex systems},
  edited by M. Tokuyama and I. Oppenheim (AIP Press, Woodbury, ADDRESS, 1999),
  pp.\ 441--451.

\bibitem{GleimKob2000}
T. Gleim and W. Kob, Eur.\ Phys.\ J. {\bf B 13},  83  (2000).

\bibitem{Franosch-Fuchs-Goetze-Mayr-Singh::PhysRevE55::1997}
T. Franosch {\it et~al.}, Phys. Rev. E {\bf 55},  7153  (1997).

\bibitem{Fuchs-Goetze-Mayr::PhysRevE58::1998}
M. Fuchs, W. G\"otze, and M. Mayr, Phys. Rev. E {\bf 58},  3384  (1998).

\bibitem{Dalnoki-VeressMurray}
K. Dalnoki-Veress, C. Murray, C. Gigault, and J. Dutcher, Phys. Rev. E {\bf
  63},  31801  (2001).

\bibitem{Fuchs::JNon-CrystSolids::1994}
M. Fuchs, J. Non-Cryst. Solids {\bf 172-174},  241  (1994).

\end{thebibliography}
\end{document}